%% file: ltrk_prd.tex
 \documentclass[aps,prd,twocolumn,superscriptaddress,showpacs]{revtex4}

\usepackage{graphicx}
\usepackage{mathrsfs}
\usepackage{feynmf}
\usepackage{subfigure}

\newcommand{\MeV}{\ensuremath{\mathrm{Me\kern -0.1em V}}}
\newcommand{\GeV}{\ensuremath{\mathrm{Ge\kern -0.1em V}}}
\newcommand{\GeVc}{\ensuremath{\mathrm{Ge\kern -0.1em V}/c}}
\newcommand{\GeVcc}{\ensuremath{\mathrm{Ge\kern -0.1em V}/c^2}}

\newcommand{\met}{\mbox{${\not\!\!E_T}$}}

\newcommand{\pb}{pb$^{-1}$}
\newcommand{\fb}{fb$^{-1}$}
\newcommand{\txtss}[1]{\textrm{\tiny{#1}}}
\newcommand{\masswin}[2]{\mbox{$#1~\GeV/c^2 < M < #2~\GeV/c^2$}}

\newcommand{\Zeemm}{$Z/\gamma^{\star}\rightarrow ee/\mu\mu$}
\newcommand{\Zee}{$Z/\gamma^{\star}\rightarrow ee$}
\newcommand{\Zmm}{$Z/\gamma^{\star}\rightarrow\mu\mu$}
\newcommand{\Ztt}{$Z/\gamma^{\star}\rightarrow\tau\tau$}
\newcommand{\pythia}{{\sc pythia}}
\newcommand{\alpgen}{{\sc alpgen}}
\newcommand{\herwig}{{\sc herwig}}
\newcommand{\SecVtx}{\textsc{secvtx}\ }

 \newcommand{\widetable}{\textwidth}

\begin{document}




\title{A Measurement of the $t\bar{t}$ Cross Section in $p\bar{p}$ Collisions at $\sqrt{s} = 1.96$ TeV using Dilepton Events with a Lepton plus Track Selection}



\input{July_2008_Authors_Visitors}
\date{\today}

\begin{abstract}
This paper reports a measurement of the cross section for the pair production of top quarks
in $p\bar{p}$ collisions at \mbox{$\sqrt{s} = 1.96$ TeV} at the Fermilab Tevatron.
The data was collected from the CDF Run II detector in a set of runs with a total integrated
luminosity of 1.1~\fb.  The cross section is measured in the dilepton channel,
the subset of $t\bar{t}$ events in which both top quarks decay through \mbox{$t \to W b \to \ell \nu b$}, 
where \mbox{$\ell = e$, $\mu$, or $\tau$}.  The lepton pair is reconstructed as
one identified electron or muon and one isolated track.  The use of an isolated track 
to identify the second lepton increases the $t\bar{t}$ acceptance, particularly
for the case in which one $W$ decays as $W\rightarrow\tau\nu$.  The purity of the sample 
may be further improved at the cost of a reduction in the number of signal events, 
by requiring an identified $b$-jet.  We present the results of 
measurements performed with and without the request of an identified $b$-jet. 
The former is the first published CDF result for which a
$b$-jet requirement is added to the dilepton selection.
In the CDF data there are 129 pretag lepton~+~track candidate events, of which 69 are tagged. 
With the tagging information, the sample is divided into tagged and
untagged sub-samples, and a combined cross section is calculated by maximizing a likelihood.  
The result is
\mbox{$\sigma_{t\bar{t}} = 9.6\pm1.2(\textrm{stat.})_{-0.5}^{+0.6}(\textrm{sys.})\pm0.6(\textrm{lum.})$ pb,}
assuming a branching ratio of \mbox{$BR(W\rightarrow\ell \nu) = 10.8$~\%} and a top mass 
of \mbox{$m_{t} = 175\ \GeVcc$}.  
\end{abstract}



\maketitle

\newpage
\section{\label{sec:intro}Introduction}

Top quark data collected at the Tevatron have been an active testing ground 
for the validity of the standard model since the discovery of the top quark 
in 1995 during {Run I~\cite{top_discovery_cdf,top_discovery_d0}.} 
The definitive observation at both the CDF and D\O\ experiments used 
data where one or both $W$'s from the top decays $t\to W^+b$ and $\bar{t}\to W^-\bar{b}$
decay in turn to a charged lepton and neutrino.  

This paper focuses on the dilepton channel, in which both $W$'s decay to leptons.
The final state contains two isolated charged leptons with large momentum
in the direction transverse to the beamline ($p_T$).  The two neutrinos also
carry large transverse momentum but escape the detector without interacting.
Their presence can be inferred by an imbalance in the total reconstructed 
transverse momentum in the detector, referred to as the missing transverse
energy (\met) because it is reconstructed from calorimeter information.
Only the momentum transverse to the beam can be used for this because in hadron 
collisions, the total longitudinal momentum of the system is not known in any 
one collision, and large longitudinal momentum may also be 
carried by very forward prongs which escape detection.
Combined with the jets produced by the hadronization of the $b$ 
quarks, the distinctive signature of the charged and neutral leptons allows the 
$t\bar{t}$ signature to be distinguished from the background.

The top quark is unique because of its large mass
\mbox{($m_t =173.1 \pm 0.6\ (\textrm{stat.}) \pm 1.1\ (\textrm{sys.})\ \GeVcc$~\cite{topmass}),}
which distinguishes it from the other fermions of the standard model and is more 
akin to the masses of the weak force carriers ($W$ and $Z$) and the expected mass range
for the proposed Higgs boson~\cite{the_pdg}.  
In Run II of the Tevatron, at center-of-mass energy of 1.96~TeV,
the CDF detector has collected well over ten times the amount of integrated luminosity obtained in Run I. 
Using these data, the study of the top quark sector continues, motivated by a desire to
better understand this unique corner of the standard model and to test for 
physics beyond what the model is able to describe.

\begin{figure*}[tbp]
\subfigure[]{ \includegraphics[width=0.35\textwidth]{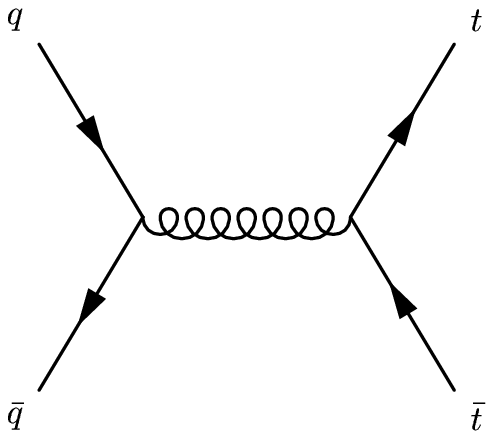}}
\subfigure[]{ \includegraphics[width=0.35\textwidth]{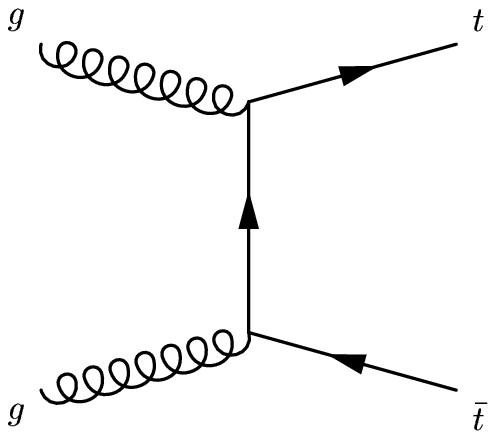}}\\
\subfigure[]{ \includegraphics[width=0.35\textwidth]{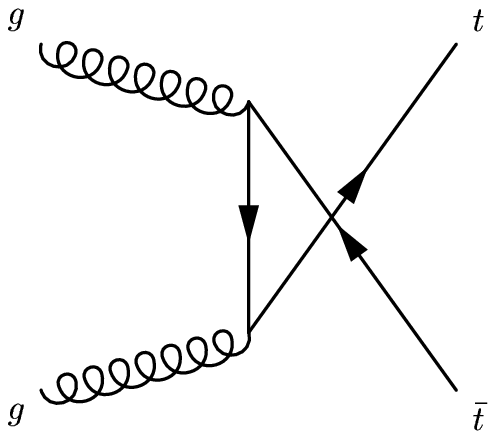}}
\subfigure[]{ \includegraphics[width=0.35\textwidth]{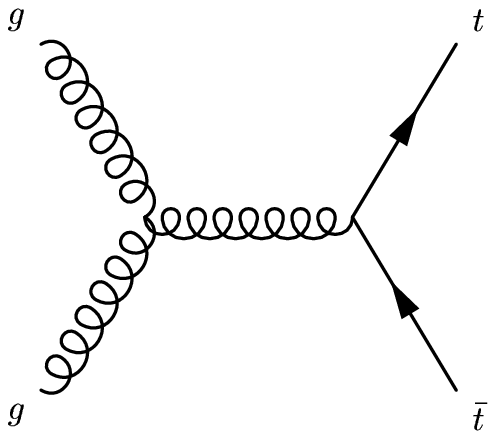}}
\vspace{0.2in}
  \caption{\label{fig:ttbar_lo}
    Leading order diagrams for top quark pair production at the Tevatron.
	}
\end{figure*}

Precise measurements of the cross section are of fundamental interest because the
top quark is one of the most recent additions to the array of particles that can
be produced in the laboratory.  The standard model predicts the production of top 
quark-antiquark pairs through the strong interaction.  The leading order Feynman
diagrams are shown in Fig.~\ref{fig:ttbar_lo}.  Approximately 85\% of $t\bar{t}$ 
pairs at the Tevatron are produced through quark-antiquark annihilation, and
the remaining 15\% through gluon-gluon fusion~\cite{ttbar_errs_cfmnr}.  Because of 
the interaction scale
involved, the cross section can be calculated using perturbative QCD techniques.
The pair production cross section has been calculated at next-to-leading order (NLO),
with the resummation of the leading logarithmic corrections due to the radiation
of soft gluons completed to next-to-leading logarithmic (NLL)
order~\cite{ttbar_nll_bcmn,ttbar_kidonakis}.  These resummations
do not change the calculated cross section by more than a few percent, but improve
the stability of the result with respect to the normalization and factorization 
scales~\cite{ttbar_errs_cfmnr}.  Recent updates to the cross section 
calculation~\cite{ttbar_new,ttbar_kidonakis_new,ttbar_moch_new} include 
newer parton distribution functions (PDFs) sets, with reduced associated uncertainties, 
and incorporate calculations of next-to-next-to-leading order (NNLO) terms.  
The predicted cross sections cited here have an accuracy of better than 10\%.  
As the accuracy of measurements improves to a comparable level, meaningful comparison 
with the theoretical prediction becomes possible.  

The measurements in this paper were completed using a reference cross section 
of $6.7_{-0.9}^{+0.7}$~pb, calculated for a top quark mass of 
\mbox{$m_t=175$~\GeVcc}~\cite{ttbar_errs_cfmnr}.  
The newer calculations give similar answers with reduced uncertainties.
For example, the similar calculation from Ref.~\cite{ttbar_new} gives
\mbox{$6.6^{+0.3}_{-0.5}$ (scale) $^{+0.4}_{-0.3}$ (PDF) pb}.  Most numbers
in this paper which depend on the theoretical calculation of the $t\bar{t}$ 
production cross section are quoted using the original reference cross section, but 
in the Results section (Sec.~\ref{sec:results}) we will compare the measured 
cross sections to the most recent predictions.

Significant deviation of the measured cross section from the predicted
value could indicate the presence of new particles or interactions.  Top quark
pair production cross section measurements can be sensitive to new physics through the 
production of a new particle or particles which then decay to top quarks. 
Examples of this include a new heavy top quark of the type predicted by ``little
Higgs'' theories~\cite{lil_higgs}, which decays to a top quark and a stable, heavy analog
to the photon, which escapes the detector, adding extra \met\ to the final state.
The resulting signature is similar to top quark decay and would enhance the measured 
cross section.  The production of $t\bar{t}$ pairs through a resonance would also 
raise the total cross section~\cite{resonance_hill,rsgluon,resonance_review}, 
although current limits on resonance production in 
the $t\bar{t}$ channel make it unlikely that it would be possible to distinguish 
the effects of a resonance on the cross section at the 
Tevatron~\cite{cdf_flame_resonance,cdf_mich_resonance,d0_resonance}.

The cross section could also be affected by a process with a 
final state sufficiently similar to the $t\bar{t}$ signature to pass the event selection.
The decay of supersymmetric particles is expected to produce multilepton, 
multijet signatures with significant missing transverse energy from the 
lightest supersymmetric particles escaping the detector~\cite{susy_nilles,susy_haber_kane}.

Finally, even in the absence of evidence of new physics at the Tevatron, a solid 
understanding of the top quark sector and the composition of the 
multilepton~+~multijet~+~\met\ sample will be a prerequisite 
to the discovery and understanding of new physics processes that may appear at the higher
energies accessible at the CERN Large Hadron Collider (LHC).  This is particularly
important because of the approximately hundredfold increase in the $t\bar{t}$ production 
cross section at the 14~TeV \mbox{center-of-mass} energy at the LHC relative to the 
Tevatron~\cite{ttbar_new}.  For example, the 
possibility of catching the decay signatures of supersymmetric partner particles
with event selection designed for top quark pairs implies the converse, that top
quark production will be an important background in searches for supersymmetry.
Searches for new physics with top quarks in the final state, motivated by models like
those referenced above which have a heavy top partner or resonance, will also rely
on thorough understanding of the top signature and associated backgrounds.


In this paper, we measure the $t\bar{t}$ production cross section 
in the dilepton channel.  The final state contains two isolated charged leptons with 
large transverse momentum, missing transverse energy from the undetected neutrinos, and
two jets from the hadronization of the $b$ quarks ($b$-jets). One or more additional jets 
may also be present, having been produced by initial or final state QCD radiation.  

The dilepton channel, because of the dual leptonic $W$ decays, has a smaller branching
ratio, about 1/9 if all $\tau$ decays are included, than the channels where one or both $W$'s
decay to quarks, which are referred to as the ``lepton~+~jets'' and ``all-hadronic'' 
channels, respectively.
The dilepton channel has the compensating advantage of a good (1:1 or better) signal 
to background ratio even without the identification of jets as possible $b$ decay
products (``tagging''), because so few
standard model processes produce two high-$p_T$ leptons and \met. 
Production of events with a $W$ and jets ($W$~+~jets) is a background
for the dilepton channel, just as it is for the lepton~+~jets channel,
but it does not overwhelm the $t\bar{t}$ signal in spite of its large cross section, 
because one of the jets must pass the lepton selection used in this analysis.
Such a misreconstructed jet is referred to a ``fake'' lepton.  
The dilepton channel also does not suffer from the same large QCD multijet background 
as does the all-hadronic channel, for similar reasons.  The background from Drell-Yan 
\mbox{($p\bar{p}\ \to Z/\gamma^{\star}+X\to \ell\ell+X$)}\footnote{In hadron-hadron
collisions, there is always the possibility of producing additional jets, photons,
or other particles in addition to the ones under direct consideration.  We will not 
explicitly write out the ``$+X$'' after this, but all final states mentioned in this 
paper are inclusive unless it is explicitly stated otherwise. }
is reduced by the requirement of multiple jets and \met. 
We also apply several criteria designed specifically to veto Drell-Yan events,
including an increased \met\ threshold when the lepton pair has a reconstructed
invariant mass close to the $Z$ resonance.  Finally, those events which may produce
both real leptons and $\met$, such as $WW$, are manageable backgrounds because 
their cross section times branching ratio is comparable to or smaller than that
for dilepton $t\bar{t}$ events after the requirement of two or more jets.

We report on two measurements of the $t\bar{t}$ cross section using 
the dilepton final state, with and without $b$-jet identification, 
using data collected between March 2002 and February 2006, corresponding to approximately 
1.1 \fb\ of integrated luminosity, using the upgraded 
Collider Detector at Fermilab (CDF~II).  These are updates 
of the previously published result in which one of leptons is reconstructed simply as an
isolated track, while the other must be identified as an electron or muon of
opposite sign~\cite{run2_dil_prl}.  The previous version did not use $b$-jet identification.  
The isolated track selection increases the acceptance by including most 
decay channels of $\tau$ leptons, thereby increasing the accessible  
branching fraction.  It also recovers acceptance for electrons or muons that 
are not within the fiducial region of the calorimetry or muon detectors.  
We will refer to the selection criteria we use (excluding the $b$-jet 
identification), and the corresponding
sample selected from the data, using the name ``lepton~+~track''.

The previous CDF publication used Run~II data corresponding to an integrated
luminosity of 200~\pb.  It included a cross section measurement in the lepton~+~track 
channel and a similar measurement where both leptons 
were fully reconstructed as electrons or muons.  The combined result was
\mbox{$7.0^{+2.4}_{-2.1}$ (stat.) $^{+1.6}_{-1.1}$ (sys.) $\pm 0.4$ (lum.) pb \cite{run2_dil_prl}}.
The D\O\ collaboration has also published a measurement in the dilepton channel using
Run~II data with an integrated luminosity of about 425~\pb.  It includes measurements 
where both leptons were fully
reconstructed as well as a measurement employing lepton~+~track and $b$-jet tagging
selection similar to that used in this analysis.  Combining the individual measurements, they find
\mbox{$7.4 \pm 1.4$ (stat.) $\pm 0.9$ (sys.) $\pm 0.5$ (lum.) pb \cite{d0_run2_dil_prd}}. 

This measurement is a substantial update of the analysis in the previous CDF publication. 
It uses more than five times the amount of integrated luminosity.  
The calculated backgrounds and associated systematic uncertainties reflect 
an improved understanding of the background composition in the lepton~+~track sample, 
with the overall systematic uncertainty decreasing from about 1.4~pb to about 0.55~pb,
i.e., from about 20\% to 6\% relative to the measured values of the cross section.
We also perform the cross section measurement using the same event 
selection, but with the added requirement that at least one jet in the event is 
$b$-tagged.  This significantly suppresses some otherwise irreducible backgrounds,
increasing the purity of the candidate sample.   The estimated signal to background ratio,
using the theoretical cross section of 6.7~pb, is about 6:1 
in the $b$-tagged sample, to be compared to about 1:1 in the pretag sample.
Finally, we divide the pretag sample into its tagged and untagged components,
in order to combine the results into a single cross section result
with smaller uncertainties than the individual measurements.

CDF and D\O\ have also measured the cross section in other $t\bar{t}$ decay modes.
In the lepton~+~jets mode, CDF has used two different methods
to identify $b$-jets.  One is based on the probability that a large number of tracks within 
a jet miss the primary vertex, and finds 
\mbox{$\sigma_{t\bar t}$ = 8.9 $^{+1.0}_{-1.0}$ (stat.) $^{+1.1}_{-1.0}$ (sys.) pb} in a data
sample with an integrated luminosity of 320~\pb~\cite{cdf_xsec_ljets_jetprob}.  
The second measurement uses the same sample, but identifies $b$ jets via a 
reconstructed secondary vertex significantly displaced from the beamline, using the 
same algorithm as is used in this paper, resulting in a cross section of
\mbox{8.7 $\pm$ 0.9 (stat.) $^{+1.1}_{-0.9}$ (sys.) pb}~\cite{cdf_xsec_ljets_secvtx}.
The D\O\ collaboration also has two recent results in the lepton~+~jets channel, both
using a data sample with an integrated luminosity of 0.9 \fb.
The first is a combined result from an analysis requiring a $b$-tagged jet and an analysis
using a kinematic likelihood discriminant, with a result of 
\mbox{$\sigma_{t\bar t}$ = 7.4 $\pm$ 0.5 (stat.) $\pm$ 0.5 (sys.) $\pm$ 0.5 (lum.) pb}~\cite{d0_xsec_ljets_likeli}.
The second is a simultaneous fit to the cross section and the relative branching ratio
$\mathcal{B}(t \to Wb)/\mathcal{B}(t \to Wq)$, where the $q$ represents any down-type quark,
resulting in a measured cross section of
\mbox{8.2 $^{+0.9}_{-0.8}$ (stat.+sys.) $\pm$ 0.5 (lum.) pb~\cite{d0_xsec_ljets_Rfit}.}
In the all-hadronic channel, both CDF and D\O\ base their measurements on events with
six or more jets, at least one of which is $b$-tagged.  The CDF collaboration applies a
neural-net-based discriminant before counting tags, and measures
\mbox{$\sigma_{t\bar t}$ = 8.3 $\pm$ 1.0 (stat.) $^{+2.0}_{-1.5}$ (sys.) $\pm$ 0.5 (lum.) pb}
in data with an integrated luminosity of 1.0 \fb~\cite{cdf_xsec_allhad}.
The D\O\ collaboration also uses a neural-net discriminant and measures
\mbox{4.5 $_{-1.9}^{+2.0}$ (stat.) $_{-1.1}^{+1.4}$ (sys.) $\pm$ 0.3 (lum.) pb}
with 0.4~\fb~\cite{d0_xsec_allhad}.
All these measurements are quoted at the reference mass of \mbox{$m_t = 175\ \GeVcc$,} and
the uncertainty on the integrated luminosity is included in the systematic uncertainty
if it is not written separately.

The cross section is determined by the number of candidate events 
$N_{\mathrm{obs}}$, the integrated luminosity $\int\mathscr{L}dt$, the acceptance
times efficiency for $t\bar{t}$ events $\mathcal{A}\epsilon$,
and the calculated number of background events $N_{\mathrm{bkg}}$.  
The acceptance $\mathcal{A}$ is defined as the fraction of $t\bar{t}$ signal events
passing the event selection, and includes the branching ratio of the
$W$ boson to a lepton pair of a particular flavor, for which we use the the measured value,
0.1080~$\pm$~0.0009~\cite{the_pdg}.  
We calculate the cross sections by maximizing the likelihood of obtaining the observed
number of candidate events given the number predicted as a function of the $t\bar{t}$
cross section, $\sigma_{t\bar{t}}$.
The number predicted, $N_\mathrm{pred}$, is the sum of the signal and background
contributions:
\begin{equation}\label{eq:npredicted}
N_\mathrm{pred} = \sigma_{t\bar{t}}\mathcal{A}\epsilon\int\mathscr{L}dt + N_\mathrm{bkg}  \mathrm{.}
\end{equation}
The uncertainties are taken from the cross section points where the logarithm of the 
likelihood decreases by 0.5, and systematic uncertainties are included as nuisance parameters
obeying Gaussian probability distributions.  The central value from the likelihood 
maximization is equal to the one obtained from the familiar formula
\begin{equation}\label{eq:xsec}
\sigma_{t\bar{t}} =
\frac{N_\mathrm{obs}-N_\mathrm{bkg}}{\mathcal{A}\epsilon\int\mathscr{L}dt} \mathrm{.}
\end{equation}
We choose to use a likelihood because it yields statistical uncertainties correctly reflecting the 
fact that the number of candidates follows a Poisson probability distribution, 
and allows extraction of a single cross section from multiple data samples.

The paper is structured as follows:  First, we briefly describe relevant features of the 
CDF~II detector (Section \ref{sec:detector}). We give details of the observed and simulated 
data samples in Section~\ref{sec:samples}. The event selection is described in 
Section \ref{sec:evtsel}, and the acceptance for that selection, including corrections,
is described in Section \ref{sec:acc}.  In Section \ref{sec:btag}, we discuss the algorithm 
to tag jets from $b$ quarks and calculate the efficiency for tagging
lepton~+~track $t\bar{t}$ events.  The background estimation methods for the pretag and tagged 
samples are described in Sections \ref{sec:pretagbg} and \ref{sec:tagbg}, respectively.  
The resulting cross section measurements, including the combination method and combined result,
are presented in Section \ref{sec:results}.


\section{\label{sec:detector}The CDF II Detector}

The CDF~II detector is described in detail elsewhere~\cite{CDFDetector}; we summarize here the
components relevant to our measurements.  We use a cylindrical coordinate system 
where $\theta$\ is the polar angle defined with respect to the proton beam, $\varphi$\ is the azimuthal
angle about the beam axis measured relative to the plane of the accelerator, 
and the pseudorapidity, $\eta$, is defined as $-\ln\tan(\theta/2)$.
Transverse energy is defined as \mbox{$E_T = E \sin \theta$}, and transverse momentum 
($p_T$) is defined similarly.

The interaction region of the detector has a Gaussian width of $\sigma_{z}$~=~29~cm.
The circular transverse cross section width is approximately 30~$\mu$m at
$z$~=~0~cm, rising to 50~$\mu$m at $z$~=~40~cm. 

\subsection{Tracking}

The charged particle tracking system of the CDF detector is contained in a solenoid 
magnet that produces a 1.4~T field coaxial with the beams, and measures the 
curvature of particle tracks in the transverse plane.  The innermost device employs silicon 
microstrip sensors, and is composed of three sub-detectors.  A single-sided 
layer of silicon sensors (L00) is installed directly onto the beryllium 
vacuum beam pipe, at an average radius of 1.5~cm~\cite{l00_nim}. 
It is followed by five concentric layers of double-sided silicon sensors (SVXII), 
located at radii between \mbox{2.5 and 10.6 cm}~\cite{svx_nim}.
The intermediate silicon layers (ISL) consist of one double-sided layer 
at a radius of 22~cm in the central region and two double-sided layers at 
radii of 20 and 28~cm in the forward regions~\cite{isl_nim}. 
Typical strip pitch in the silicon sensors is \mbox{55 - 65 $\mu$m} for axial strips,
\mbox{60 - 75 $\mu$m} for small-angle stereo strips (1.2${^{\circ}}$), 
and \mbox{125 - 145 $\mu$m} for 90$^{\circ}$ stereo strips. 
The axial-position resolution of the SVXII sensors is about 12~$\mu$m. For the 
ISL sensors, it is about 16~$\mu$m.

Surrounding the silicon sensors is the central outer tracker (COT), a 
3.1~m long open-cell cylindrical drift chamber covering radii from 
\mbox{40 to 137 cm}~\cite{cot_nim}.  
The COT has 96 measurement layers arrayed in eight alternating axial and $2^\circ$ stereo 
superlayers of 12 wires each.  The COT provides coverage for 
\mbox{$|\eta| < 1$}, and the fiducial region of the \mbox{SVXII-ISL} system extends 
out to \mbox{$|\eta| \sim 2$}.  The resolution of the combined tracker 
for tracks with \mbox{$\theta = 90^\circ$} is \mbox{$\sigma(p_T) / p_T^2$ = 0.15 \%/GeV.}

\subsection{Calorimetry}

Outside of the tracking systems and the solenoid coil are the
electromagnetic and hadronic calorimeters which measure the energy of
particles that interact electromagnetically or hadronically,
respectively.  The
central electromagnetic calorimeter (CEM) is a lead-scintillator
sampling calorimeter which covers the range $|\eta| \leq 1.1$.
The CEM has an energy resolution of 
\mbox{$13.5 \% / \sqrt{E_T}$} for electrons and photons~\cite{cem_nim}. 
The electromagnetic calorimeter in the forward regions (the ``plug'') is 
of similar design, covers the region \mbox{$1.2 \leq |\eta| \leq 3.6$}, and has
an energy resolution of \mbox{$(16 \% / \sqrt{E_T}) \oplus 1 \%$}~\cite{pem_nim}.

Crucial to electron and photon identification are the shower 
maximum detectors, placed at a depth of about six radiation
lengths in the electromagnetic calorimeter.  
The shower maximum detectors allow detailed measurement, in the plane
approximately transverse to the incident particle direction, of the shower shape
at the expected peak of its development. 
The precision two-dimensional position measurements are made by orthogonal
wire proportional chambers and resistive strips in the central calorimeter, and
stereo layers of scintillator in the plug calorimeter~\cite{ces_nim,pes_nim}.

The hadronic calorimeter is an iron-scintillator sampling calorimeter, and is
between 4.5 and 7 interaction lengths deep, depending on the pseudorapidity.  
It surrounds the electromagnetic calorimeter and is divided into three sections:
the central section covers \mbox{$|\eta| \lesssim 0.8$}, the forward (plug) section
covers \mbox{$1.2 \leq |\eta| \leq 3.6$}, and the ``wall'' section covers the 
intermediate range.   

The entire calorimetry system covers the pseudorapidity range $|\eta| < 3.6$.
All calorimeters are segmented into projective towers which 
point at the nominal center of the interaction region.

\subsection{Muon Detectors}

In the pseudorapidity range \mbox{$|\eta| \lesssim 0.6$}, two sets of planar
drift chambers are used to identify muons.  The inner layer (the CMU, for 
``Central Muon'') is located just outside of the central hadron calorimeter towers.
The outer layer (the CMP, for ``Central Muon Upgrade'') is also instrumented 
with scintillation counters for trigger and timing information. 
The CMP has a square profile and lies outside the 
CMU, behind an additional 60~cm of iron shielding.
Muons in the region \mbox{$0.6 < |\eta| < 1.0$}
are detected with the Central Muon Extension (CMX), a layer of
drift chambers between layers of scintillator counters.  The geometry of
the CMX is that of a pair of truncated cones, opening from the interaction point at 
the center of the detector.  The CDF muon system is described in more detail in 
\mbox{Refs.~\cite{muon_nim1} and \cite{muon_nim2}}.
By convention, muons are named according to the muon detector in which they are reconstructed. 
A CMUP muon has a track segment (``stub'') in both the CMU and CMP detectors.

\subsection{Online Event Selection (Trigger)}

The 2.5 MHz nominal bunch crossing-rate of the Tevatron far exceeds the rate at which data
can be written to permanent storage (75 Hz).  CDF uses a three-level trigger system to select 
a subset of the events to record~\cite{level12_trig,level3_trig}.  
Each successive level of processing reduces the event rate and 
refines the criteria used for event selection.  

The first level, is implemented entirely through custom hardware.
It uses information from the calorimeter, the axial layers of the COT, and the muon 
detectors to quickly reconstruct simple objects.  Tracks are built from COT
axial hits using a predefined set of patterns, and electron and muon candidates
are built from tracks matched to energetic towers in the electromagnetic calorimeter and 
hit segments in the muon detectors, respectively.  

Level 1 accepts events and passes them to the next level of processing, level 2, 
at a rate of up to 50 kHz.  Level 2, also built of custom hardware, 
performs further reconstruction.  In particular, clustering of calorimeter 
towers is performed, for photon, electron, and jet identification.  

Events satisfying level 2 criteria are passed to level 3, where they are 
directed to one of about 300 dual-processor Linux computers.  Level 3 applies
the full event reconstruction, using the same software that is used for offline analysis,
including the application of preliminary calibration constants.  This allows more stringent event
selection to be made, improving background rejection while maintaining efficiency for signal.   
Selected events are written to tape for offline analysis.

\subsection{Luminosity Determination}

Luminosity is measured at CDF by a pair of conical Cerenkov detectors 
surrounding the beam pipe, at \mbox{$3.7 < |\eta| < 4.7$}, on each side of the interaction 
region.  Each detector contains 48 smaller mylar cones filled with 
isobutane at about 1.5 atmospheres of pressure.  Photomultiplier tubes at the large 
ends of the mylar cones collect Cerenkov light produced by particles emerging from inelastic $p\bar{p}$ 
scattering.  The mean number of interactions per beam crossing is inferred from the number
of interactions in which no particles are observed in either of the detectors (the
``zero-counting method'').  The instantaneous luminosity is calculated from the mean 
number of interactions, the total inelastic $p\bar{p}$
cross section, and the bunch crossing rate.  The uncertainties on the luminosity are 
from the understanding of the acceptance for the detectors as well as the 4\% uncertainty 
on the value of the total $p\bar{p}$ cross section.  The combined uncertainty of 6\%
contributes to the total uncertainty on the cross section.



\section{\label{sec:samples}Collision Data and Monte Carlo Samples}

We measure the $t\bar{t}$ cross section in the subset of the $p\bar{p}$ collision 
data which appear to have at least one high-$p_T$ lepton, as determined by the trigger
system.  To quantify the signal acceptance, we use a sample of $t\bar{t}$ events which have
been simulated using Monte Carlo algorithms.  Numerous other observed and simulated data samples
are needed to refine the estimated acceptance and estimate the background in the 
lepton~+~track sample.  In this section we describe the various
samples used in this measurement.  

\subsection{Data Quality Requirements}

Because the lepton~+~track event selection relies on many detector subsystems for the 
reconstruction of electrons, muons, tracks, jets, and \met, as well as the measurement
of the luminosity, we use only the CDF data in which all of the relevant parts of the 
detector -- the calorimetry, tracking, shower maximum, muon, and luminosity detectors -- are 
fully operational.  For the measurement requiring a $b$-tag, 
we also require the silicon tracking detector to be functioning because high-precision
position measurements are necessary for the reconstruction of a displaced 
secondary vertex.  
The integrated luminosity of the data sample including information from the silicon 
detector is 1000~$\pm$~60~\pb.  For the pretag measurement, we include an additional
$70$~\pb\ which has no silicon information but which is otherwise acceptable.
For these data, PHX (forward) electrons cannot be reconstructed and some tracking 
requirements are changed, as will be specified in Section~\ref{sec:evtsel}. 

\subsection{Data Samples}

We select lepton~+~track $t\bar{t}$ candidates from events passing the high-$p_T$ lepton triggers.  
There are high-$p_T$ central and forward electron triggers, as well as triggers for both the
CMUP and CMX regions of the muon detectors.
The central electron trigger selects events containing a cluster with transverse energy greater than 18~$\GeV$ in the central
electromagnetic calorimeter and a matched track with \mbox{$p_T > 9\ \GeVc$}.  Track matching is not
available online for forward electrons. To reduce the background trigger rate from jets,
the electron candidate $E_T$ threshold is raised to 20~$\GeV$, 
and the events are required to have at least 15~$\GeV$ of \met. These requirements 
maintain efficiency for selecting electrons from $W$ decays, where the mean neutrino
$p_T$ is above 20~$\GeVc$.
Both electron triggers require the ratio of the energy in the hadronic calorimeter to the
energy in the electromagnetic calorimeter to be less than 0.125 in order to reject hadronic jets.
There are separate triggers for CMUP and CMX muons.  Each requires a track with 
\mbox{$p_T > 20\ \GeVc$} to be matched with a muon track segment (``stub'') in the relevant detector(s).

Most of the data samples used in this measurement are derived from the 
set of events passing the high-$p_T$ lepton triggers.
This includes the $Z$ events used to study lepton identification and the modeling of 
jet production by QCD radiation, as well as the $W$~+~jets sample used in the 
calculation of the background from events with a fake lepton and the $Z$~+~jets 
sample used in the calculation of the background from \Zeemm~+~jets events.

To estimate the background from events with a fake lepton, we need a sample with
a large number of jets.  We use the events passing a photon trigger with
a transverse energy threshold of 25~\GeV.  


\subsection{Monte Carlo Samples}

To calculate the acceptance of the lepton~+~track selection for the $t\bar{t}$ signal, we 
apply the event selection to a sample of simulated $t\bar{t}$ events generated using 
\pythia\ version 6.216~\cite{pythia_gen} for event generation and parton showering. 
The leptonic branching fraction for the $W$ boson is set to the measured value of 
0.1080~$\pm$~0.0009~\cite{the_pdg}.  For the central value of the cross section, we use 
a sample generated with a top mass of \mbox{$m_t = 175\ \GeVcc$.}  Identical samples generated
at other values of the top quark mass are used to recalculate the cross section at
those mass points.  We
also use a sample of $t\bar{t}$ events generated using \herwig\ version 6.510~\cite{herwig_gen}
to check the dependence of the calculated acceptance on the event generator.  

To estimate the contribution of backgrounds to the lepton~+~track sample, 
we use other Monte Carlo samples, which will be described in the relevant sections.
Most of them are generated using \pythia, in the same version as the signal.  For some studies, 
we use a $W$~+~jets sample with matrix elements calculated by 
\alpgen\ version~$2.10^\prime$~\cite{alpgen_gen} and \pythia\ used for parton showering.  

In the Monte Carlo samples in this paper, we use the \textsc{cteq5l} parton distribution 
functions to model the momentum distribution of the initial state partons~\cite{cteq_pdf}.
The interactions of particles with the detector are modeled using \textsc{geant} version 3~\cite{geant}, 
using the \textsc{gflash} parameterization~\cite{gflash} for showers in the calorimeter.
Details on the implementation and tuning of the CDF detector simulation may be found
in Ref.~\cite{cdfsim}.  


\section{\label{sec:evtsel}Lepton + Track Event Selection}

The lepton~+~track sample is drawn from the set of events with one or more fully 
reconstructed electron or muon candidates and at least one isolated track which is 
distinct from the first lepton and has the opposite sign.
We also require candidate events to have significant missing transverse energy (\met), a key 
discriminant between the $t\bar{t}$ signal and 
backgrounds, particularly Drell-Yan events where the final state
leptons are electrons or muons. The \met\ in such events is generally the result of
mismeasurement of the energies of leptons or jets and the resulting distribution falls 
off rapidly with increasing \met.  For this reason, we make a 
series of corrections to the \met\ and place restrictions on the final-state kinematics 
to reduce residual contributions from such events.  

The requirement that the isolated track has the opposite charge of the fully reconstructed
lepton candidate reduces the contribution from events where, due to a fluctuation of 
fragmentation and hadronization, a jet has reproduced the signature
of a lepton candidate.  This requirement is nearly 
100\% efficient for the signal and all other backgrounds, but only 61\%
efficient for the background from events with jets producing a lepton-like signature.

Finally, we require events to have two or more jets.  The $t\bar{t}$ signal contains two 
$b$ jets at leading order, while the cross sections of the backgrounds are
significantly reduced by requiring two or more jets in the final state.

\subsection{Electron Selection}

The electron and muon identification criteria used in this analysis are
very similar to those described in Ref.~\cite{CDFDetector}.  
Electron selection is based on a reconstructed track, energy deposition 
in the electromagnetic calorimeter, and the quality of the match between the track and the 
energy signature in the calorimeter.  This analysis uses two classes of electrons.  
Central (``CEM'') electrons, in the range $|\eta| \lesssim 1.1$, have tracks in the 
central tracker and deposit their energy in the central electromagnetic calorimeter.  
Forward (``PHX'') electrons are identified in the range $1.2 \lesssim |\eta| \lesssim 2.0$, and
deposit their energy in the plug electromagnetic calorimeter.  Forward 
electrons have tracks that use information from the silicon tracker,
and derive their abbreviated name ``PHX'' from 
``Phoenix'', the name of the tracking algorithm~\cite{wassym_phx}.

\subsubsection{Calorimeter Requirements}
First, the calorimeter cluster of the electron must have $E_T > 20$~\GeV,
calculated after the electron energy has been corrected for calorimeter nonuniformities
and the absolute energy scale.
The cluster must also be isolated, in the sense that the total energy in the 
towers in a cone surrounding the tower containing the candidate electron shower is 
required to be less than 10\% of the candidate electron energy.  
The cone is defined to include objects within 
\mbox{$\Delta R = \sqrt{\Delta\eta^2 + \Delta\varphi^2} < 0.4$} around the candidate,
but the towers in the electron cluster are excluded.
The distribution of energy between the towers in the cluster and the shape of the 
shower in the shower maximum detector are required to be consistent with expectation
as determined, for instance, in test beam and studies of electrons from $W$ and $Z$ decays.
Finally, the amount of energy deposited in the hadronic part of the calorimeter 
must be significantly less than the amount deposited in the electromagnetic part.
For central electrons, we require that the energy in the hadronic calorimeter be less than
5.5\% of the energy in the electromagnetic calorimeter, with a small energy-dependent
correction to allow for the fact that showers from more energetic electrons extend 
farther into the hadronic calorimeter.  For plug electrons, we require that the 
energy in the hadronic calorimeter be less than 5\% of the energy in the electromagnetic
calorimeter.

\subsubsection{Track Reconstruction and Requirements}

Central electron candidate tracks are three-dimensional helices reconstructed
from COT hit information.  
If there are silicon hits in the path of the track through the silicon tracking 
system, the hits are added and the track is refitted.  This makes the measurement of
track parameters more precise, but we do not require silicon hits, to maintain
efficiency and allow use of data where the silicon tracking detector was not in use.
Candidate tracks must have at least three axial and two stereo 
segments in the COT, where each segment is a set of at least five of twelve 
possible hits contained in a single superlayer.  

Forward electron candidate tracks are reconstructed in the 
silicon tracker.  The track reconstruction algorithm builds seed track helices
from plug calorimeter information, taking a point from the shower maximum cluster
centroid and another from the interaction vertex. The curvature is estimated by equating the
momentum to the energy in the 
calorimeter.  This yields two track hypotheses, one for each choice of sign.
A road-based search algorithm attempts to attach silicon hits to each of the track
hypotheses, and helices with attached hits are refit for a more precise measurement
of the track parameters.  The track fit is considered successful if three or more silicon 
hits are attached and the fit has a $\chi^2$ per degree of freedom less than 10.
If there are multiple tracks found for an electron candidate, the one
with the best fit quality, as measured using the $\chi^2$ per degree of freedom, is taken.   
For both central and plug electron candidates,
we require the track to originate from a point along the beam line that is less than
60~cm from the nominal center of the detector (\mbox{$|z_0| <$ 60 cm}).  

\subsubsection{Conversion veto}
Central electrons may be flagged as having originated from a photon conversion 
if there is a second track near to the electron track with opposite sign.  
We do not use central electrons which have been flagged as 
conversions.  There is no explicit conversion veto for forward electrons, but the
silicon tracking algorithm suppresses tracks from conversions.  The algorithm
creates a track hypothesis assuming that the electron track is prompt and has 
momentum equal to the energy in the calorimeter, but these assumptions are wrong
for most conversion electrons.  Silicon hits from conversion electron tracks
will not generally be close enough to the track hypothesis to be attached, and 
the track finding fails.

\subsection{Muons}

Muon candidates are defined as a track in the COT with \mbox{$p_T > 20\ \GeVc$} matched to a track 
segment in one or more of the muon drift chambers.  We require either a 
stub in both the CMU and CMP detectors, or a stub in the CMX detector, and refer to 
the resulting muon candidates as CMUP or CMX muons, respectively.  
Requiring muon signatures in both the CMU and 
CMP detectors reduces the probability of reconstructing a muon from a hadron that reaches
the CMU as a result of a particle shower that is not fully contained in the 
hadronic calorimeter.

\subsubsection{Calorimeter Signature}
The energy deposited in the region of the calorimeter intersected by the 
candidate muon track is required to be consistent with the expectation
for a minimum-ionizing particle.  Specifically, there must be no more than
2~\GeV\ in the electromagnetic calorimeter and 6~\GeV\ in the hadronic calorimeter,
with a small correction for muons with momentum over 100~$\GeVc$ to allow for
the expected rise in ionization. We also require muon candidates to be 
isolated in the sense that the total sum $E_T$ in the calorimeter towers
in a cone of $\Delta R < 0.4$ around the one intersected
by the extrapolated muon track is less than 10\% of the muon $p_T$.
 
\subsubsection{Tracking Requirements}
Muon candidates use the same tracks and track quality requirements as central 
electron candidates.  We make a few additions to the quality requirements from muons,
motivated by backgrounds particular to muons, such as cosmic rays and kaon decays-in-flight.  
In addition to the COT track and $z_0$ requirement, the candidate track must
have a small impact parameter ($d_0$).  The impact parameter is the two-dimensional
distance, in the plane transverse to the beam direction, between the beamline and 
the point of closest approach of the track helix to the beamline.  We require that
the impact parameter for muon tracks be less than 20 (200) $\mu$m for tracks with (without) attached
silicon hits.  We also require that the $\chi^2$, given the number of degrees of freedom 
in the track fit (i.e., the number of hits on the track minus the number of fit parameters)
is such that the probability to have found a larger $\chi^2$ for that track by chance is 
greater than $10^{-8}$.  This in essence requires that the track be well reconstructed.
It is similar in spirit to a requirement that the $\chi^2$ or
$\chi^2$ per degree of freedom be less than a specified value, 
but it removes the dependence of the efficiency for good tracks on
the number of degrees of freedom.

\subsubsection{Track-Stub Matching}
We check the quality of the spatial match between the COT track and the muon stub(s). 
The quantity used is the distance between the track stub in the muon detectors 
and the point at which the extrapolated COT track crosses the front plane of the corresponding
detector element.  The distance is measured in the plane of the muon detector, transverse 
to the measurement wires.  A CMUP muon track must extrapolate to within 7~cm of the 
CMU stub and within 5~cm of the CMP stub.  For CMX, the maximum allowed displacement is 6~cm.

\subsection{Track Lepton Selection}\label{sec:tracksel}

We use an isolated high-$p_T$ track to identify the second lepton in the event.  To 
qualify, the tracks must have \mbox{$p_T >$ 20 $\GeVc$}, pass certain quality requirements, and be 
isolated in $\eta-\varphi$ space from other energetic track activity.
The track may be left by either a charged lepton or a charged hadron from the 
decay of a $\tau$ lepton, but it is in either case indicative of the presence of a lepton.  
The isolated track, in this role, is also referred to as a ``track lepton'' because its 
identification relies entirely on information from the tracking detectors, and also to 
distinguish it from the fully reconstructed electron and muon candidates.
The added acceptance for $t\bar{t}$ signal events where one and sometimes even both 
$W$'s have decayed to a $\tau$ and $\nu_{\tau}$ is discussed in more detail in Sec.~\ref{sec:acc}.

\subsubsection{Track quality}
As is the case for muons, it is important for the track to be well-measured,
both to reject background and because the track momentum is the only measure of 
the particle's energy.  The track must have at least 24 hits in the axial layers 
of the COT and at least 20 hits in the stereo layers and satisfy the same $\chi^2$
probability requirement as muons.  
The requirement of a minimum number of track hits limits the acceptance for 
track leptons to $|\eta| \lesssim 1.15$, according to the geometry of the COT.
There is also a maximum allowed impact parameter, 0.025~cm,
but unlike the muon case, 
the requirement is independent of the presence of silicon hits.  
We also require silicon hits to be present if they are expected, to reduce the 
incidence of fake tracks reconstructed from accidental combinations of hits.
Specifically, if the track passes through three or more layers of the 
silicon tracker known to be 
functional, it must have at least three silicon hits attached.  

\subsubsection{Track isolation}
Track isolation is crucial to the rejection of backgrounds from jets.
We sum the $p_T$ of every track with \mbox{$p_T > 0.5~\GeVc$}, including the candidate track,
within a cone of \mbox{$\Delta R < 0.4$} around the candidate.
The ratio of the candidate track $p_T$ to the sum 
$p_T$ in the cone is required to be at least 0.9.  To be included in the $p_T$ sum, tracks must 
pass quality requirements similar to, but less stringent than, those for the track lepton.
No $\chi^2$ probability or impact parameter restrictions are made, and only 20 axial 
and 16 stereo hits are required.

\subsection{Jet Definition}\label{ss:jet_defn}

Jet reconstruction is based on a calorimeter tower clustering cone algorithm 
with a cone size of $\Delta R = 0.4$.  
Towers corresponding to identified electrons according to the definition 
above are removed before clustering.  The $E_T$ values of the jets are 
corrected for the effects of jet fragmentation, calorimeter non-uniformities 
and the calorimeter absolute energy scale~\cite{jetenergies}.  

We extend the jet definition to facilitate the calculation of the rate for 
a jet to be reconstructed as an isolated track, and use this jet definition 
everywhere in the analysis for consistency.  
The details of the fake lepton background calculation are described in 
Section~\ref{sec:fakes}, but the core idea is to ensure that any object which could 
be identified as a track lepton is included in the jet collection, because 
that jet collection forms the denominator of the measured probability for
an object of hadronic origin to be identified as a lepton.

This requires modification of the jet definition.  For each track
passing all of the track lepton requirements, but ignoring the isolation requirement,
we check whether it is within  
\mbox{$\Delta R = 0.4$} of the axis of a jet.  Here, we consider all jets from 
the cone algorithm with \mbox{$E_T > 10$~\GeV}.  If the track is
not matched, we add it to the jet collection.  If it is matched, we check whether the 
$p_T$ of the track exceeds the corrected $E_T$ of the jet.  If it does, 
we substitute the kinematic information of the track for the kinematic 
information for the jet.
If the $p_T$ of the track is less than the corrected $E_T$ of the jet, 
we leave the jet kinematic information as is.  
Inclusion of track information in this manner ensures counting of the products
of parton fragmentation where most of the momentum is carried by a single 
charged particle which does not deposit all of its energy in the hadronic calorimeter.  
In extreme cases, jet energy corrections will not account for all of the 
unmeasured energy and the track momentum is the best measure of the parton
energy.

The final jet collection thus includes standard jets clustered with a cone 
size of 0.4, jets with kinematic information from tracks, and unaffiliated tracks.  
For event selection we count the number of jets with $E_T > 20$~\GeV\ and 
$|\eta| < 2.0$, excluding those jets within $\Delta R \leq 0.4$ 
of either the lepton candidate or the isolated track.  When making a $W$~+~jets selection,
such as is used in the fake lepton background estimates for both the pretag
and tagged samples, only the fully reconstructed lepton is excluded 
from the jet counting.

\subsection{Missing Transverse Energy Reconstruction}

\begin{figure}[tbp]
\centering
\includegraphics[width=0.5\textwidth]{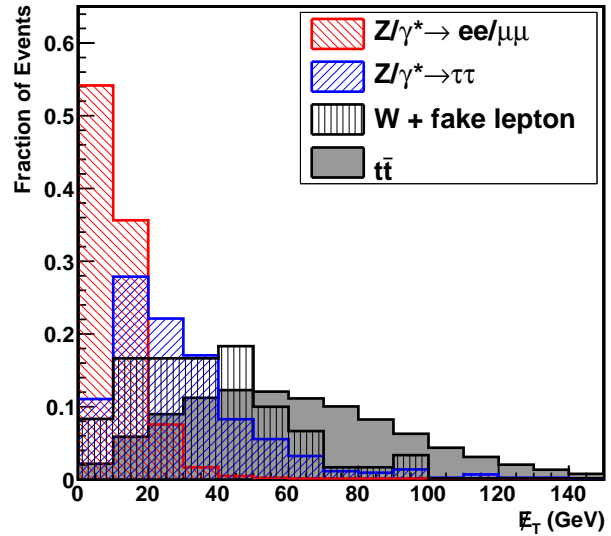}
\caption{\label{fig:basic_met}
  Corrected missing transverse energy from several lepton~+~track backgrounds, compared to
    the $t\bar{t}$ signal.  The distribution for $WW$ events is similar to the $t\bar{t}$ 
    distribution but slightly softer, and is omitted for legibility.  
    Candidate events are required to have a fully 
    reconstructed lepton and track and two or more jets, but no other event selection is applied. 
    For the cross section measurements, we require events to have $\met > 25$~\GeV.
    Distributions taken from events generated with \pythia\ and normalized to have unit
    area.
}
\end{figure}

A transverse momentum imbalance in the detector indicates that particles 
have exited the detector without interacting.  Dilepton $t\bar{t}$ events
have two high-$p_T$ neutrinos in the final state, leading to a considerable
amount of \met\ in signal events.  Figure~\ref{fig:basic_met} shows the simulated \met\ 
distributions for the $t\bar{t}$ signal and some of the backgrounds.  
Comparison of these distributions shows that using a \met\ threshold to 
select events reduces the contribution from many of the backgrounds considered, particularly
Drell-Yan events, but is quite efficient for $t\bar{t}$ events.

The missing transverse energy is defined as:
\begin{equation}
\not\!\! \vec{E}_T = - \sum_{i} E_T^i \hat{n}_i, 
\end{equation} 
where $i$ is an index that runs over all calorimeter towers with $|\eta| < 3.6$ and
$\hat{n}_i$ is a unit vector perpendicular to the beam 
axis and pointing at the i$^{th}$ calorimeter tower. 
The scalar, \met, is then defined as $\not\!\!E_T=|{\not\!\!\vec{E}_T}|$. 

Some care must be taken with the \met\ calculation, because
a transverse momentum imbalance can also be generated by incorrect measurements 
of objects in the event and the energy resolution of the calorimeter towers 
themselves.  To reduce the inclusion of events where the \met\ is produced by 
energy mismeasurements, the \met\ is adjusted
in those cases where the calorimeter information is not the best 
measure of an object's energy.

\subsubsection{Muon Correction}
Muons are minimum-ionizing particles and deposit very little energy in the 
calorimeter. Thus, if the fully reconstructed lepton in the event is a muon 
(CMUP or CMX), we subtract the transverse components of the muon momentum 
from the corresponding components of the \met.  No correction is made to 
the calorimeter energy for the small amount of energy deposited by the muon.

\subsubsection{Track Correction}
We also correct for all tracks (excepting the fully reconstructed lepton if it is a muon)
pointing at a 3 by 3 block of calorimeter towers where the $E_T$ measured is less 
than 70\% of the $p_T$ of the track.  
All tracks with \mbox{$p_T > 10\ \GeVc$}, \mbox{$|d_0| < 250 \mu$m,} at least 24
(20) hits on the axial (stereo) wires of the COT, and appearing to come from the 
same interaction vertex (\mbox{$|\Delta z_0| < 5$ cm}) as the primary lepton, 
are considered.  
The 70\% threshold excludes normal fluctuations in the
energy and momentum resolution, so that we correct only for tracks where the energy
deposit measured in the calorimeter is clearly not consistent with the momentum
measured in the tracker.  This correction accounts for tracks pointing at cracks in 
the calorimeter, minimum-ionizing particles such as muons,
and cases where showers produced by hadronic particles in the calorimeters
have unusually low light yields.  

\subsubsection{Jet Correction}
We also correct the \met\ for the jet energy calibrations by subtracting the 
difference between the corrected and uncorrected jet energies.  By doing this we
use the best estimate of the 
energies of those objects which are identifiable as jets.  Jets with corrected
\mbox{$E_T > 10\ \GeV$} and \mbox{$|\eta| < 2.0$} are included, except for objects 
in the jet 
collection which are tracks or have had their kinematic information replaced by 
that of an associated track.  These will have already been 
accounted for in the track correction.

\subsection{Event Selection}

\subsubsection {Basic event selection}

Having defined our basic analysis objects, we can select events with features
typical of $t\bar{t}$ dilepton events.  
First, there must be at least one fully reconstructed electron or muon
with \mbox{$p_T > 20\ \GeVc$} in the event.  
Once a primary lepton is identified, we take as the track lepton the 
highest $p_T$ isolated track with \mbox{$p_T > 20\ \GeVc$.}  To qualify, the track must 
appear to be from the same 
interaction vertex as the primary lepton, (\mbox{$|\Delta z_0| < 5$ cm}). 
If there is no such isolated track, we try the event selection again with the 
next fully reconstructed lepton, if another has been identified. 
The leptons are considered in the following order: central (CEM) 
electrons, CMUP muons, CMX muons, and finally forward (PHX) electrons.  Within a 
particular lepton type, the leptons are tested in order of descending $E_T$ or $p_T$.  
In the CDF data, for a lepton of a particular type to be considered, 
the trigger corresponding to
that category must have fired for that event, and the relevant parts of the 
detector must be known to be fully functional at the time the event occurred.  

After a track lepton is found, we
correct the \met, and require the corrected \met\ to be greater than 25~$\GeV$.  
Each fully reconstructed lepton in an event is considered 
in turn until the event has passed all of the selection criteria, or has failed 
them for all leptons. 

\subsubsection{$\Delta \varphi$ Requirements}

Drell-Yan events may appear to have \met\, in spite of the absence of neutrinos in the
final state.  If the energy of
one lepton or jet is measured incorrectly, false \met\ appears,
pointing along or opposite to the direction of that object.  
For this reason we require that no lepton
or jet in the event be pointing directly at the \met.  The requirements for different
objects are determined by their respective angular size and potential for mismeasurement.
Studies of Drell-Yan events in simulation show that although \met\ may be generated
either pointing near or away from a track lepton, most \met\ associated with fully
reconstructed leptons or jets is pointing in the same direction as the lepton or
jet.  These studies also show that it is uncommon for \met\ associated with a jet
to exceed 50~\GeV.  Therefore, we veto events where the primary lepton points within 
$5^\circ$ of the \met\ or 
the track lepton is within $5^\circ$  of parallel or anti-parallel to the \met.  
Also, all jets in the event must be more than $25^\circ$ away from the direction of 
the \met, unless the event has \met\ $> 50$~\GeV.  

\subsubsection{$Z$ Boson Veto}

To further reduce background from Drell-Yan events,  
the \met\ threshold is raised to 40~\GeV\ if the invariant mass of the lepton~+~track 
pair is in the range of the $Z$ boson resonance (\masswin{76}{106}). 
This requirement is referred to as the ``$Z$ veto''.

\subsubsection{Candidate Events}

For the cross section measurements we count events with at least two jets with 
corrected \mbox{$E_T > 20\ \GeV$} and \mbox{$|\eta| < 2.0$}.  The jets used for this are the 
extended collection described in Sec.~\ref{ss:jet_defn}, which is based on a calorimeter clustering 
algorithm with a cone size of $\Delta R < 0.4$.  Any jets within $\Delta R < 0.4$
of either the lepton candidate or the isolated track are excluded from the jet counting.
The final requirement is that the fully reconstructed lepton candidate and the 
track lepton candidate have opposite sign.

Applying this selection to 1.1 \fb\ of CDF Run II data, we find 129 pretag 
lepton~+~track $t\bar{t}$ candidate events with two or more jets.


\section{\label{sec:acc}$t\bar{t}$ Dilepton Acceptance}

We determine the geometric and kinematic acceptance for $t\bar{t}$ dilepton events by 
applying the lepton~+~track event selection to the \pythia\ $t\bar{t}$
sample described in Section~\ref{sec:samples}.
The acceptance is defined as the number of simulated $t\bar{t}$ 
events passing the selection criteria, divided by the total number of 
$t\bar{t}$ events in the sample.  To be included in the numerator, the event must
be identified as a dilepton decay at the generator level, where the $W$'s may decay to any 
of $e\nu_e$, $\mu\nu_{\mu}$, or $\tau\nu_{\tau}$.  Other $t\bar{t}$ events passing 
the event selection are accounted for as background (see Section~\ref{sec:fakes}).
Corrected for discrepancies between observed and simulated data, the
acceptance is 0.84~$\pm$~0.03~\%, where the uncertainty includes 
the systematic uncertainties.  In the rest of this section, we discuss the 
acceptance, the corrections made to it, and the systematic uncertainties on it.  

\subsection{Contributions to the Acceptance}

One of the advantages of identifying the second lepton only as a track is the
enhanced acceptance for $\tau$ leptons from $W$ decays.  Standard electron and 
muon selection will accept a fraction of $\tau$ decays, since 35\% 
are through leptonic channels. There will be some inefficiency, because a portion of 
the momentum of the original $\tau$ will be lost to the two neutrinos 
produced.  On the other hand, if ``single-prong'' hadronic decays are included,
85\% of $\tau$ decays have a single charged track in the final state.
About 20\% of the total lepton~+~track acceptance is from
events where one or both of the $W$'s decays to a $\tau$ lepton, and 65\% of that
(13\% of the total) is from events where at least one of the $\tau$ 
leptons decays hadronically.  Table \ref{tab:genacc} shows how the 
acceptance is distributed among the different lepton types.
\begin{table*}[tbp]
 \centering
%
 \begin{tabular*}{\widetable}{c*{7}{@{\extracolsep{\fill}}r@{\extracolsep{0in}}@{ $\pm$ }l}}
\hline\hline
	& \multicolumn{2}{c}{$ee$} & \multicolumn{2}{c}{$e\mu$} & \multicolumn{2}{c}{$\mu\mu$} & \multicolumn{2}{c}{$e\tau$} & \multicolumn{2}{c}{$\mu\tau$} & \multicolumn{2}{c}{$\tau\tau$} & \multicolumn{2}{c}{total} \\
\hline
$e$+track	& 17.4 & 0.2	& 29.5 & 0.3	&  0.0 & 0.0	& 7.6 & 0.1	& 2.3 & 0.1	& 0.5 & 0.0	& 57.4 & 0.4	\\
$\mu$+track	&  0.0 & 0.0	&  5.9 & 0.1	& 15.5 & 0.2	& 0.5 & 0.0	& 4.8 & 0.1	& 0.3 & 0.0	& 26.9 & 0.2	\\
\hline            
total		& 17.4 & 0.2	& 35.4 & 0.3	& 15.5 & 0.2	& 8.1 & 0.1	& 7.1 & 0.1	& 0.8 & 0.0	& 84.3 & 0.5	\\
\hline\hline
 \end{tabular*}
 \caption{\label{tab:genacc} Acceptance for \mbox{opposite-charge} lepton~+~track events with two 
   or more jets, for each possible pairing of generated charged leptons from the $W$ decays.
   Numbers have been multiplied by 10000 for legibility.  The first row shows the acceptance in the 
   channels where the fully reconstructed lepton is an electron, and the second row shows the
   acceptance for the fully reconstructed muon channels.
   The majority of events are accepted as electron plus track because there is more
   geometric acceptance for electrons, and because the ordering of primary leptons
   means that events generated as electron-muon events will be preferentially
   accepted as electron+track.  
   The uncertainties quoted only include the statistical uncertainty.}
\end{table*}

\subsection{Corrections to the Acceptance}\label{sec:acc_corr}

To understand the discrepancies in lepton reconstruction between observed and simulated data, we 
study the performance of the reconstruction in large control samples and derive appropriate
corrections.  Real and simulated $Z$ boson events are used, 
because the available samples are large and the reconstruction of the invariant mass peak 
allows selection of a very pure sample of dilepton events, even with minimal identification 
requirements placed on the second lepton. We also correct the acceptance for the 
small inefficiency of the high-$p_T$ lepton triggers. 

These corrections are also used in some of the background calculations.

\subsubsection{Trigger Efficiencies}

We measure single lepton trigger efficiencies with a combination of $Z$ data and data
taken using an independent trigger.  The $Z$ sample is especially useful when
the two lepton candidates are found in sections of the detector corresponding to different 
triggers.  Independent triggers designed to share some, but not all, of the 
requirements of the trigger of interest enable measurement of the efficiency of the 
omitted requirements.

What we need for the cross section measurements is the probability for a lepton~+~track
candidate to fire one of the high-$p_T$ lepton triggers.  This probability is higher than the 
single lepton trigger efficiency since each event has two chances to fire one of the triggers,
one for each lepton.  On the other hand, the second lepton 
is not fully reconstructed in our event selection, so the event trigger efficiency is not just a 
simple combination of single-lepton trigger efficiencies.  
To determine the per-event trigger efficiency for a particular process and fully 
reconstructed lepton type, we count the number of events in a simulated sample
of that process that have one lepton of that type and the number with two of that type.
For events with one fully reconstructed lepton, we use the single-lepton trigger efficiency 
as the event trigger efficiency.  For events with two, we use the probability for at least 
one of the two leptons to fire the trigger, given by $1 - (1 - \epsilon)^2$ where
$\epsilon$ is the single-lepton trigger efficiency.  We then take the average of the
two per-event efficiencies, weighted by the relative number of events with one and 
two fully reconstructed leptons.  The plug electron trigger also includes a $\met$
threshold, so the trigger efficiency we use for those events also depends on the 
value of the $\met$ as it would be calculated for the trigger decision.  Note that we 
include the electron $E_T$ and \met\ dependence of the trigger efficiencies where applicable,
by convoluting the single-lepton trigger efficiencies with the $E_T$ and or \met\ distributions
for the class of events in question.

The per-event trigger efficiency is also needed for background estimates that use an 
acceptance calculated from simulation.
For a given lepton type, the per-event efficiencies are very similar across different 
physics processes, so the $t\bar{t}$ value is used.  The one exception is PHX~+~track \Ztt\ 
events.  For those events the typical plug electron $E_T$ and \met\ fall in the middle of
the turn-on curves for the trigger, and the trigger efficiency, about 66\%, is lower than those typical
of $t\bar{t}$ and diboson events.  

The single-lepton and total per-event trigger efficiencies 
are given in Table~\ref{tab:lepid}.

\subsubsection{Fully Reconstructed Electrons and Muons}

Identification efficiencies for fully reconstructed leptons
are measured in a sample of $Z$ candidates.  These candidates consist of one 
fully reconstructed lepton candidate and one \mbox{opposite-charge} lepton candidate of the same 
flavor which meets minimal kinematic and identification criteria.  The fully reconstructed 
candidate must pass the corresponding high-$p_T$ lepton trigger, and the invariant mass 
of the lepton candidate pair is required to be close to the central value of the $Z$ resonance peak.  

For central (CEM) electrons, the minimally-identified lepton candidate is an electromagnetic 
cluster fiducial to the central calorimeter with \mbox{$E_T > 20\ \GeV$} and a matched track with 
\mbox{$p_T > 10\ \GeVc$} and \mbox{$|z_0| < 60$ cm.}  The electron candidate pair must have an 
invariant mass
in the interval \masswin{76}{106}.  Electromagnetic clusters fiducial to the forward calorimeter 
are used to measure the forward (PHX) electron efficiency.  No track requirement is made, so the
efficiencies measured include the tracking efficiency.  	
The invariant mass window used for this candidate pair is \masswin{81}{101}.  

For muons, the total reconstruction efficiency is the product of the efficiency to find a track
stub in the muon chambers and the efficiency for a muon candidate with a track and stub to pass
all of the remaining identification requirements.  To measure the efficiency to find a track stub, 
the second muon candidate in the $Z$ pair is a track pointing at the fiducial region of the 
muon detectors and 
meeting the same requirements on the energy deposition in the calorimeter as fully reconstructed
muon candidates, except with the maximum scaled up by 50\%.  To measure the identification efficiency,
the second muon candidate is a track with \mbox{$p_T > 20\ \GeVc$} matched to a track stub in the CMU 
and CMP, or in the CMX.  We accept only events where the muon candidate pair invariant mass 
is in the range \masswin{81}{101}.

The denominator of the efficiency is the number of leptons in the $Z$ candidates
passing the minimal requirements, and the numerator is the subset of those 
also passing all lepton selection requirements.   We measure the efficiency in both observed
and simulated data, because the full lepton selection is applied in calculating the acceptance.
We therefore use the ratio of the efficiency in observed data to the efficiency in 
simulated data as a ``scale factor'' which is multiplied by the acceptance to correct it. 
Scale factors for the four primary lepton types are given in Table \ref{tab:lepid}.

\subsubsection{Track $\chi^2$ Probability}

The $\chi^2$ probability requirement, imposed on fully reconstructed muons and on track
leptons, is intended to reject hadron decays-in-flight that can be mistaken for 
prompt high-$p_T$ muons.  Tracks reconstructed from a particle that decays in the tracker
have a worse track fit because the track is constructed from hits from both the original 
hadron and the secondary muon, some of which will be far from the single
reconstructed trajectory. 

Because the requirement is made only in observed data, the acceptance is multiplied 
by the efficiency as measured in observed data, rather than by a scale factor.
We measure this efficiency in a sample of $Z$ candidates identified from a fully reconstructed 
lepton and an isolated track.  One subtlety here is that the $\chi^2$
is correlated between the tracks of the two objects, through the hit timing information 
in the COT, so the efficiency to apply it to both is not equal to the product of 
efficiencies of the individual objects.  Thus, for electron~+~track
events, where the requirement applies only to the track lepton, the efficiency is the 
number of tracks that pass the requirement, divided by the total number of tracks. 
In contrast, for muon~+~track events, where the requirement applies to both, the relevant 
efficiency is the ratio of muon~+~track $Z$ events where both 
leptons pass the requirement to all muon+track $Z$ events.  The measured efficiencies are
0.962~$\pm$~0.001 for electron+track events, 0.944~$\pm$~0.001 for CMUP~+~track events, 
and 0.951~$\pm$~0.002 for CMX~+~track events, and are included in Table \ref{tab:lepid}.

\subsubsection{Isolated Tracks}

Efficiencies for the track isolation and impact parameter requirements differ between observed and 
simulated data.
To quantify the efficiency of the track isolation requirement, we use
$Z$ candidates from a fully reconstructed electron or muon and an \mbox{opposite-charge} track 
passing all of the track lepton requirements except isolation, where the lepton~+~track pair have an
invariant mass in the interval \masswin{76}{106}.  
To reduce background from jets, we accept only events where the track appears to be from
a lepton of the same flavor as the fully reconstructed one, using information from the 
calorimeter towers at which the track points.  
The efficiency of the isolation requirement is the ratio of the number of tracks passing it 
to the total number of tracks.  The efficiencies drop from about 95\% for events with 
zero jets to about 90\% for events with two or more jets.  Taking the ratio of the
efficiency from observed data to the efficiency from simulated data, the resulting scale factors are 
1.004~$\pm$~0.001 for events with zero jets, 1.002~$\pm$~0.003 for events with one jet, 
and 0.965~$\pm$~0.011 for events with two or more jets.

We measure the efficiency of the impact parameter requirement similarly.
The total observed efficiency in data is 0.909~$\pm$~0.003, calculated as the weighted combination of
0.940~$\pm$~0.002 for data including silicon detector information
and 0.53~$\pm$~0.02 for the rest of the data.  The corresponding 
efficiency is 0.9185~$\pm$~0.0007 in simulation: 0.947~$\pm$~0.001 for data
including silicon detector information and 0.55~$\pm$~0.01 for the rest of the data. 
Taking the ratio of the results yields a scale factor of 0.989~$\pm$~0.003.

\begin{table*}[tbp]
\centering
\begin{tabular*}{\widetable}{l*{4}{@{\extracolsep{\fill}}c}}
\hline
\hline
		& Reconstruction	& $\chi^2$ probability	& Single lepton		& Event		\\
Event type	& scale factor		& efficiency		& trigger efficiency	& trigger efficiency	\\
\hline
\end{tabular*}
\begin{tabular*}{\widetable}{l*{4}{@{\extracolsep{\fill}}r@{\extracolsep{0in}}@{.}l@{~~~~~~~~}}}
CEM + track	& 0 & 981 & 0 & 962 & 0 & 971  & 0 & 975 \\
PHX + track	& 0 & 935 & 0 & 962 & 0 & 918* & 0 & 918 \\
CMUP + track	& 0 & 926 & 0 & 944 & 0 & 908  & 0 & 916 \\
CMX + track	& 0 & 984 & 0 & 951 & 0 & 910  & 0 & 937 \\
\hline
\hline
\end{tabular*}
\caption{
  Correction factors applied to the calculated acceptance.  
    Uncertainties on these numbers are about a percent or smaller.  
    CEM and PHX are the central and forward electrons, and CMUP and CMX 
    are muons.  The $\chi^2$ probability efficiency applies to just the isolated
    track in electron~+~track events, but to both the muon and the isolated track
    in muon~+~track events.
  *For the forward (PHX) electron trigger, this efficiency
    is for $W$ events, since the trigger also has a \met\ requirement.  
    This also means that the per-event efficiency is identical to the single-lepton efficiency.
  }\label{tab:lepid}
\end{table*}

\subsection{Systematic Uncertainties on Acceptance}

The systematic uncertainties on the acceptance reflect the limits on 
experimental understanding of the final-state objects used to identify $t\bar{t}$ events,
as well as our ability to model $p\bar{p}$ interactions with Monte Carlo simulations.
The first category includes uncertainties on lepton identification and the jet energy scale.
The second includes uncertainties on QCD radiation, parton density functions, and the Monte 
Carlo generator used to calculate the acceptance.  

The systematic uncertainties on the signal acceptance are discussed
individually below and summarized in Table~\ref{tab:summary_syst}. 
\begin{table}[tp]
  \begin{center}
    \begin{tabular*}{8.6cm}{l@{\extracolsep{\fill}}c} 
      \hline \hline
      Source & Uncertainty \\  
      \hline
      Fully rec. lepton identification 		& 1.1\%\\ 
      Track lepton identification 		& 1.1\%\\ 
      Jet energy scale			& 1.3\%\\
      Initial-state QCD radiation 	& 1.6\%\\
      Final-state QCD radiation 	& 0.5\%\\
      Parton density functions 		& 0.5\%\\
      Monte Carlo generator 		& 1.5\%\\
      \hline
       Total	                      & 3.1\%\\
      \hline \hline
    \end{tabular*}
    \caption{Summary of systematic uncertainties on the signal acceptance.}
    \label{tab:summary_syst}
  \end{center}
\end{table}

\subsubsection{Primary Lepton Identification Efficiency}

The dominant uncertainty on the identification efficiency for fully reconstructed leptons
is associated with isolation and our ability to model additional activity in the event, 
such as jets or unclustered low-$p_T$ tracks, using Monte Carlo simulations.  
As described in Section~\ref{sec:acc_corr}, the lepton identification
efficiencies are derived from real and simulated $Z$ data, in which most events have zero
jets. In the $t\bar{t}$ sample, where most events have two or more jets, 
and nearby jet activity can reduce the efficiency to identify isolated electrons and muons.  

To quantify these effects, we measure the scale factor in the $Z$ samples
as a function of the distance $\Delta R$ between the lepton candidate and the nearest jet.  
We calculate the correction appropriate to $t\bar{t}$ events by folding 
this function with the $\Delta R$ distribution for simulated $t\bar{t}$ candidate events.
For each primary lepton 
type, the statistical uncertainty on the re-weighted scale factor exceeds the
difference between the original and re-weighted scale factors.  Therefore, we 
take the statistical uncertainties on the re-weighted scale factors as the 
uncertainties on the scale factors.
The total systematic uncertainty is the weighted average of the uncertainties on the individual
lepton types, where the weights are the acceptances for each lepton category.
The resulting uncertainty is 1.1\%.

\subsubsection{Track Lepton Identification Efficiency}

This uncertainty quantifies how well the simulation models the track isolation requirement in
an environment with many jets, in analogy to the uncertainty on well-reconstructed 
leptons.  In this case, we base the uncertainty on the behavior of the correction as 
a function of the number of jets.   We correct the acceptance with the
scale factor measured in events with two or more jets, and take the 1.1\% statistical uncertainty
as the uncertainty on track lepton identification.

\subsubsection{Jet Energy Scale}

The jet energy scale influences the $t\bar{t}$ acceptance because if the jet energies are over-corrected, 
more events will have two or more jets and pass the event selection, and vice versa.  
It also influences the acceptance through the jet energy corrections 
to the \met\ and the restriction on the $\Delta \varphi$ between the jets and the \met\ for events with
\mbox{\met\ $<50\ \GeV$.}  To estimate the uncertainty on the acceptance from 
the jet energy scale, we recalculate the signal acceptance twice.  
First, we vary the jet energy corrections up by the uncertainties from 
Ref.~\cite{jetenergies} and recalculate the energies of all the jets in the event,
and then recalculate the acceptance. 
We repeat the exercise, varying the jet energies down by their uncertainties, and then take   
half the difference between the two recalculated acceptances, 1.3\%, as 
the systematic uncertainty.

\subsubsection{Initial and Final State Radiation}\label{sec:isrfsr}

Additional jets can be produced in association with the $t\bar{t}$ pair through 
radiation of one or more gluons from the initial or final state particles.  
We can measure the dependence of the acceptance on the rate of QCD radiation by comparing 
the central value of the acceptance to values calculated in simulated \pythia\ 
$t\bar{t}$ samples identical to those used to calculate the central value, except that
the \pythia\ parameters governing the rate of initial and final state radiation 
via parton showering have been varied.  
The range of allowed values is set by study of
the reconstructed $p_T$ and $M^2$ of the $Z/\gamma^*$ in Drell-Yan events with 
electrons or muons in the final state~\cite{isrfsr}.  
Drell-Yan events allow isolation of initial-state radiation effects, because the dilepton
final state is colorless.  The range of parameters found to cover the variation in the observed
initial-state radiation can then also be used to generate samples with more and less
final-state radiation, because the same parton shower algorithm is used.

The acceptance increases for the sample with more initial-state radiation, and decreases
for the sample with less.  We take half the full difference, 1.6\%, as the systematic 
uncertainty.  The results for final-state radiation 
are less conclusive, as the measured acceptances in the modified samples differ from the 
nominal value by less than their statistical uncertainties of 1\%. 
We therefore take the larger of the two observed differences, 0.5\%, as the 
systematic uncertainty.



\subsubsection{Parton Distribution Functions}

The parton distribution functions (PDFs) describe the probabilities for each 
type of parton to carry a given fraction of the proton momentum. Variations of 
the PDFs can have a significant effect on the 
$t\bar{t}$ cross section~\cite{ttbar_new}.
The PDFs also have a smaller effect on the acceptance through the kinematics of the 
$t\bar{t}$ decay products.  Twenty independent sources of uncertainty 
identified for the \textsc{cteq5l} PDF set are considered~\cite{cteq_pdf}.  In evaluating the
total uncertainty, we also include the
difference between the \textsc{cteq5l} and \textsc{mrst}~\cite{mrst_pdf} PDF sets and
the effect of lowering $\alpha_s(M_Z^2)$ from the preferred value of 0.1175
by 0.005, the uncertainty on the world average measured value at the time
the PDF set was calculated~\cite{mrst_pdf2}.

To quantify the effect of PDFs on the lepton~+~track acceptance, we recalculate the acceptance
twice for each variable of interest: once each for the upper and lower bounds on that variable.
Information about the types and momenta of generated particles are stored
when Monte Carlo events are produced, allowing the incoming partons and their 
momenta to be identified.  
The corresponding probabilities for those values are found in both the 
nominal PDF and the variation under study.  The event weight is the ratio of the product 
of the altered probabilities to the nominal:
\begin{equation}
{\rm weight} = \frac{p(x_1,Q^2)\, p(x_2,Q^2)}{p'(x_1,Q^2)\, p'(x_2,Q^2)} \textrm{ ,}
\end{equation}
where $p$ is the nominal PDF and $p'$ is the modified PDF.  The PDFs depend on the
momentum transfer $Q$ and the fraction $x_i$ of the hadron's momentum carried by the parton,
where the index $i$ specifies one of the two incoming partons.
To calculate the acceptance as a ratio of accepted to total events, 
each event contributes the calculated weight to the denominator of the ratio but the 
weight is only added to the numerator if the event passes the selection.
We repeat this process for each PDF variation and record the resulting change in acceptance.
Adding the results of all the variations in quadrature and averaging the
positive and negative uncertainties, we find a total uncertainty of 0.5\%.

\subsubsection{Monte Carlo Generator}

To account for a possible dependence of the measured acceptance on the choice of Monte Carlo event
generator, the $t\bar{t}$ acceptance is remeasured, again for \mbox{$m_t = 175\ \GeVcc$,} 
using the \herwig\
Monte Carlo and compared to the nominal value obtained using the \pythia\ Monte Carlo.  
In calculating the difference, we exclude the effect of the different $W\to\ell\nu$ 
branching ratios used by the two generators: \pythia\ uses the measured value, 
\mbox{0.1080 $\pm$ 0.0009 (\cite{the_pdg})}, and \herwig\ uses 1/9.  
The remaining difference between the acceptances measured with the two generators is 1.5\%,
which we include as a systematic uncertainty.


\section{\label{sec:btag}Identification  of Jets from $b$ Quarks}

The CDF \SecVtx algorithm identifies $b$-jet candidates based on the determination
of the primary event vertex and the reconstruction of one or more secondary vertices
using displaced tracks associated with jets~\cite{Run2_SecVtx, Run2_SecVtx_proc}. 
If a secondary vertex is found that is significantly displaced from the 
primary vertex in the plane transverse to the beam, the jet is 
said to be ``tagged'' as a $b$-jet candidate.

\subsection{Determination of the Primary Vertex}
A primary vertex in an event is defined as the point from which all
prompt tracks originate. The location of the primary vertex in an 
event can be found by fitting well-measured tracks to a common 
point of origin.  In high instantaneous luminosity 
conditions, more than one primary
vertex may exist in an event, but these are typically separated in $z$. The $z$ coordinate
for each vertex is found by taking the weighted average of 
the $z$ coordinates of all tracks within 1~cm of the first iteration vertex. 
The $z$ position measurement of this first vertex
has a resolution of 100~$\mu$m~\cite{Run2_SecVtx, Run2_SecVtx_proc}. 
The location of the primary
vertex is then refined by the above information, along with
constraints of the beamline position, and some tracking information.

\subsection{The {\sc secvtx} Algorithm}
\label{ss:SVXAlg}
The \SecVtx algorithm starts from the primary interaction vertex for each
event. In the present application, this is the vertex that is associated with 
the lepton~+~track. 
It then examines the tracks associated with each jet
and applies basic quality criteria to them. These include 
the number of silicon layers associated with the track, minimum and maximum 
allowed impact
parameters, and the track $\chi^2$ per degrees of freedom. 
The algorithm then attempts to resolve a secondary vertex that is 
significantly displaced from the primary vertex using tracks with 
large impact parameter significance, $d_0 / \sigma_{d_0}$, where $\sigma_{d_0}$
is the uncertainty on the impact parameter.

The \SecVtx algorithm is based on a two-pass system.
The first pass of the algorithm 
builds an initial vertex, known as the ``seed'',
from the two most displaced tracks. The seed vertex initiates the \SecVtx
algorithm.
Pairs of tracks with invariant masses consistent with the $K_s^0$ and $\Lambda$ mass
are removed from the track list. The algorithm then seeks to add tracks to the seed 
vertex. The additional tracks must pass quality
requirements on the impact parameter and $p_T$ and must not result in a poor $\chi^2$ for 
the resulting three track vertex. 
If no such vertex is found, then another seed vertex, made of the next two
most displaced tracks is tried. This continues 
until a vertex is resolved, or the seed list is exhausted. In the latter case,
the algorithm moves on to the second pass, in which it attempts to
find a vertex using only two tracks for which 
the quality requirements of the tracks are made more stringent. 
Again, pairs of tracks whose invariant mass is
consistent with the $K_s^0$ and $\Lambda$ masses are removed.

\begin{figure*}[tbp]
\subfigure[]{ \includegraphics[width=0.48\textwidth]{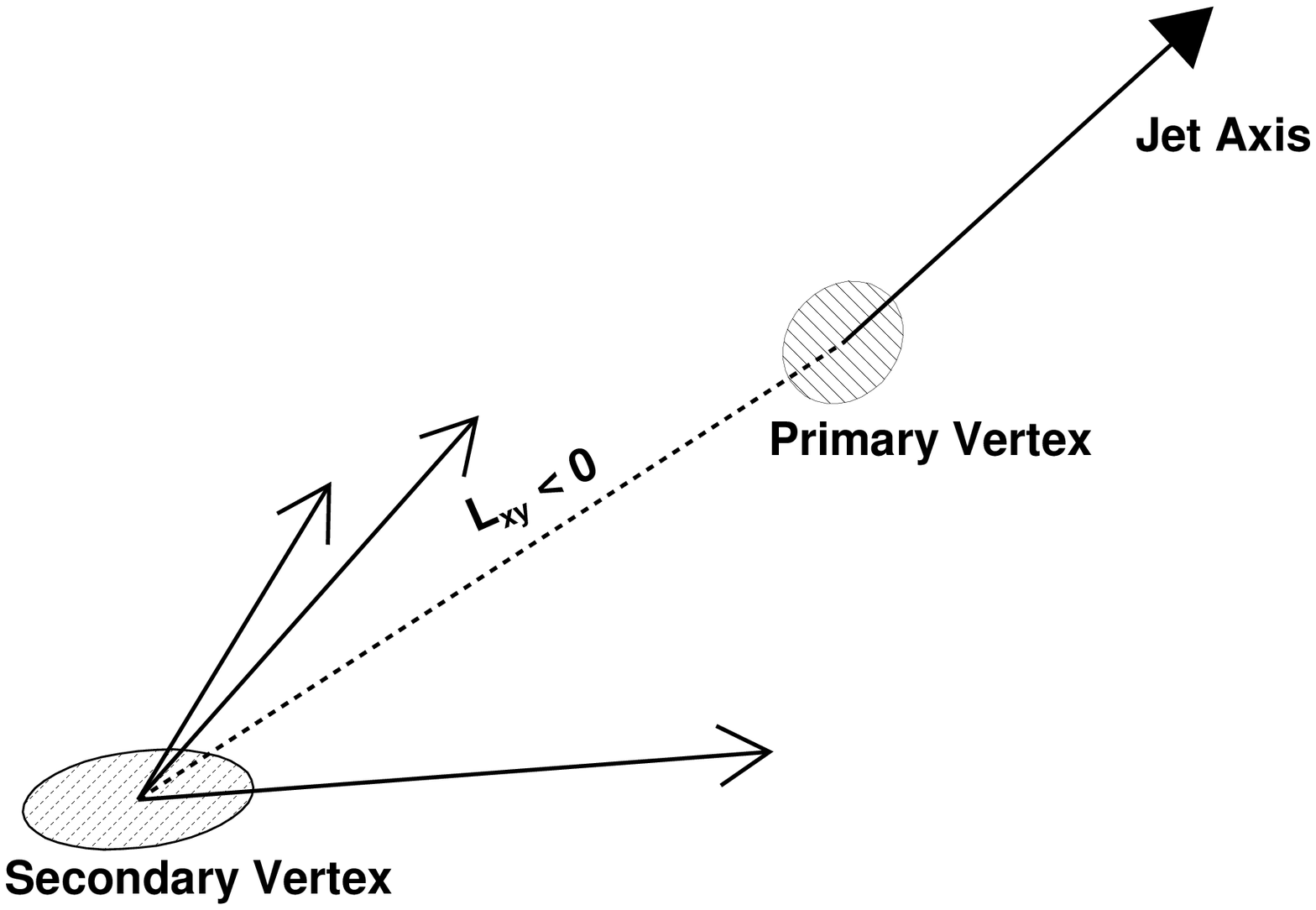}}
\subfigure[]{ \includegraphics[width=0.48\textwidth]{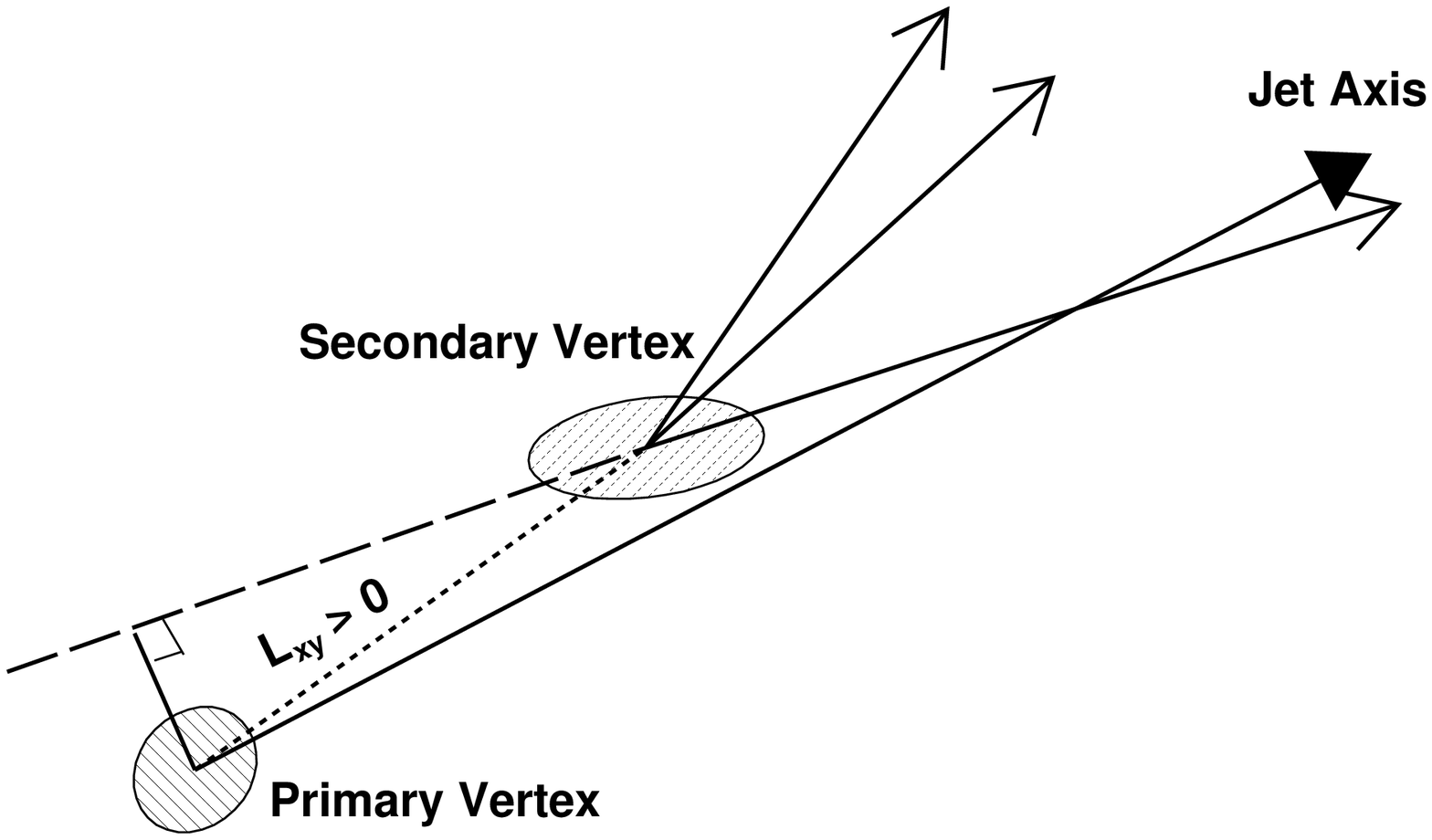}}
  \caption{\label{fig:tagger} 
Drawing showing reconstructed secondary vertices, illustrating example cases
 where (a) $L_{xy} < 0$ and (b) $L_{xy} > 0$.}
\end{figure*}

With a secondary vertex in hand, \SecVtx calculates the length of the vector
between the primary
and secondary vertices  in the plane perpendicular to the beamline. This
vector is then projected onto the jet axis:
\begin{equation}
L_{xy} = (\vec{r}_\txtss{PV} - \vec{r}_\txtss{SV}) \cdot \hat{p}_\txtss{jet}	\rm{,}
\end{equation}
where $\vec{r}_\txtss{PV}$ is the position of the primary vertex, $\vec{r}_\txtss{SV}$ is the position
of the secondary vertex, and $\hat{p}_\txtss{jet}$ is the jet direction. $L_{xy}$
is the two-dimensional decay length along the jet axis, and $\sigma_{L_{xy}}$ the 
associated uncertainty.
\SecVtx defines a ``displaced'' (or ``tagged'') vertex as one with significance 
$|L_{xy} / \sigma_{L_{xy}}| >$~3.0. 
A long-lived hadron will generally travel in roughly the same direction as the jet 
formed from the fragmentation and hadronization process. As a result, the cosine 
of the angle between the jet axis and the vector extending from the primary 
to the secondary vertex will be positive, and so will $L_{xy}$; see Fig.~\ref{fig:tagger}.
A negative
value of $L_{xy}$ can result from resolution smearing of the track parameters and
poorly reconstructed tracks. Depending on the sign of $L_{xy}$, tags will be
referred to as \emph{positive} or \emph{negative}. 
The $L_{xy}$ distribution for negative tags will be interpreted as the result of 
``mistags'', or tags from non-$b$-jets.

\subsection{Event Tagging Efficiency}

The event tagging efficiency is the efficiency for
tagging at least one of the two $b$-jets in a $t\bar{t}$ lepton~+~track
event using the \SecVtx tagger. To find the event tagging efficiency
we use a $t \bar{t}$ \pythia\ Monte Carlo sample with \mbox{$m_{t} = 175~\GeVcc$,} the same
sample used to calculate the pretag acceptance. 
Corrections to the event tagging
efficiency are made for two effects. The first correction accounts for our
ability to reconstruct jets which correspond to a $B$ hadron decay. The second
correction is for the possibility of mistakenly tagging light quark jets as heavy
flavor jets.

The event tagging efficiency  is given by the formula
\begin{equation}
\epsilon_\txtss{tag} = {\epsilon_b^\prime}\, F_{1b} + 2 {\epsilon_b^\prime} (1- \epsilon_{b}^\prime )\, F_{2b} + (\epsilon_b^\prime)^{2}\, F_{2b} .
\end{equation} 
where $\epsilon_b^\prime$ is the corrected single jet tagging efficiency (see below), 
and $F_{1b}$ and $F_{2b}$ are the taggable jet fractions.
The taggable jet fractions
describe the fraction of events with one or two jets which originate with the
hadronization of a $b$ quark and might be tagged.  The denominator 
contains events from the simulated $t\bar{t}$ sample which pass the lepton~+~track
selection, including the $\geq 2$ jet requirement.  The numerator of $F_{1b}$ is 
the number of those which have one jet which is matched to a $b$ hadron at
generator level and contains two or more tracks passing the \SecVtx quality requirements
described in Section~\ref{ss:SVXAlg}.  The numerator of $F_{2b}$ is the number
with two such jets.
See Table~\ref{tab:Method2} for the values of the taggable jet fractions.

The single jet tagging efficiency $\epsilon_b$ is the ratio of the number of 
taggable jets with a positive \SecVtx tag to the number of taggable jets.
We multiply $\epsilon_b$ by a scale factor, $\epsilon_\txtss{SF}$
to account for differences in the single jet tagging efficiency between 
observed and simulated data.  We also apply corrections for the efficiency to tag light
quark jets $\epsilon_q$,
and the efficiency to match a jet to a $b$ hadron decay $\epsilon_\txtss{bmatch}$.
The corrected single jet tagging efficiency is:
\begin{equation}
\epsilon_b^\prime = \frac{\epsilon_b \, \epsilon_\txtss{SF} \, \epsilon_\txtss{bmatch}}{(1-\epsilon_q)}\	\rm{.}
\end{equation}

The $\epsilon_\txtss{bmatch}$ correction accounts for 
the situation in which a $b$ quark fragments to produce a jet which does not
pass the jet selection criteria employed in this analysis.
We measure the efficiency for matching a $B$ hadron to a 
reconstructed jet in simulated $t\bar{t}$ events.
We find  \mbox{$\epsilon_\txtss{bmatch} = (98.89 \pm 0.04)$ \%}. We multiply the
per jet tagging efficiency obtained above by the matching efficiency 
to account for this small $b$-jet reconstruction inefficiency.

The last correction, $1/(1-\epsilon_q)$, accounts for tags of light
quark jets, which results in an enhancement to the single jet tagging efficiency.
As stated in Section~\ref{ss:SVXAlg}, negative tags are interpreted as mistakes
made by the tagging algorithm, and are due to resolution effects. The
negative tagging rate is similar for long-lived $b$-jets and for light
quark jets. So, to find the efficiency for tagging light quark jets
in $t\bar{t}$ decays, we find the efficiency for negative
tags in $b$-jets in \pythia\ $t\bar{t}$ simulation events, which
is equivalent to the rate of tagging of light quark jets.  We find
\mbox{$\epsilon_q = (1.3 \pm 0.1)$ \%}.
To correct the single jet tagging efficiency for the tagging of
light quark jets we divide by $(1-\epsilon_q)$.

Table~\ref{tab:Method2} gives a summary of the inputs used to calculate
the final event tagging efficiency and
we obtain a value of 0.669$\pm$0.037. This translates into
5.5\% systematic uncertainty due to event tagging.

Applying the \SecVtx tagging algorithm to the jets in the 129 lepton~+~track candidates,
we find 69 events with one or more tagged jets.

\begin{table}[tp]
  \begin{center}
    \begin{tabular*}{8.6cm}{@{\hspace{0.05\textwidth}}@{\extracolsep{\fill}}cr@{\extracolsep{0in}}@{.}l@{ $\pm$ }l@{\hspace{0.05\textwidth}}}
        \hline
        \hline
	Quantity	& \multicolumn{3}{c}{Value~~~~~~}	\\
	  \hline
        $F_{1b}$ 		& 0&321  & 0.003 \\
        $F_{2b}$ 		& 0&611  & 0.003 \\
        $\epsilon_{b}$ 		& 0&591  & 0.002\\
        $\epsilon_\txtss{SF}$ 	& 0&94   & 0.06 \\
        $\epsilon_q$	& 0&013  & 0.001  \\
        $\epsilon_\txtss{bmatch}$	& 0&9889 & 0.0004 \\
        \hline
        $\epsilon_\txtss{tag}$	& 0&669  & 0.037 \\
        \hline
        \hline
    \end{tabular*}
    \caption{Inputs and results of calculation of event tagging efficiency.  Note that the 
    	uncertainty on the event tagging efficiency is dominated by the systematic uncertainty 
	  on $\epsilon_{SF}$.  The other quoted uncertainties are all statistical uncertainties
	  from simulation and are negligible.
    	 }
    \label{tab:Method2}
  \end{center}
\end{table}


\section{\label{sec:pretagbg}Background Estimation in Pretag Sample}

Background events in the $t\bar{t}$ dilepton sample generally have one or 
two massive vector bosons decaying to leptons.
Non-negligible background processes are $W$~+~jets and similar events where one of the jets 
is misidentified as a lepton, diboson production, and Drell-Yan 
events where \met\ is produced by a combination of $\tau$ decays and the mismeasurement
of the energy of one or more objects in the event.  
Each of these processes requires the production of extra jets to satisfy 
event selection criteria.

\subsection{Backgrounds with a Jet Misidentified as a Lepton (``Fakes'')}\label{sec:fakes}

$W\to\ell\nu$ events with extra jets can pass the lepton~+~track selection if one of the jets is
misidentified as a lepton.  This can happen if the fragmentation of a parton 
results in a single charged hadron carrying most of the momentum of the 
original parton.  If a single charged particle carries more than about 90\% of the total
momentum of all the charged particles produced by fragmentation, it may satisfy
the criteria for an isolated track.  This is a relatively rare occurrence, but the 
inclusive $W$ cross section times the branching ratio to leptons is about 
2700~pb~\cite{CDFDetector}, as compared to the 6.7~pb cross section for $t\bar{t}$ production.  
Even though only about one in 500 $W$+jets events will have enough (three) jets to
produce a fake lepton and still pass the event selection, it remains the largest 
single source of background events.  

The estimation of the background from events with a fake lepton has three primary components:
the rate for the production of $W$~+~jets events with the right kinematic features, 
the probability for a jet to be misidentified as a lepton (the ``fake rate''), and the 
fraction of events in which the fake and true leptons have opposite
charge.  All present difficulties for the simulation of physics events.  The rate 
for the production of multiple extra jets in addition to a vector boson can in principle
be calculated perturbatively, but for large numbers of jets, the complexity of the 
calculation grows prohibitively, although progress has been made in recent 
years~\cite{alpgen_gen}.  The fake rate is affected by parton fragmentation, a 
non-perturbative QCD process which is not currently modeled with the needed accuracy.  
The fragmentation model will also affect the predicted charge correlation between the true 
and fake leptons.  Inaccuracies in the detector simulation further complicate the picture.
Therefore, we rely primarily on observed events for the estimate of the fake 
lepton background, and use simulated events only when it is impossible to isolate the relevant
effect in data.  

We summarize the calculation of the expected number of background events before 
describing the individual components in detail.
The total number of lepton~+~track events with $n$ jets where one of the leptons is fake, 
$N_\txtss{fake}^n$, is the sum of the number of events $N_t^n$ with a fake track 
lepton and the number $N_\ell^n$ with a fake fully reconstructed lepton, where we use the 
subscript $t$ to indicate numbers relating to track leptons and $\ell$ to indicate numbers 
relating to fully reconstructed leptons.  The estimates $N_t^n$ and $N_\ell^n$ are 
calculated separately using similar procedures, and then corrected for the efficiencies 
$\epsilon_{\txtss{OS}}^n$ and $\epsilon_Z^n$ (explained in more detail at the end of this section)
of the remaining lepton~+~track event selection:
\begin{equation}\label{eq:fakes}
	N_\txtss{fake}^n = \epsilon_{\txtss{OS}}^n\, \epsilon_Z^n\, (N_\txtss{t}^n + N_\ell^n)
\end{equation}
where
\begin{equation}
	N_t^n = \sum_{E_T,|\eta|}\left(f_t^{(n+1)}(E_T,|\eta|)\right) \left(N_j^{(n+1)}(E_T,|\eta|)\right)
\end{equation}
and
\begin{equation}
	N_\ell^n = \sum_{i=1}^4 \sum_{E_T,|\eta|}\left(f_\ell^i(E_T,|\eta|)\right) \left(N_j^{(n+1)}(E_T,|\eta|)\right) (A_\ell^i/A_t)\ \textrm{.}
\end{equation}
To predict the number of events $N_t^n$ with $n$ jets and a fake track lepton, we multiply 
the number of jets $N_j^{(n+1)}(E_T,|\eta|)$ in the lepton~+~\met~+~$(n+1)$ jet sample, 
binned in jet $E_T$ and $|\eta|$, by the track lepton fake rate $f_t^{(n+1)}(E_T,|\eta|)$
for the same number of jets and range of jet $E_T$ and $|\eta|$.  The result is summed over jet 
$E_T$ and $|\eta|$.  The selection for lepton~+~\met~+~jets events is described in 
Section~\ref{sec:fakes_wjetssel}, and the fake rate for track leptons is defined in 
Sections~\ref{sec:fakes_tlfr} and~\ref{sec:fakes_zjets} after motivating the choice of the 
$\gamma$~+~jets sample for the fake rate calculation in Section~\ref{sec:fakes_jets}.  
We test the performance of the track lepton fake rate by comparison among relevant jet 
samples in Section~\ref{sec:fakes_checkfr}.
To include the contribution from events with a fake fully reconstructed lepton, 
$N_\ell^n$, we multiply the same jet distributions $N_j^{(n+1)}(E_T,|\eta|)$ by the fully 
reconstructed lepton fake rates $f_{\ell}^{i}$, where the index $i$ indicates the type of the 
lepton identification criteria, and both the fake rate and jet counts are binned in jet
$E_T$ and $|\eta|$ as for the fake track leptons.   This yields the predicted numbers of 
events with two fully reconstructed leptons where one is real and one is fake.
We rescale the result by the ratio of the $W\to\ell\nu$~+~jets acceptance $A_t$ for track leptons
to the acceptance $A_\ell^i$ for fully reconstructed lepton type $i$, to find the number where 
the track lepton is the lepton from the $W$ and the fully reconstructed lepton is fake. 
Summing over the four fully reconstructed lepton types ($i$ runs from 1 to 4) gives 
the total result.
Details on the inclusion of fake fully reconstructed leptons are given in 
Section~\ref{sec:fakes_tightl}. 

The dominant contribution to the lepton~+~\met~+~jets sample is $W$~+~jets, but for
the sample with three or more jets, which is used in the background prediction for the
cross section measurement, there is also a significant contribution from $t\bar{t}$.
This happens in the ``lepton~+~jets'' decay channel, in which the $W$ that decays
to a pair of quarks adds one or two jets to the final state
in addition to the two jets from the $b$ quarks.
If the lepton from the decay of the other $W$ is reconstructed as an electron or muon
and one of the jets in the final state produces an isolated track (or vice versa),
such events can pass the full lepton~+~track event selection.  The treatment of
the $t\bar{t}$ contribution to the candidates with a fake lepton is discussed in 
Section~\ref{sec:fakes_ljets}.  These events require some special care, as they 
introduce a dependence on the cross section we are measuring.  That is 
treated by explicitly including the dependence in the likelihood used to calculate the 
cross section (see Section~\ref{sec:likeli}).  

The remaining component of the estimate is the efficiency for the selection criteria 
that cannot be applied in selecting the lepton~+~\met~+~jets sample.  
First is the \mbox{opposite-charge} requirement.  We calculate the efficiency 
$\epsilon^n_{\txtss{OS}}$ for this requirement using a combination of observed data
and Monte Carlo simulations in Section~\ref{sec:fakes_os}.  The other two requirements are
the increased \met\ threshold for events with a lepton~+~track invariant mass 
close to the $Z$ resonance, and the track lepton-\met\ opening angle veto.  
The efficiencies for these, collectively labeled $\epsilon_Z^n$ in Eq.~\ref{eq:fakes},
are given in Section~\ref{sec:fakes_effs}.

Finally, we tally the systematic uncertainties on this background estimate in 
Section~\ref{sec:fakes_sys}.

\subsubsection{Selection of $W$~+~jets Events}\label{sec:fakes_wjetssel}

The selection for $W$~+~jets events in the data is based on the event selection
described in Section~\ref{sec:evtsel}, with all requirements involving the track
lepton omitted.  That is, we select events from the same high-$p_T$ electron and 
muon trigger sample containing the signal candidates, with one
fully reconstructed electron or muon.  We also require that these events have
\mbox{\met\ $>25\ \GeV$} and pass the $\Delta \varphi$ criteria for the fully reconstructed 
lepton and the jets.  Since one of the jets will be reconstructed as a lepton in 
events with a fake lepton, we predict the number of events passing the full lepton~+~track
selection and having $N$ jets by using events with $N+1$ jets.  We count the number
of events with one jet, two jets, or at least three jets.  

The largest contribution to this sample is $W$~+~jets, where the $W$
decays leptonically.  There is also a significant contribution from $t\bar{t}$,
which will be discussed in more detail below.  There is also
a small contribution from pure-QCD multijet events where one of the jets has
been wrongly identified as a fully reconstructed electron or muon.  

\subsubsection{Jet Properties Influencing the Fake Rate}\label{sec:fakes_jets}

To motivate the choice of the $\gamma$~+~jets sample for the fake rate, we focus our 
attention on the largest contribution to the fake lepton background, $W$~+~jets events.
The lepton fake rate is determined by parton fragmentation, and fragmentation is determined
by the energy and type of the parton.  The energy dependence will be included by parameterizing
the fake rate as a function of jet $E_T$ and $|\eta|$.  Here we consider the possible 
influence of the parton type.

\begin{figure*}[tbp]
\subfigure[]{ \includegraphics[width=0.35\textwidth]{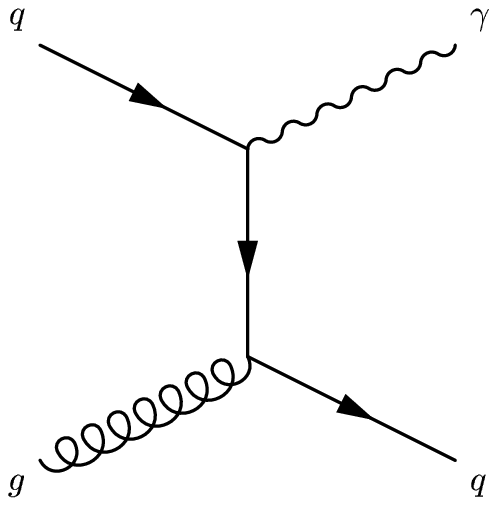}}
\subfigure[]{ \includegraphics[width=0.35\textwidth]{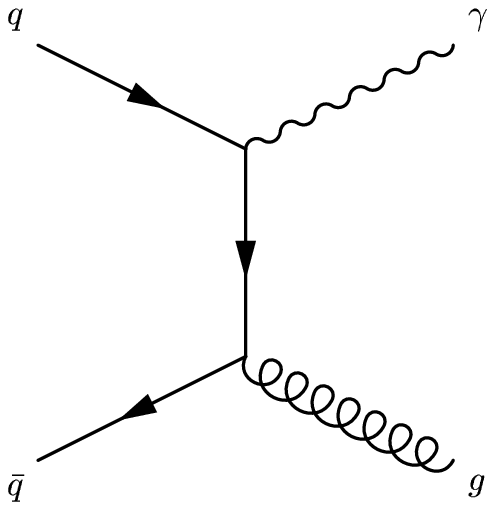}}\\
\subfigure[]{ \includegraphics[width=0.35\textwidth]{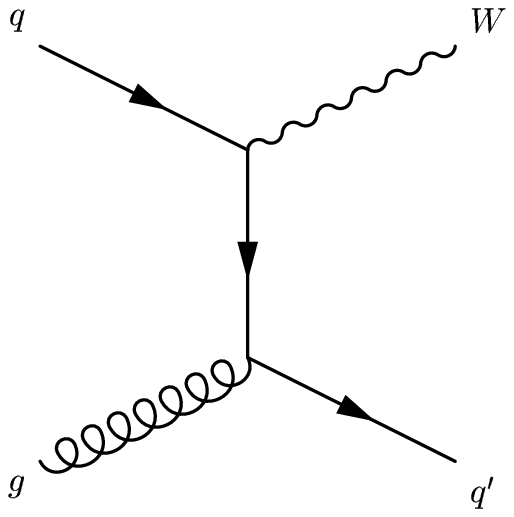}}
\subfigure[]{ \includegraphics[width=0.35\textwidth]{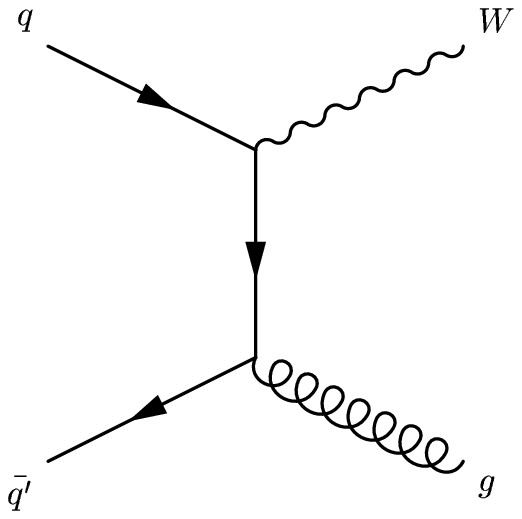}}
\vspace{0.2in}
  \caption{\label{fig:Wdiagrams} 
	Leading diagrams for $W$ and $\gamma$ production with one associated jet
	at the Tevatron.  Diagrams (a) and (b) are for $\gamma$~+~1 jet production, and 
	(c) and (d) are for $W$~+~1 jet production.
	Diagrams (a) and (c) dominate
	because of the relative strength of the gluon PDF at low $x$.
	}
\end{figure*}

A jet produced in association with a $W$ has a higher probability of being a quark jet,
meaning a jet which originates from a quark, than
a jet of the same energy produced in a generic QCD multijet event.  To understand this, 
note that most $W$ and multijet production at the Tevatron takes place at relatively low
$x$, where $x$ is the fraction of the proton momentum carried by an individual parton.  
For $Q^2$ values typical of $W$~+~jets or multijet production at the Tevatron, the gluon 
PDF is strong relative to the valence quark PDFs for $x \lesssim 0.15$.  
The two leading diagrams for $W$~+~1 jet production 
(see Fig.~\ref{fig:Wdiagrams}) at the Tevatron have the same amplitude.  The preference 
for diagrams with incoming gluons implies that the diagram with an outgoing quark
will prevail, so the jets associated with a $W$ boson are more likely to be quark jets.  
This preference for quark jets is even more marked in $\gamma$~+~jets samples because
the photon is massless, so we sample an even lower-$x$ part of the PDF.
In the case of multijet production, the leading order diagram for $gg\to gg$ has a larger 
amplitude than any of the other leading order $2 \to 2$ diagrams containing other 
permutations of light quarks and gluons.  Taking that in combination with the strength
of the gluon PDF, one expects jets produced in pure strong
interaction events to be predominantly gluon jets, that is, jets originating from gluons.  

The different partonic origins of jets matter because jets from light quarks appear to have 
a different fake rate than jets from gluons.  Gluon fragmentation results, on average, 
in a larger number of charged particles than quark fragmentation.  This behavior has been 
verified experimentally, and can be seen in simulation~\cite{cdf_trackmult}.  
An immediate consequence is the fact that a quark jet will be more likely 
than a gluon jet to contain a single charged track carrying most of the parton's energy.  
This also implies an increased probability to produce a fake lepton.   

\begin{figure*}[tbp]
\subfigure[]{ \includegraphics[width=0.46\textwidth]{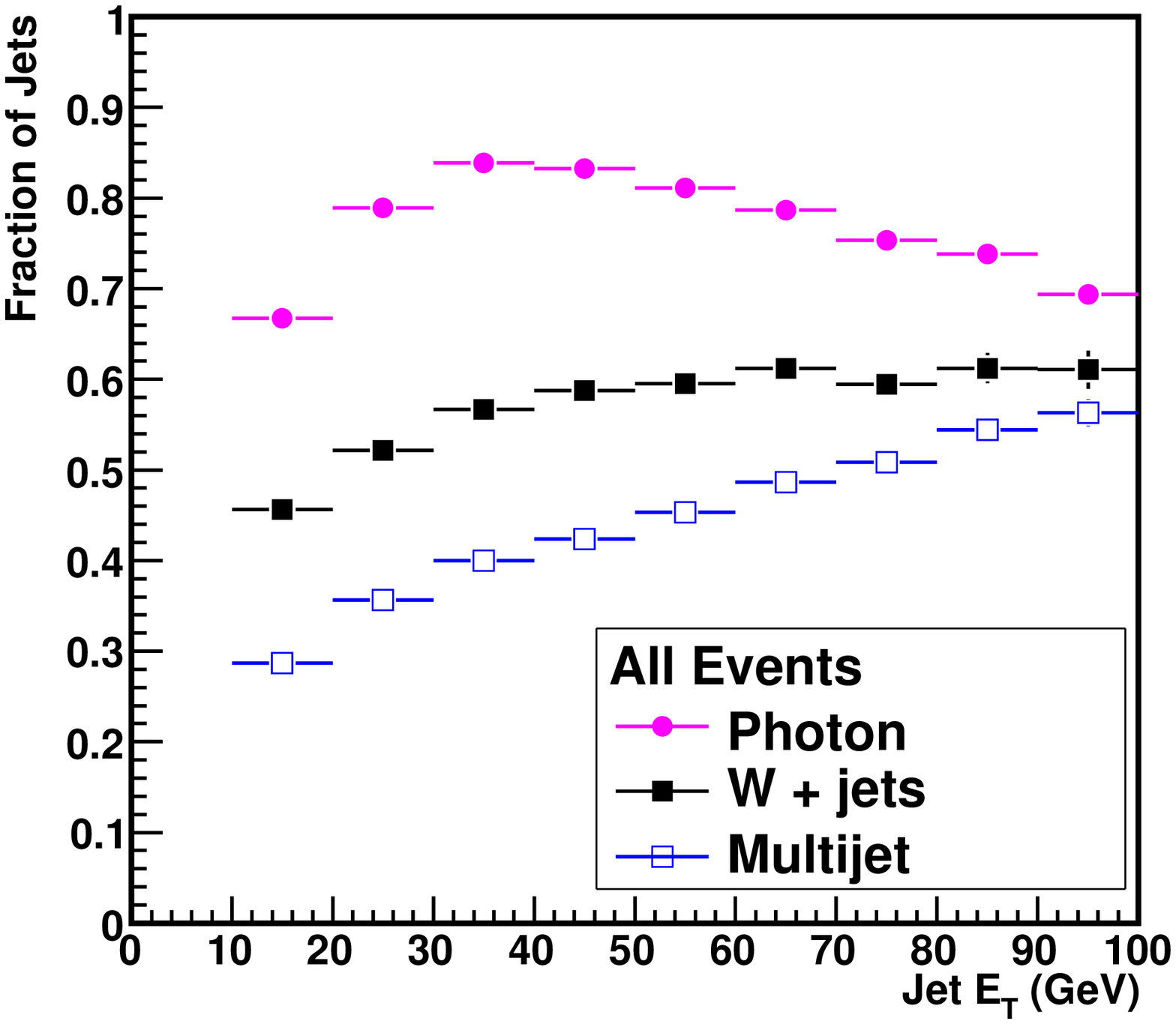}}
\subfigure[]{ \includegraphics[width=0.46\textwidth]{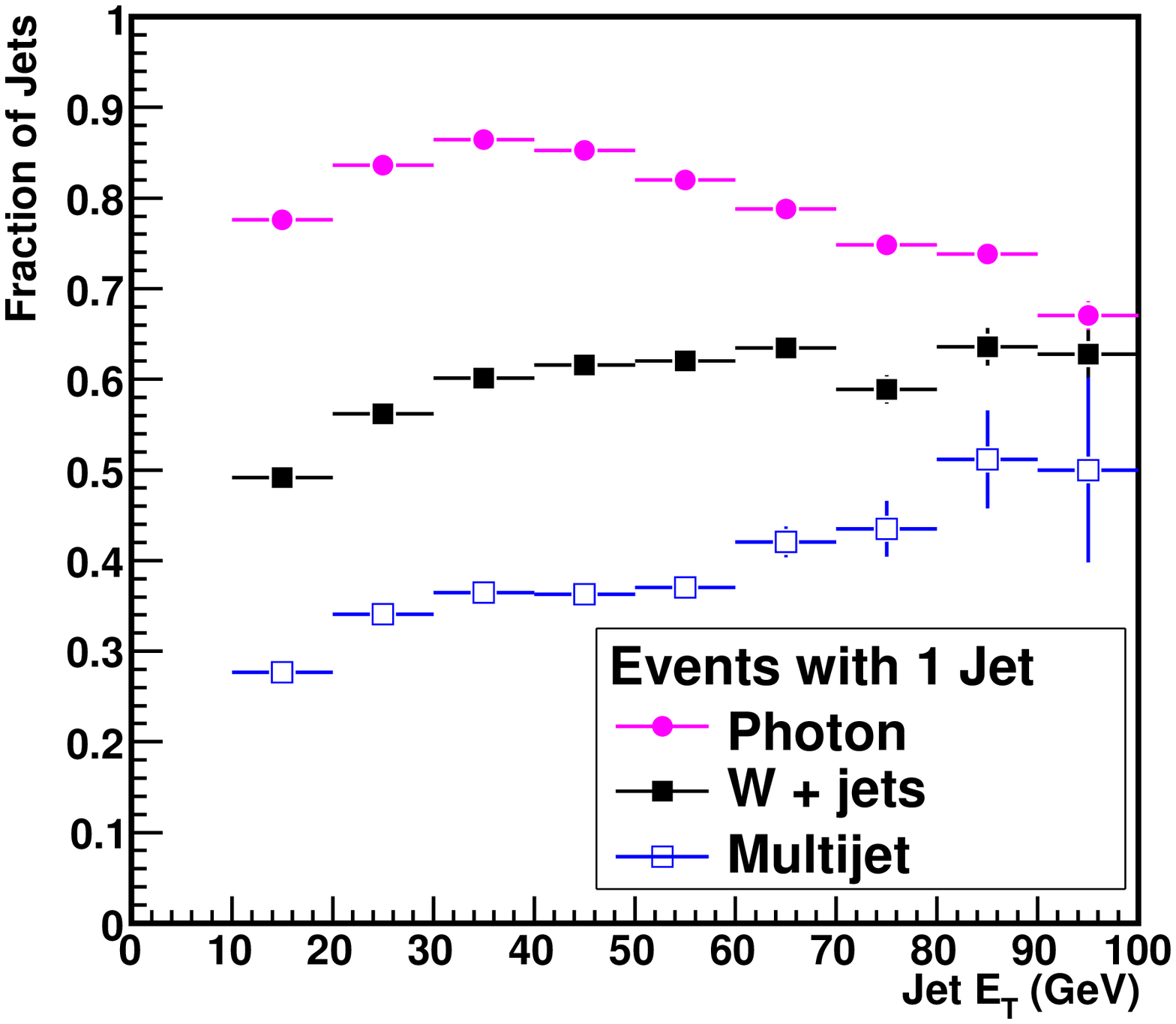}}
\subfigure[]{ \includegraphics[width=0.46\textwidth]{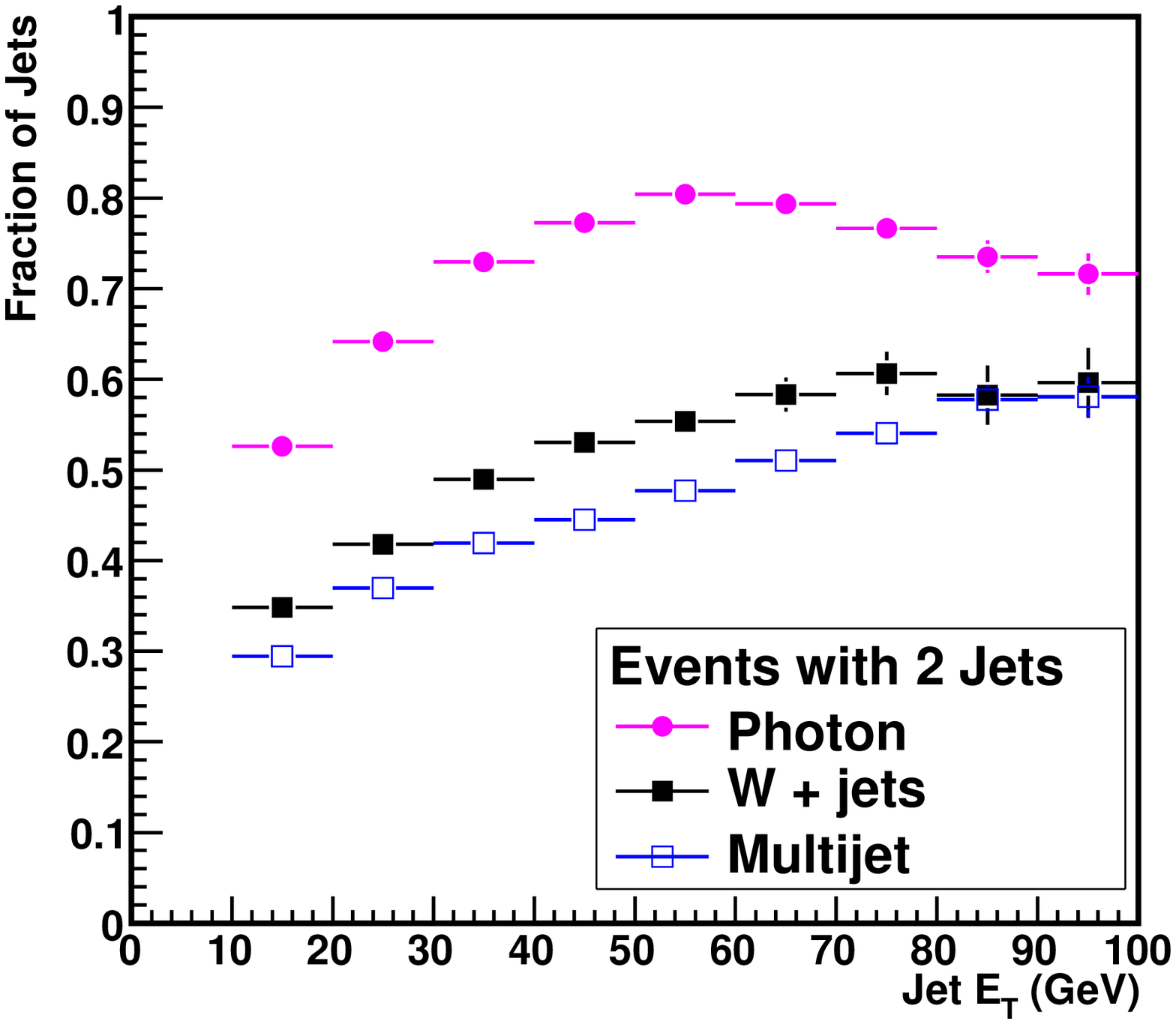}}
\subfigure[]{ \includegraphics[width=0.46\textwidth]{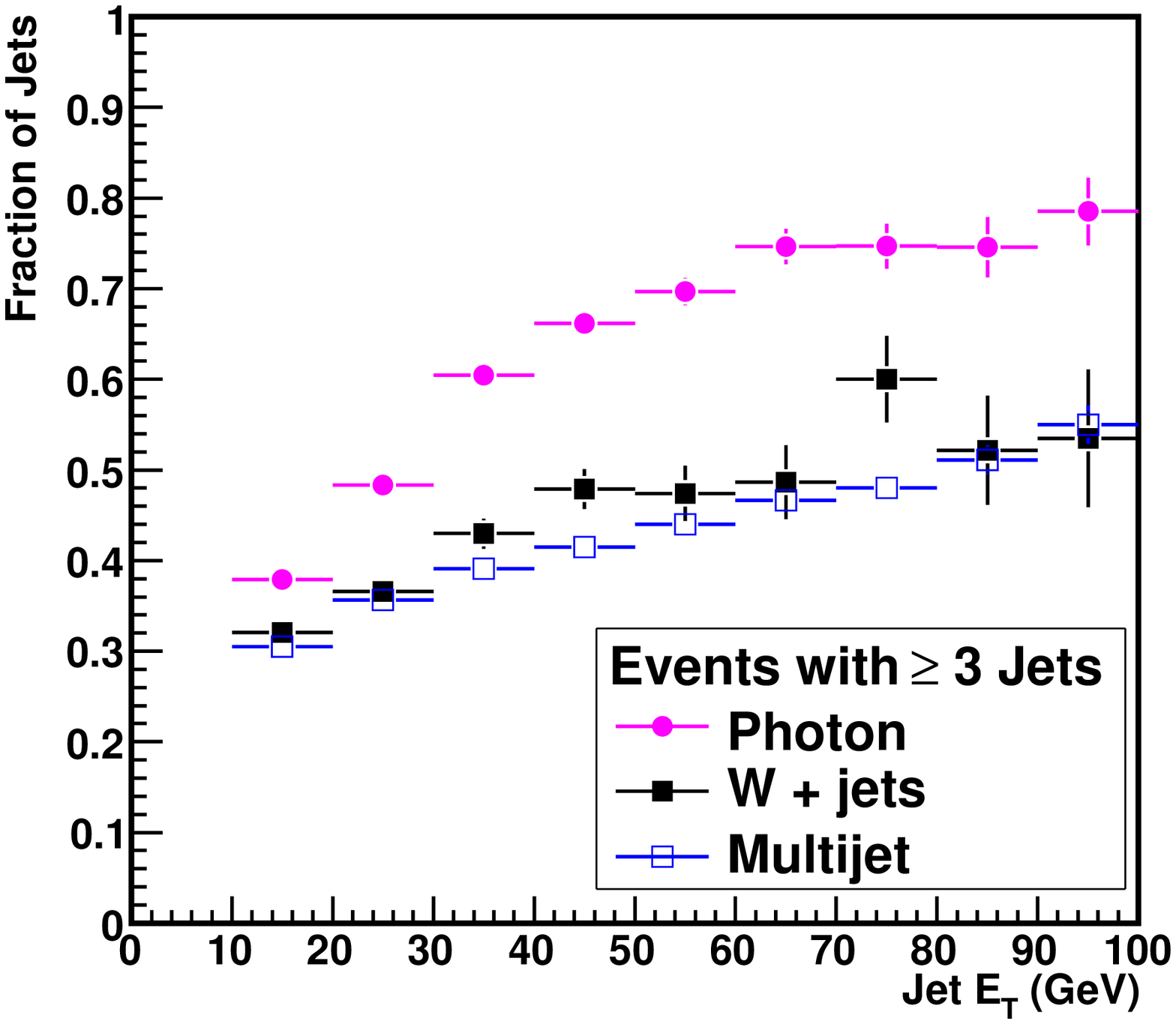}}
  \caption{\label{fig:partons_by_et} 
    The fraction of jets matched to a quark in simulated photon (filled circles), 
    $W$ (filled squares) and multijet (open squares) events,
    as a function of jet $E_T$. 
    The fraction is shown for events with any number of jets (a), and then
    separately for events with one (b), two (c), and at least three (d) jets.
  }
\end{figure*}

We can test the partonic origins of the jets in different processes and their effect
on fake rates using simulated data.  In simulation, it is possible in most cases to match 
a reconstructed
jet to the quark or gluon from which it most likely developed.  While we do not trust the 
absolute value of fake rates in simulation, relative comparisons are still meaningful.
First, we find the fraction of jets matched to a quark in simulated $\gamma$~+~jets, $W$~+~jets, 
and QCD multijet events.  All of these samples are generated using \pythia, as described
in Section~\ref{sec:samples}.  The minimum photon $E_T$ in the photon sample is 22~$\GeV$
and we select events with reconstructed photon \mbox{$E_T > 25\ \GeV$}.  The minimum parton $p_T$
in the jet sample is 18~$\GeVc$.  The quark jet fractions for all of these processes as a function 
of jet $E_T$ are shown in Fig.~\ref{fig:partons_by_et}.  For any choice of jet 
multiplicity, the fraction of jets which are from quarks is highest in the photon sample, next highest 
in the $W$ sample and smallest in the multijet sample.  Also note that in the $W$ and 
photon samples with higher jet multiplicities, the tendency 
toward quark jet dominance persists but is diminished by the enhanced impact of higher-order processes.  
Finally, one also sees that with increasing jet $E_T$, the preference for quark jets in 
multijet events increases. This is also to be expected given 
the increased $Q^2$ of the interaction and hence a reduced probability of an initial $gg$ interaction.

\begin{figure*}[tbp]
\subfigure[]{ \includegraphics[width=0.46\textwidth]{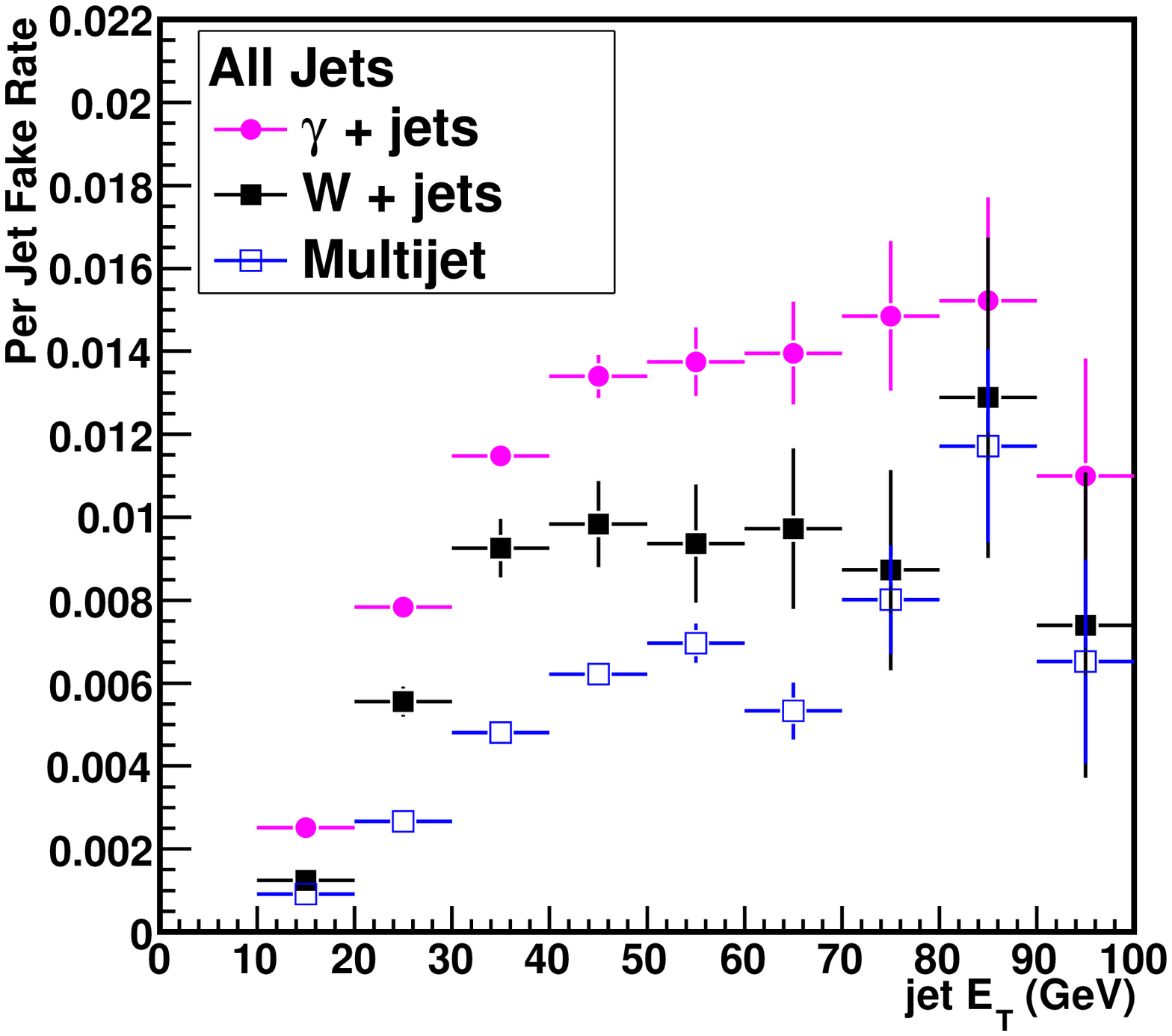}}
\subfigure[]{ \includegraphics[width=0.46\textwidth]{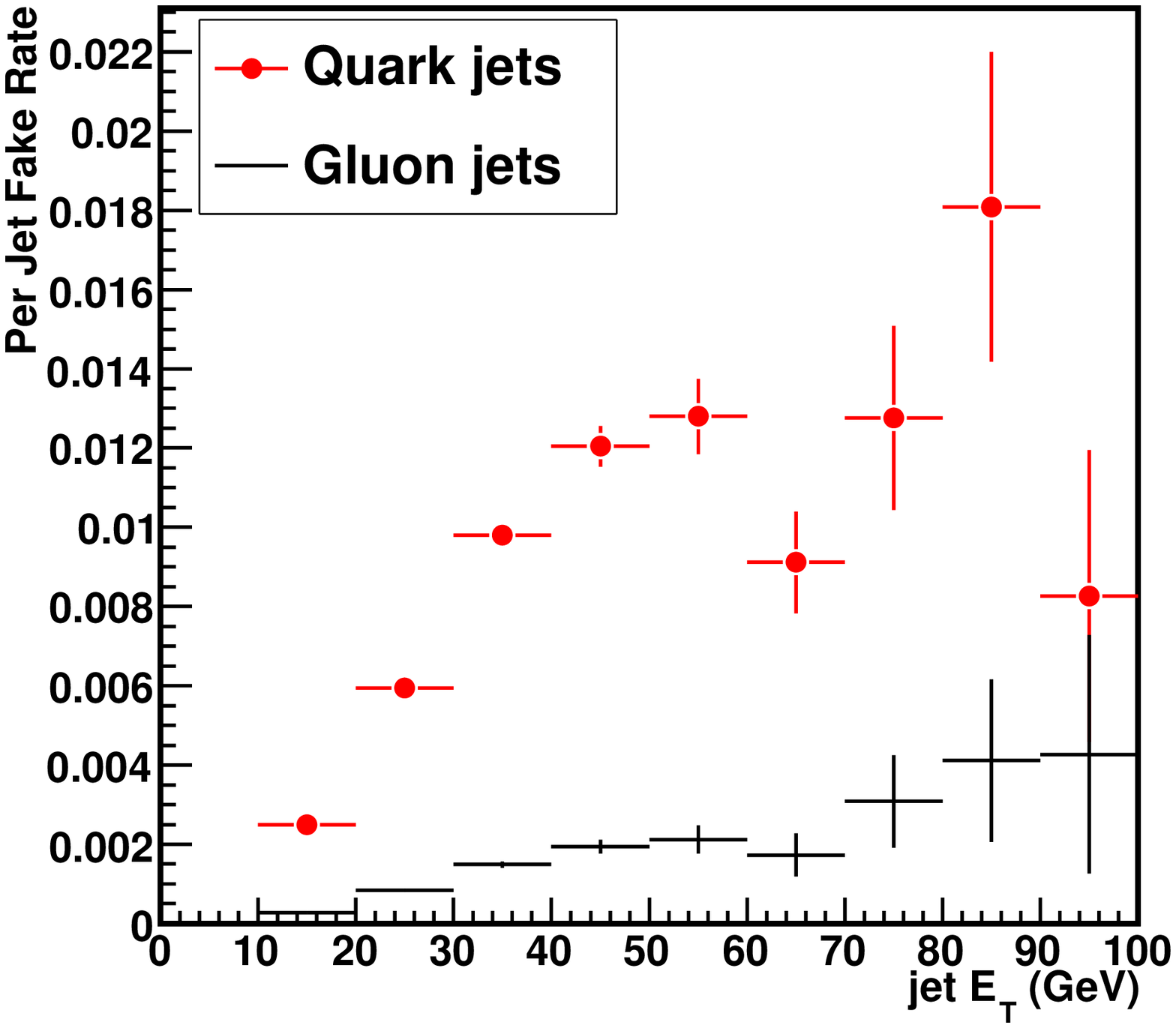}}
\subfigure[]{ \includegraphics[width=0.46\textwidth]{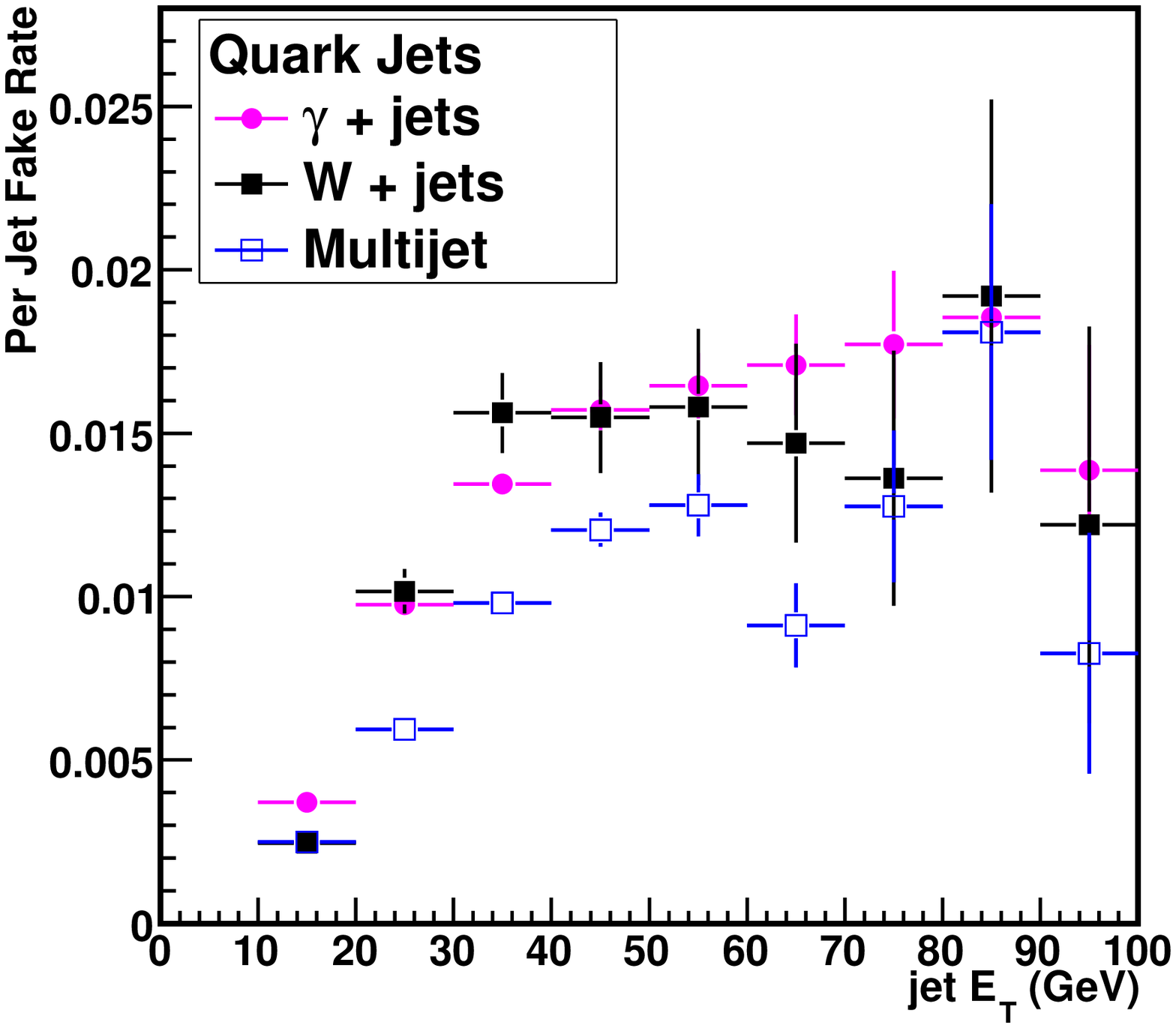}}
\subfigure[]{ \includegraphics[width=0.46\textwidth]{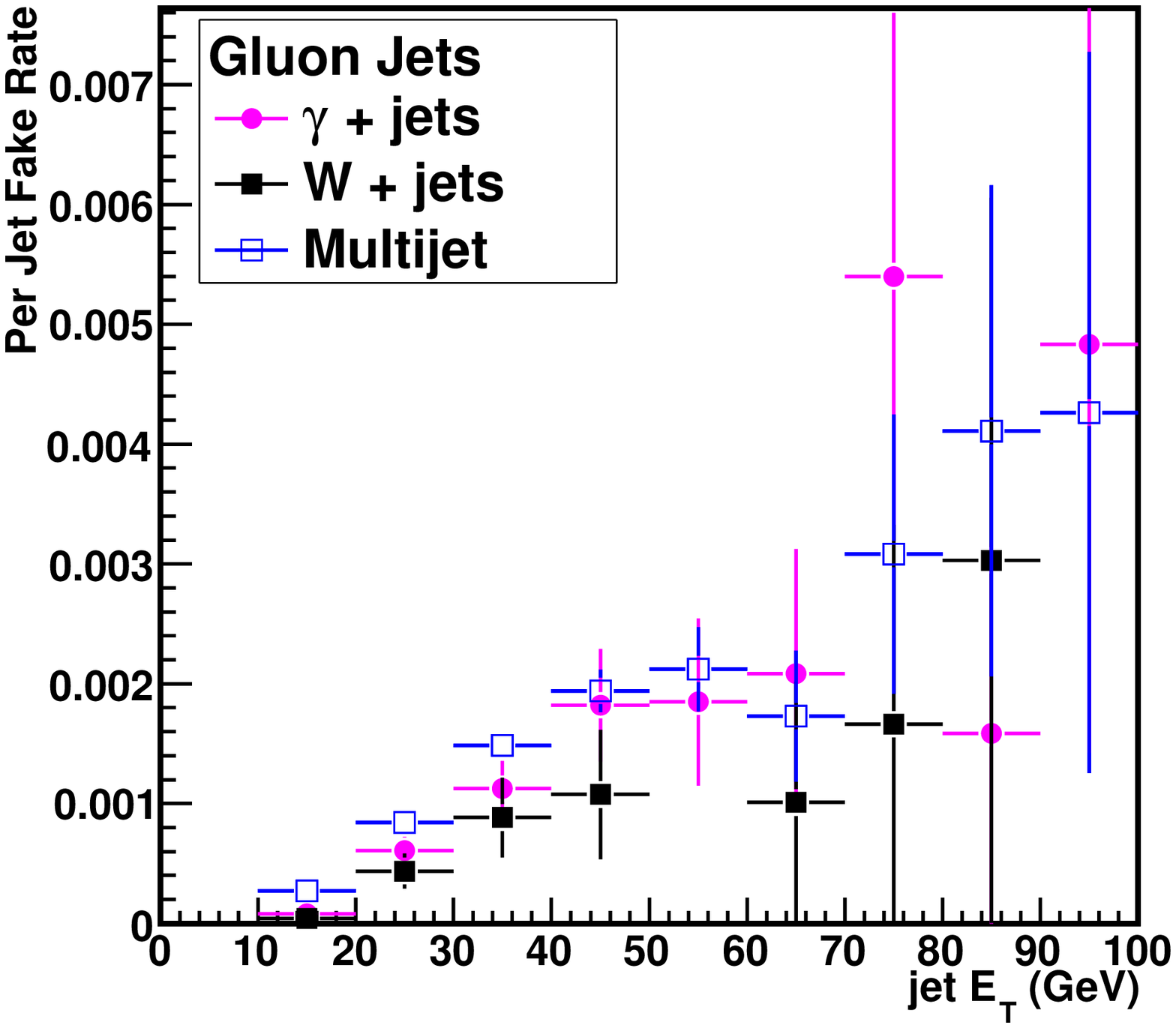}}
  \caption{\label{fig:fakes_by_parton} 
    Track lepton fake rates in simulation as a function of denominator jet $E_T$.  
      (a) Fake rate for all jets in simulated $\gamma$~+~jets, $W$~+~jets, and multijet events.
      The fake rate in multijet events is lower than the fake rate in 
      $W$~+~jets events, and both are lower than that in $\gamma$~+~jets events.
      Note that this $\gamma$~+~jets sample does not have the \mbox{$E_\gamma > 80\ \GeV$} 
      requirement applied. 
      (b) Fake rate for jets matched to quarks (``quark jets'') compared to the fake rate for
      jets matched to gluons (``gluon jets''), showing the disagreement between them.
      Fake rates taken from the simulated multijet sample.
      (c) Fake rates for quark jets in the three samples.  
      The agreement is better than that observed in (a).  
      (d) The same, but for jets matched to gluons.  
  } \end{figure*}

Figure~\ref{fig:fakes_by_parton} shows the difference in the fake rates between the three jet 
samples considered earlier: $\gamma$~+~jets, $W$~+~jets, and QCD multijet.  The $\gamma$~+~jets
sample has the highest fraction of quark jets (See Fig.~\ref{fig:partons_by_et}), followed by 
the $W$~+~jets and multijet samples, and their fake rates follow the same pattern.  
Figure~\ref{fig:fakes_by_parton} also shows that, in simulated multijet events, the quark jet fake rate is 
nearly an order of magnitude higher than the gluon jet fake rate. Combined with
the different propensities for producing quark jets, this leads to different fake rates in different
samples.  We also compare the separated quark and gluon jet fake rates for the three samples in
Fig.~\ref{fig:fakes_by_parton}, and observe that the agreement between the separated fake rates
is better than the agreement between the inclusive ones.  

\subsubsection{Fake Rate Definition}\label{sec:fakes_tlfr}

The track lepton fake rate is the number of isolated tracks, divided by the number 
of jets, in a sample containing no true leptons.  The numerator is the number of track leptons 
according to the definitions in Section~\ref{sec:tracksel}.  The denominator 
is the number of jets according to the definition in Sec.~\ref{ss:jet_defn}.
Recall that in addition to the standard calorimeter cluster-based
jets, this jet collection includes tracks not associated with a jet 
as well as jets containing a high-$p_T$ track which otherwise would
have fallen below the jet selection $E_T$ threshold.  The jets used in constructing 
the fake rate are identical to those used to count jets for candidate event selection.
Note that a fake track lepton is a true isolated track, but one that does not
originate from a lepton.

Ideally, the fake rate is measured in a data sample where the contamination from true leptons 
is negligible.  For this analysis, we use a sample of events triggered by a photon 
with $E_T > 25$~\GeV.  Photons are restricted to the central calorimeter and selected
with criteria similar to those used for electrons, except that there must be no 
track pointing at the energy deposited in the electromagnetic calorimeter~\cite{photon_sel}. 
To further reduce the possibility of lepton contamination, we exclude events in which 
the invariant mass of the photon and any jet in the event is close 
to the $Z$ resonance (\masswin{76}{106}) when the jet is a track or has 
more than about 90\% of its energy in the electromagnetic calorimeter.  Finally, we require the 
photon to have at least 80~\GeV\ of energy (not $E_T$), to strengthen the analogy to $W$ 
production through the required $Q^2$. 

\begin{figure*}[tbp]
\subfigure[]{ \includegraphics[width=0.48\textwidth]{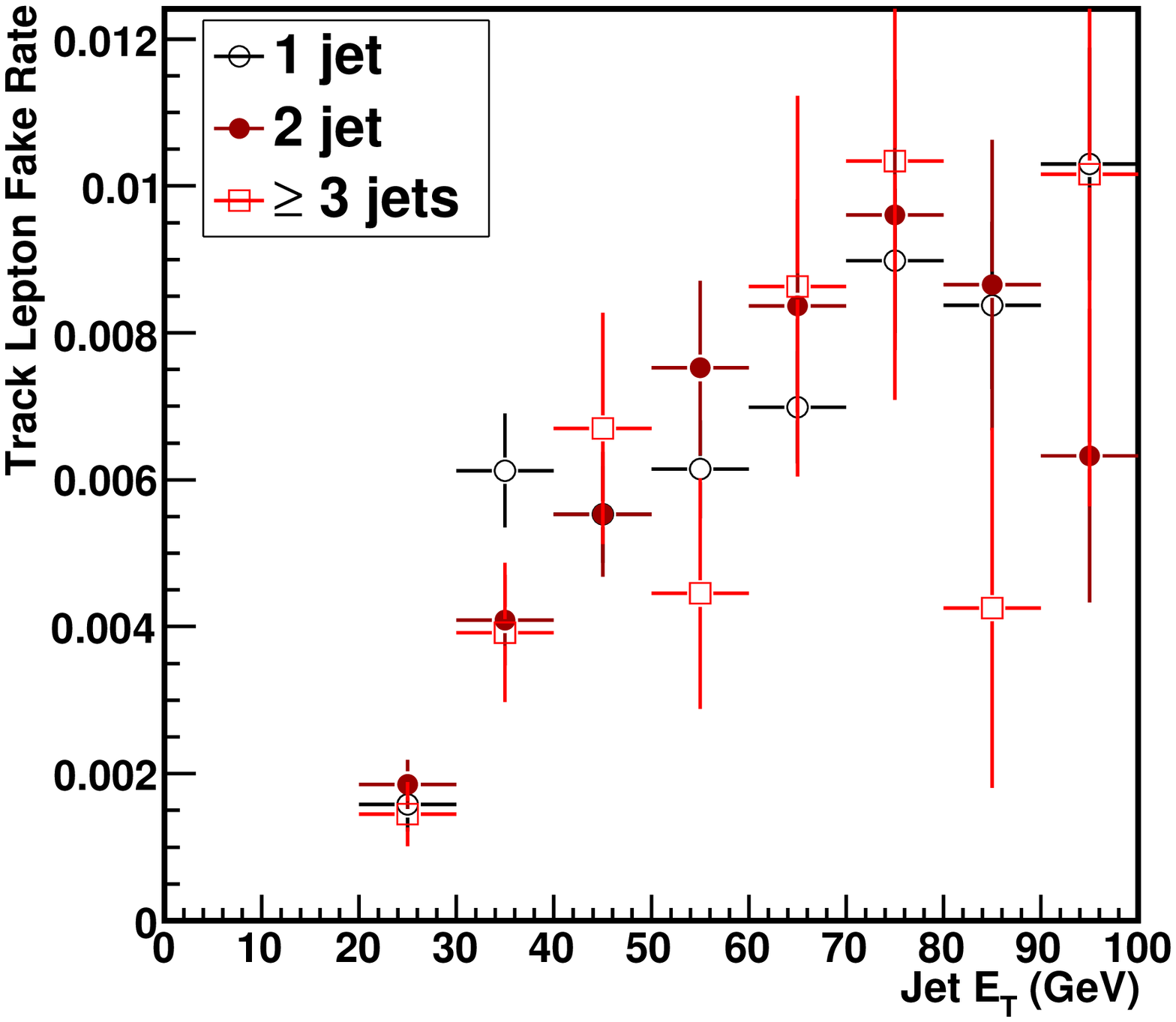}}
\subfigure[]{ \includegraphics[width=0.48\textwidth]{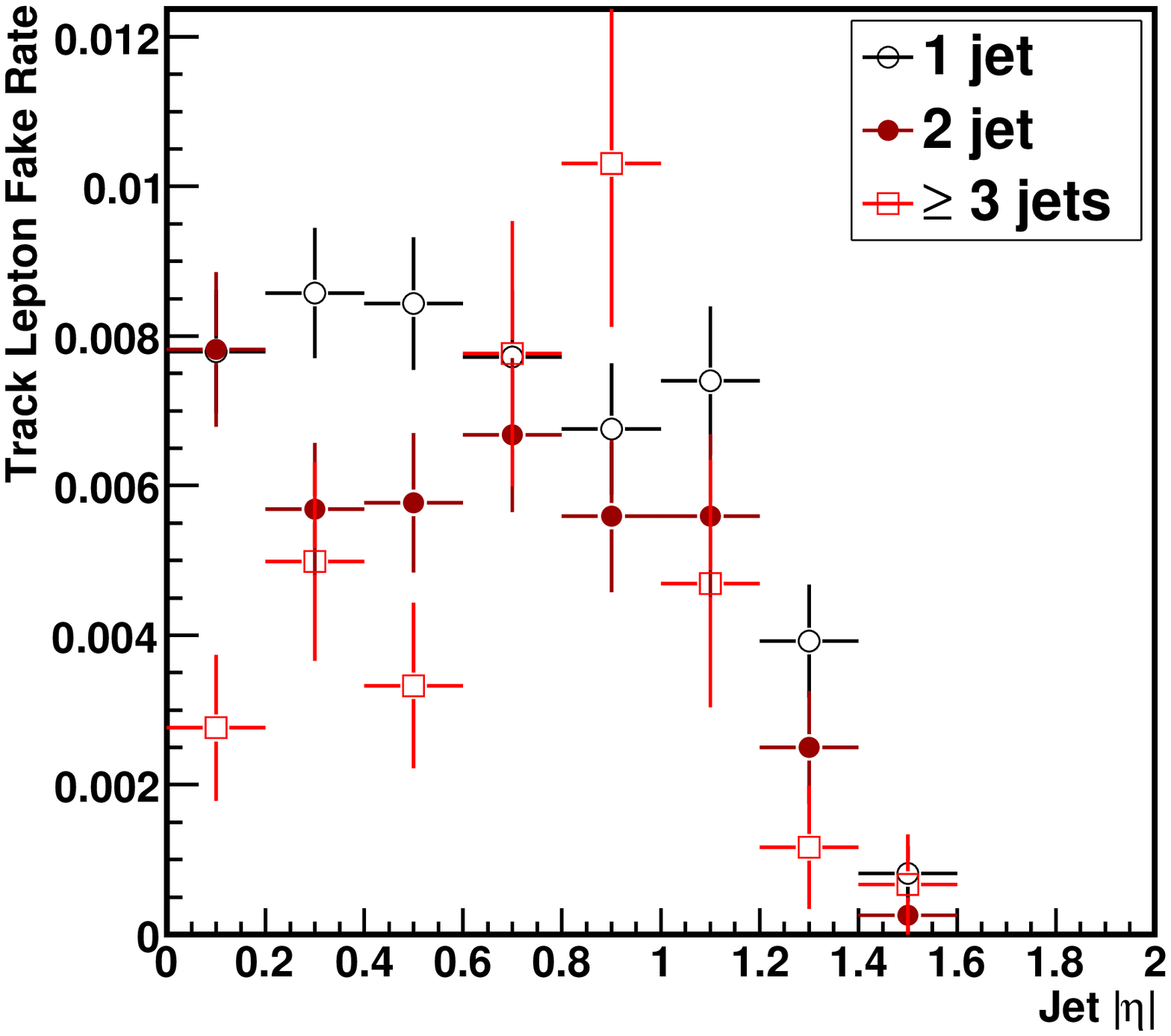}}
  \caption{\label{fig:isotl_fakerates} 
    Track lepton fake rates as a function of $E_T$ (a)
    and $|\eta|$ (b) of the faking jet.  The fake rate is the probability for a 
      jet to pass the track lepton identification requirements, including the track isolation.
      The fake rate is measured in jets from photon~+~jets data, and includes $Z$~+~1 jet
      events in the 1 jet fake rate.  The uncertainty is statistical only.  See 
      Section~\ref{sec:fakes_tlfr} for details.
  } \end{figure*}

The fake rate depends strongly on the $E_T$ and $|\eta|$ of the jets in the denominator, so 
we parameterize it as a function of these quantities.  The fake rate for 
track leptons for each jet multiplicity is shown as a function of the $E_T$
and $|\eta|$ of the denominator jets in Fig.~\ref{fig:isotl_fakerates}.

Taking the jets for the lepton fake rate from a photon-triggered data sample instead of a 
jet-triggered has not been done before in a dilepton $t\bar{t}$ cross section measurement.
The Run~I dilepton cross section measurement~\cite{run1_cdf_dil_prl} and the Run~II 
measurement with 200~\pb\ of integrated luminosity~\cite{run2_dil_prl} both used samples triggered by high-$E_T$ jets.
The Run~I measurement placed an uncertainty of 62\% on the fake lepton background.  
In the two measurements of the previous Run~II result, the uncertainty on 
the fake lepton background is 30\% for the earlier version of this analysis, and 
51\% for the analysis using two fully reconstructed leptons.  The uncertainty on the
fake lepton background for this analysis will be described in detail at the end of this
section, but it is a total of 20\%, with 6\% from the statistical uncertainty.  
The previously published measurements have a factor of 5 to 10 less integrated 
luminosity, so the statistical uncertainty would have made this technique impractical
in earlier measurements.  As an aside, it is because of the availability of larger data 
samples that the failings of fake rates calculated with 
jet-triggered samples have started to become apparent.
We also note that our initial attempts to improve the fake lepton background estimate
were based on adding a third parameter, such as the number of tracks per jet, to the
fake rate.  This technique was dropped in favor of the one presented here because it
was less successful when tested in simulation.


\subsubsection{Use of $Z$+jets Data}\label{sec:fakes_zjets}

In photon plus one jet events, conservation of momentum 
implies that events where the measured $E_T$ of the balancing jet is significantly 
lower than the photon $E_T$ are rare.  Because of this, there are very few
events in the lowest $E_T$ bins of the fake rate.  This is not crucially important since this 
is an input to the zero jet event count prediction, which does not enter into the cross section
calculation.  Nevertheless, it is possible to fill the gap in the one jet fake rate 
by including the $Z$~+~1
jet sample.  The $Z$~+~jets sample is a near-perfect analog to the $W$~+~jets sample, up to the 
slight difference in mass scale.  Most $Z$~+~jets events have at most one jet, typically 
near the $E_T$ threshold, so there is not enough data to make useful measurements of 
lepton fake rates for higher jet multiplicities.  But, one-jet events where the jet is in 
the lower $E_T$ range is exactly what $\gamma$~+~jets events are lacking.  Therefore, 
the total rate used to predict the number of fake leptons in the zero-jet lepton~+~track 
sample is the combined rate from the $\gamma$~+~1 jet and $Z$~+~1 jet samples.  We combine the two fake rates by
adding the jets from the $Z$~+~1 jet numerator (denominator) to the $\gamma$~+~1 jet 
numerator (denominator) before calculating the fake rate.

\subsubsection{Validation of Track Lepton Fake Rate}\label{sec:fakes_checkfr}

We test the accuracy of the track lepton fake rate in both real and simulated 
data.  To increase the size of the sample for validation, a lower
kinematic threshold of 15~\GeV\ is used for both the track leptons and
jets.  This adds jets to the sample because of the steep falloff of the $E_T$ distribution.
Also, the fake rate for 15~\GeV\ track leptons is higher than for 20~\GeV\ 
track leptons, because it corresponds to a larger portion of the 
fragmentation spectrum. 

\begin{figure*}[tbp]
\subfigure[]{ \includegraphics[width=0.32\textwidth]{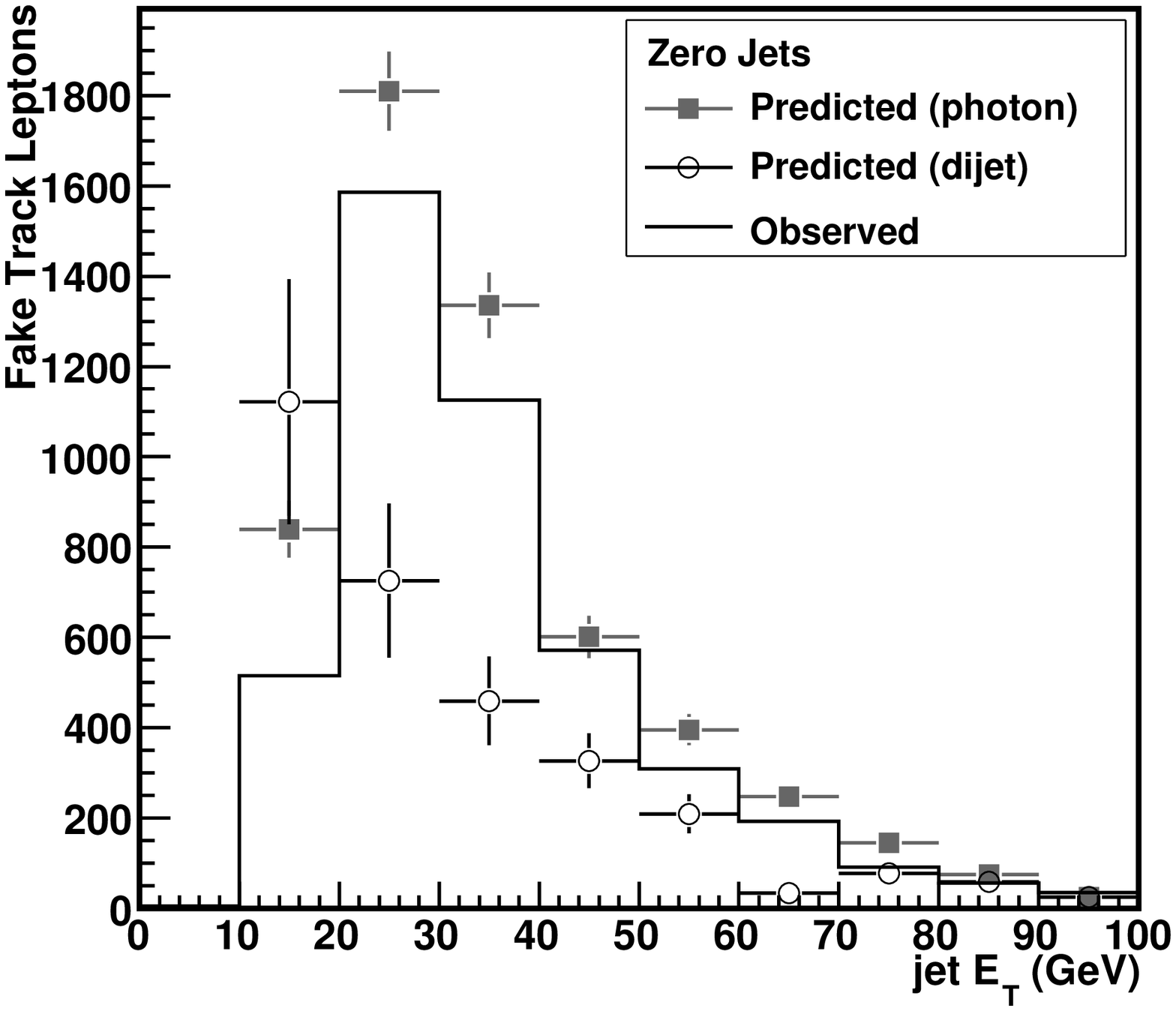}}
\subfigure[]{ \includegraphics[width=0.32\textwidth]{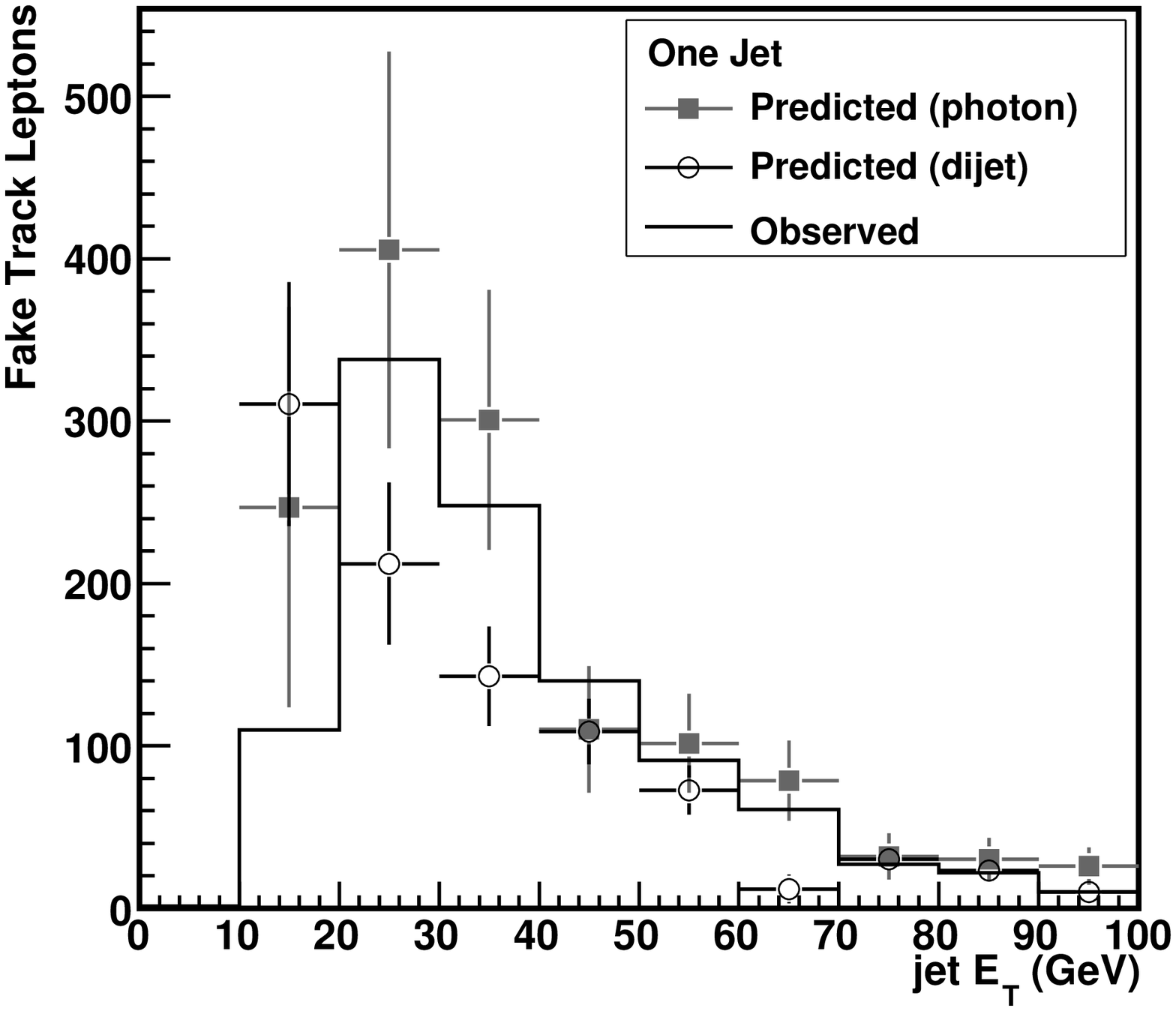}}
\subfigure[]{ \includegraphics[width=0.32\textwidth]{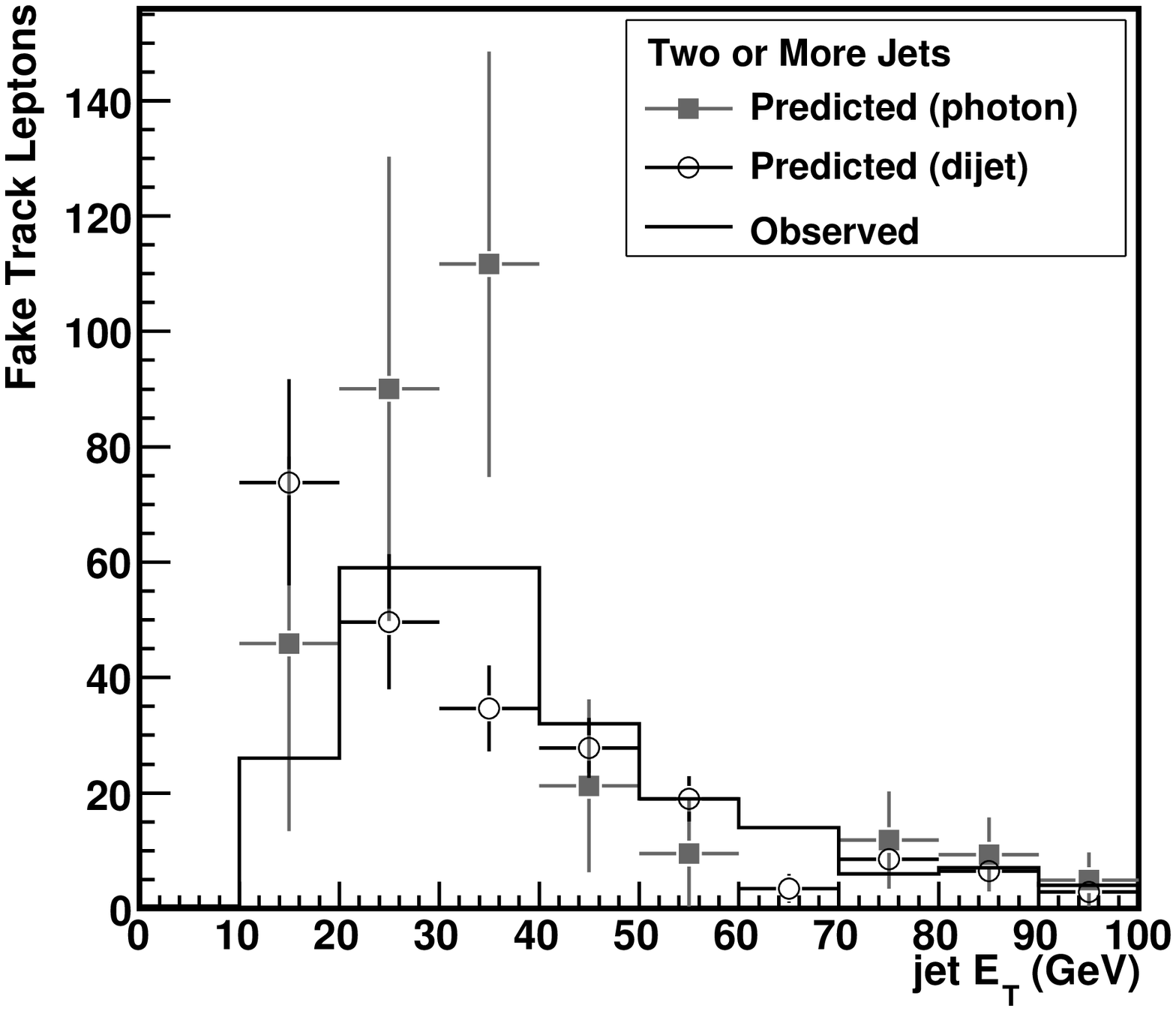}}
  \caption{\label{fig:mc_faketests} 
		Predicted and observed number of fake track leptons in simulated $W$~+~jets
		events, as a function of jet $E_T$.  The comparison is made for events
	        with zero (a), one (b), and at least two (c) jets in addition to the one that
		is reconstructed as an isolated track.  
		One prediction, shown as solid squares, is made using the a fake rate derived 
		from simulated $\gamma$~+~jets and $Z$~+~jets events, exactly as the fake rate 
		is constructed from the observed data.  The other, shown as open circles, 
		is made using a fake rate derived from simulated QCD multijet events.
		Uncertainties on the predictions are statistical only.  
  }
\end{figure*}

Using simulated CDF data, it is possible to test the fake rate estimation procedure 
by using the fake rates obtained from simulated $\gamma$~+~jets events to predict 
the number of fake leptons in simulated $W$~+~jets events.  Events with 0, 1, 
and $\geq$ 2 jets in addition to the fake lepton are considered.
Figure \ref{fig:mc_faketests} shows the predicted and observed number of
fake track leptons as a function of the $E_T$ of the misidentified jet, for each jet
multiplicity. The integrated results are provided in Table \ref{tab:faketests}.
The fake rate from jets associated with an 80~\GeV\ photon is seen to overestimate the number 
of fake leptons observed in jets associated with a $W$.  This effect is 
only statistically significant in events with one jet,
and we will include this 18\% discrepancy in the systematic uncertainty on this background.  
Omitting the $Z$~+~1 jet data from the fake rate only exacerbates the disagreement  
(The previous section describes the use of the $Z$~+~jets data in the one jet fake rate).  

\begin{figure*}[tbp]
\subfigure[]{ \includegraphics[width=0.32\textwidth]{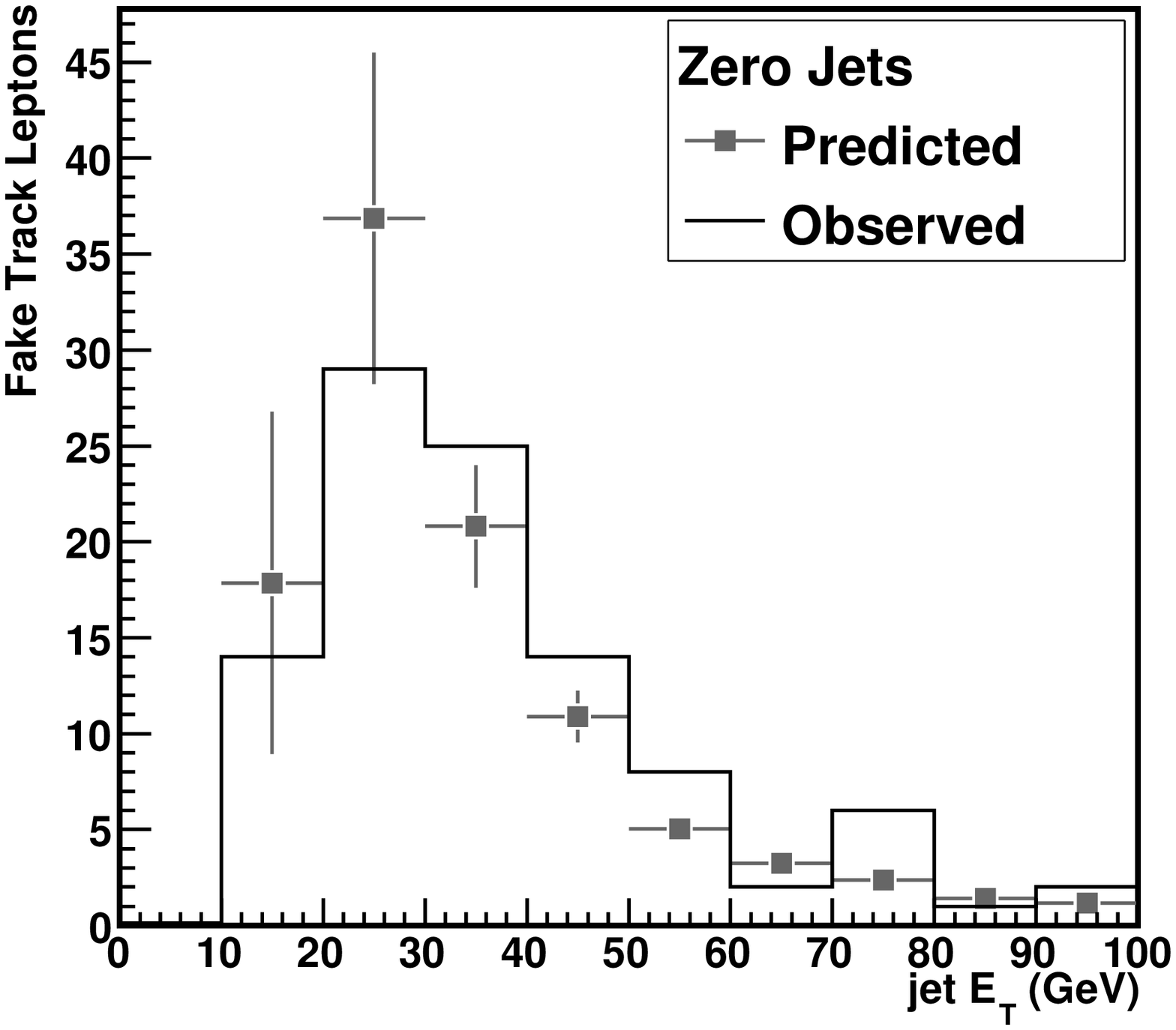}}
\subfigure[]{ \includegraphics[width=0.32\textwidth]{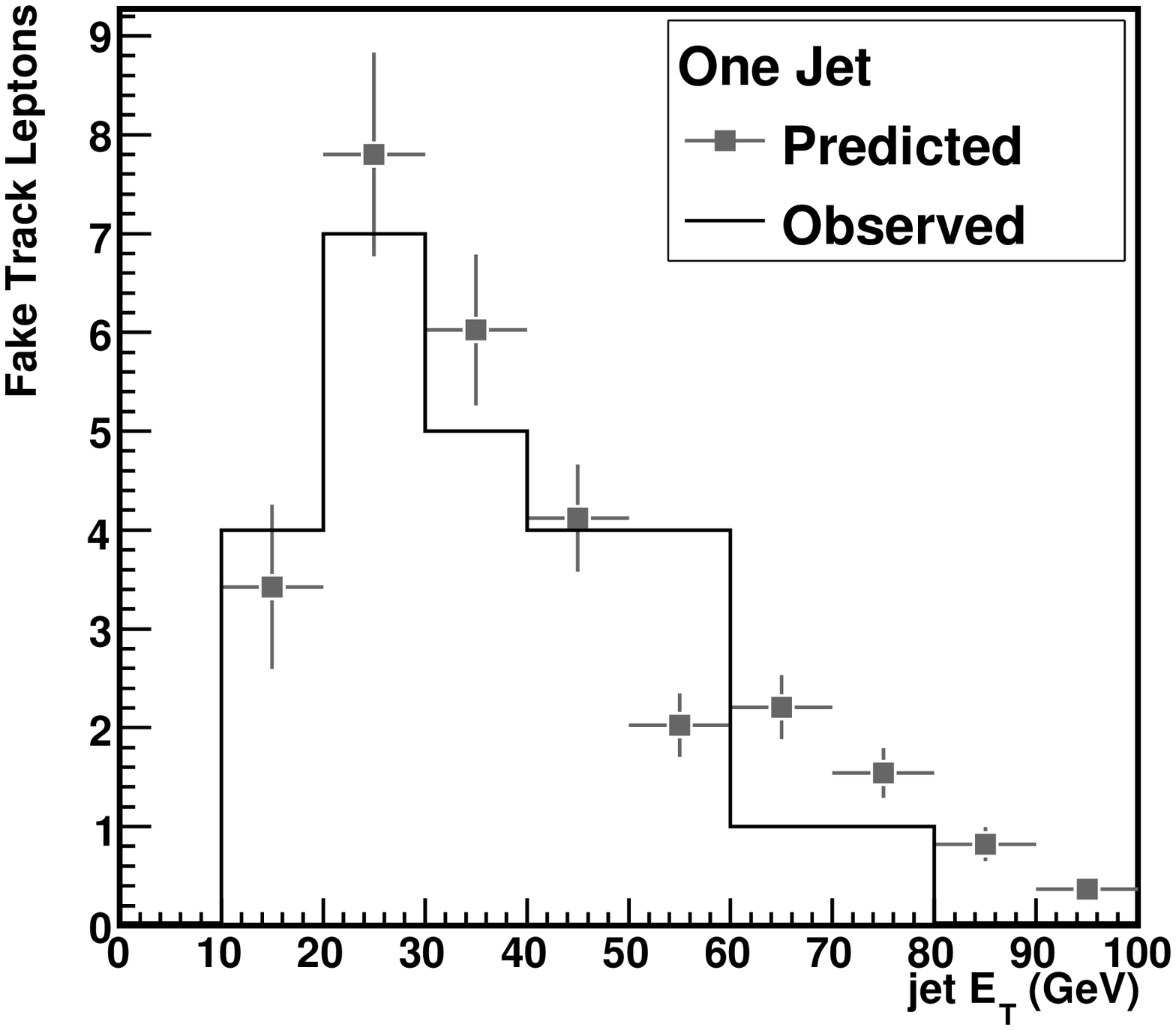}}
\subfigure[]{ \includegraphics[width=0.32\textwidth]{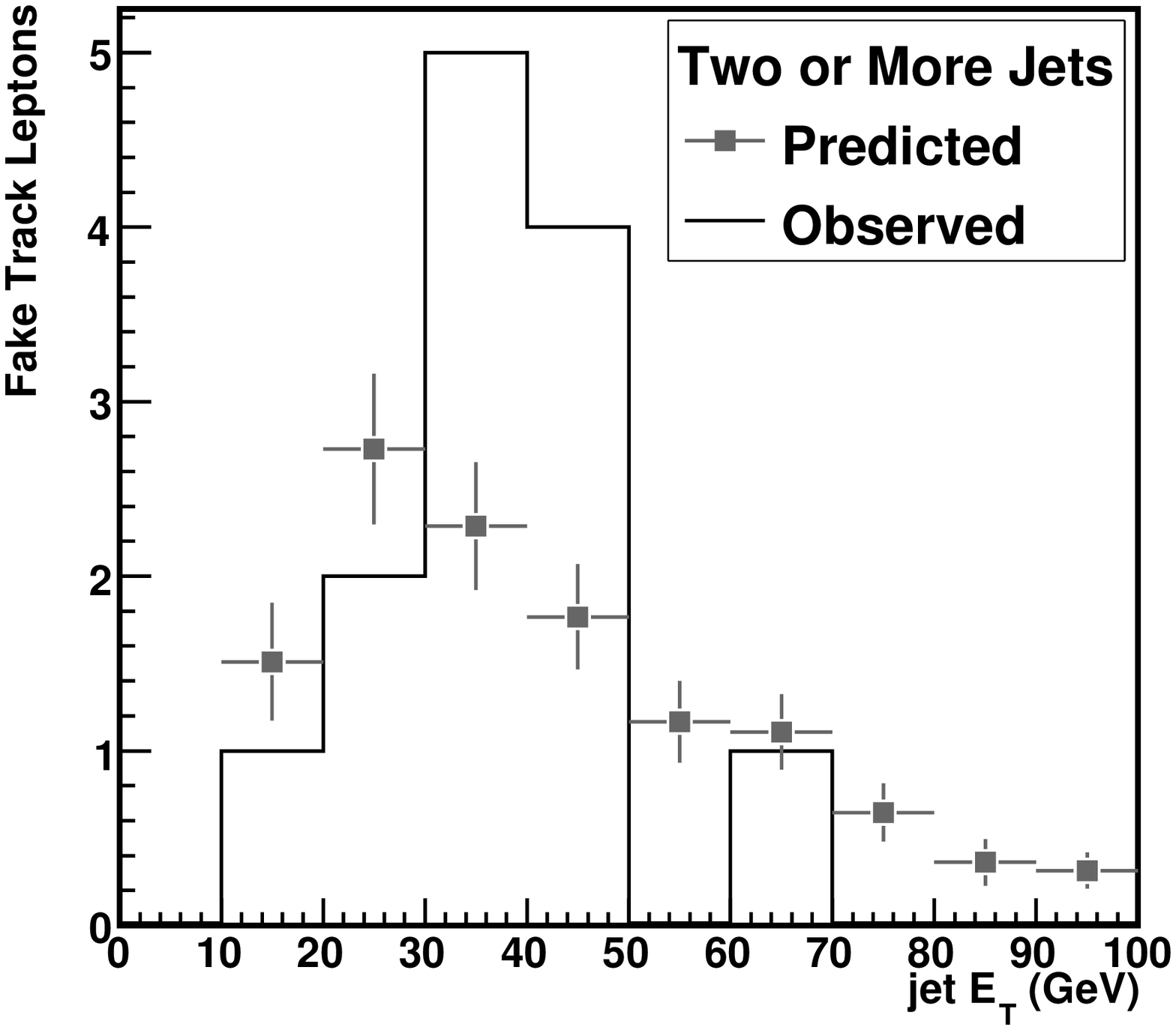}}
  \caption{\label{fig:data_faketests} 
		Predicted and observed number of isolated tracks observed in CDF $Z$~+~jets 
		data, as a function of jet $E_T$.  The comparison is made for events
	        with zero (a), one (b), and at least two (c) jets in addition to the one that
		is reconstructed as an isolated track.
		The fake rate used is derived from $\gamma$~+~jets data, and the uncertainty
		on it is statistical only. 
  }
\end{figure*}

We directly test the fake rate obtained from the photon data using $Z$~+~jets data.  
The number of isolated tracks predicted in the $Z$~+~jets data is compared
to the number observed for events with 0, 1, and $\geq 2$ jets in addition to the 
isolated track.  For this test, only the fake rate from $\gamma$~+~1 
jet events is used to predict
the number of fake leptons coming from the $Z$~+~1 jet sample.  Although the event sample
is small, no statistically significant discrepancy is observed for any jet multiplicity in this test.  
Figure \ref{fig:data_faketests} shows the predicted and observed number of
isolated tracks as a function of the $E_T$ of the jet, for each jet
multiplicity. Table \ref{tab:faketests} provides the integrated results.  The integer number
of isolated tracks is well-predicted, and the shape of the isolated track $p_T$ 
distribution is well-modeled for events with zero or one jets in addition to the isolated track.
The agreement between the predicted and observed distributions in the two-or-more
jet case is more difficult to assess.  There are 2358 $Z$ events with
three jets, so the predicted distribution is smooth, but there are only thirteen
$Z$ candidates with two jets and an isolated track, so the distribution
of the $p_T$ of the isolated tracks is highly prone to statistical fluctuations.

We use the same framework to test fake rates obtained from multijet events.
To mimic the fake rate used in the previous published version of this analysis,
we simulate the requirements of the 50~\GeV\ jet trigger 
at CDF on a \pythia\ multijet sample produced requiring a minimum parton $p_T$ of 18~$\GeVc$. 
The fake rate is constructed as described above, except that events with any 
number of jets are included.  The inclusive jet fake rate is then applied to jets
from simulated $W$~+~jets events.  The results are shown next to the prediction from the 
photon~+~jets fake rate in Fig.~\ref{fig:mc_faketests}.  For $W$~+~1 jet events,
3040~$\pm$~350 events are predicted, and 4480 are observed.  Using the same logic as 
used to derive the 18\% systematic uncertainty quoted above, this corresponds to a 47\% 
systematic uncertainty for the fake rate from data collected using a jet trigger. 
This motivates the choice to use the jets from the photon trigger sample to build the fake
rate. 




\begin{table}[tbp]
\begin{center}
\begin{tabular*}{8.6cm}{l}
\hline\hline
Predict $W$~+~jets with $\gamma$ and $Z$~+~jets:\\
Test in simulated data
\end{tabular*}
\begin{tabular*}{8.6cm}{@{\extracolsep{\fill}}ccc}
		& Predicted	& Observed	\\
\hline
1 jet		& 5473 $\pm$ 147	& 4480		\\	
2 jets		& 1332 $\pm$ 200	& 1047		\\	
$\ge$ 3 jets 	&  304 $\pm$ 67		& 226		\\
\hline
\end{tabular*}
\begin{tabular*}{8.6cm}{l}
Predict $Z$~+~jets with $\gamma$~+~jets:\\ 
Test in CDF data
\end{tabular*}
\begin{tabular*}{8.6cm}{@{\extracolsep{\fill}}ccc}
		& Predicted	& Observed	\\
\hline
1 jet		& 100 $\pm$ 13	& 101		\\	
2 jets		& 28 $\pm$ 2	& 26		\\	
$\ge$ 3 jets 	& 12 $\pm$ 1	& 13		\\
\hline\hline
\end{tabular*}
\end{center}
 \caption{
	Predicted and observed number of isolated tracks of hadronic origin in 
	tests performed in observed and simulated data.  The number of jets quoted is in 
	addition to the jet which is reconstructed as an isolated track.  The only 
	statistically significant discrepancy observed
	is in the one jet category in the simulation, and is taken as
	part of the basis of the systematic uncertainty on this background estimate.
	Note that in the column headings, ``Observed'' refers to the directly counted
	isolated tracks, regardless of whether the study is done in real or simulated data.
}
 \label{tab:faketests}
\end{table}

\subsubsection{Fake Rates for Fully Reconstructed Leptons}\label{sec:fakes_tightl}

\begin{figure*}[tbp]
\subfigure[]{ \includegraphics[width=0.48\textwidth]{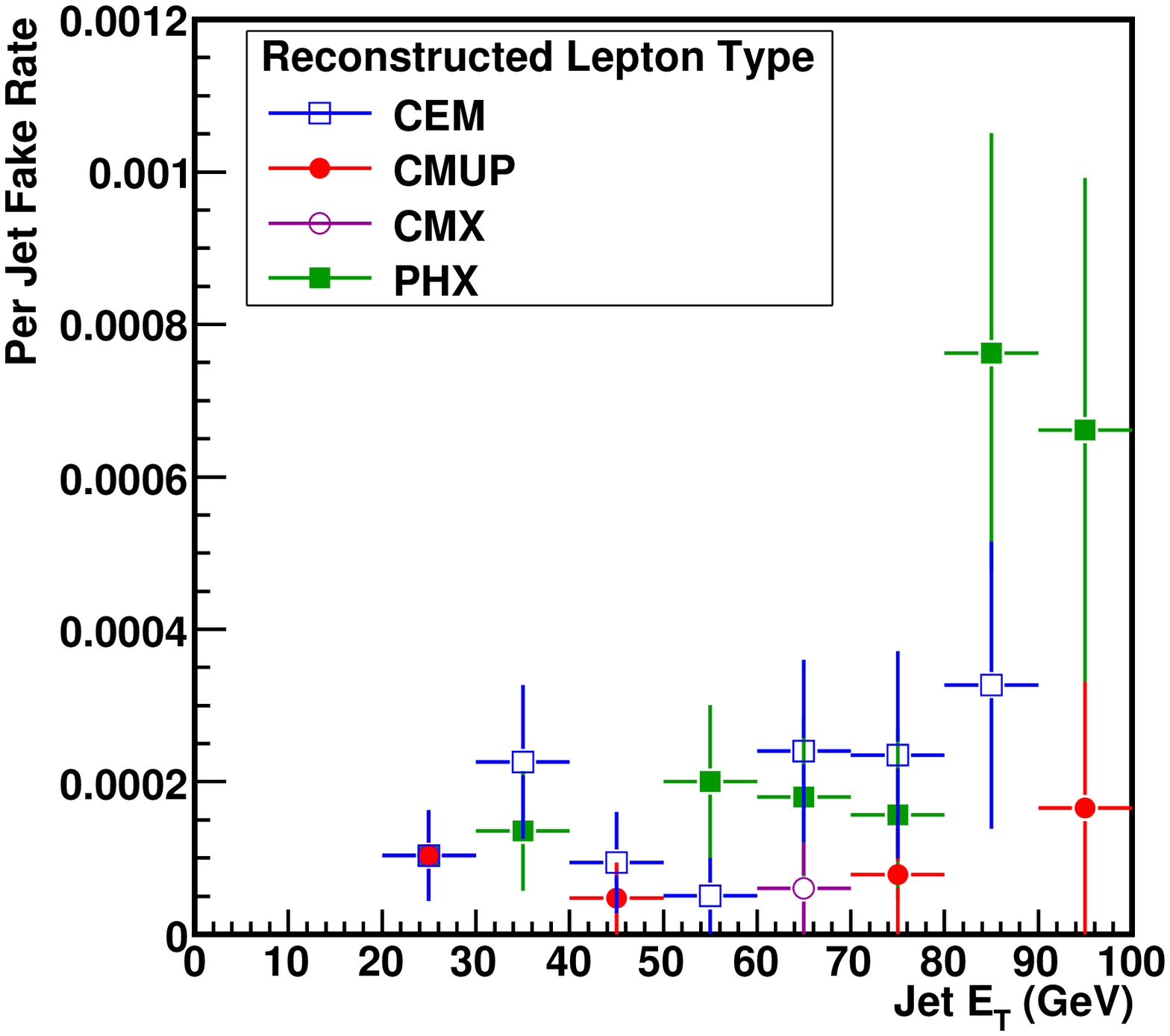}}
\subfigure[]{ \includegraphics[width=0.48\textwidth]{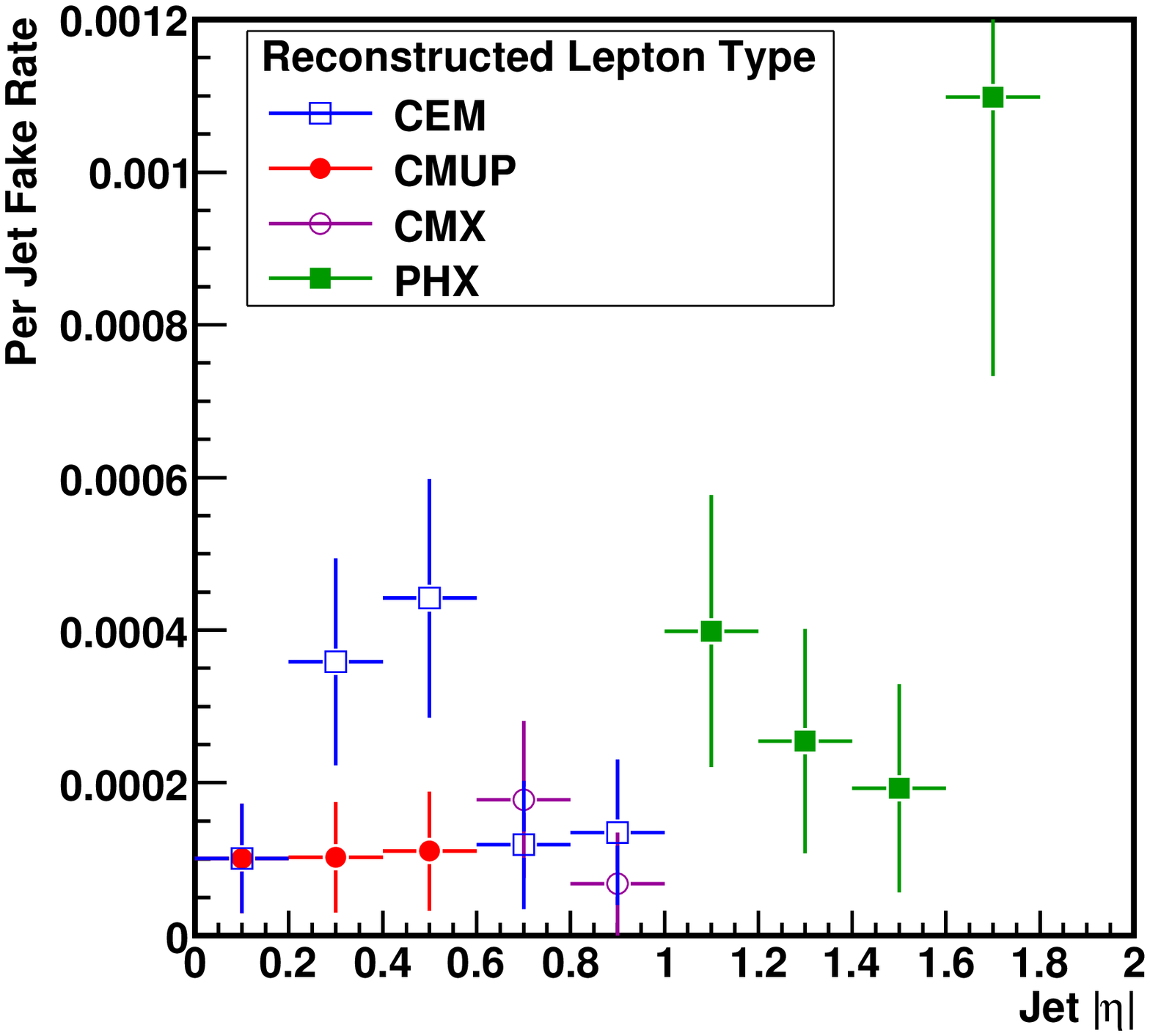}}
  \caption{\label{fig:tight_fakerates} Fully reconstructed lepton fake rates as a 
    function of $E_T$ (a) and $|\eta|$ (b) of the jet in the denominator.  The fake rate is the 
      probability for a jet to pass the lepton identification requirements and
      is measured in jets from photon~+~jets data.
      The shown uncertainty is statistical only.  
  } \end{figure*}

We also measure fake rates for all four primary lepton types, using the same
method and data as is used for the track leptons.
The fake rate is at least an order of magnitude smaller for primary leptons 
than for track leptons.  Because there are so few events in the numerator 
of the fake rate for the primary leptons, we use an inclusive fake rate 
instead of calculating it separately for events with different numbers of jets. 
The fake rates for the fully reconstructed leptons, as a function of the $E_T$
and $|\eta|$ of the denominator jets, are shown in in Fig.~\ref{fig:tight_fakerates}.

To find the number of events where the track lepton is the lepton from the $W$ and the fully 
reconstructed lepton is fake, we multiply the fully reconstructed lepton fake rates by 
the jet $E_T$ and $|\eta|$ distributions from the lepton~+~\met~+~jets data 
described in Section~\ref{sec:fakes_wjetssel}.  That gives the number of events with two
fully reconstructed leptons where one is fake.  To find the number where the true lepton 
is the track lepton, we scale the number by the ratio of the $W\to\ell\nu$~+~jets acceptance for 
track leptons to the acceptance for fully reconstructed leptons. 
This ratio is measured in $W$~+~jets events simulated using \pythia.  There are other sources
of fake leptons in the sample, but $W$~+~jets is the dominant contribution, and the ratio of the
acceptances for the sub-dominant contribution from $t\bar{t}$ should be similar, since the real lepton
is still from a $W$.  Summing over all lepton types yields the total contribution from fake 
fully reconstructed leptons, which are 6\% of the total fake lepton background.

\subsubsection{Contribution from $t\bar{t}$ Events with a Fake Lepton}\label{sec:fakes_ljets}

Other than $W$~+~jets, the only process contributing significantly 
(more than 5\%) to the lepton~+~\met~+~jets data sample is $t\bar{t}$ lepton~+~jets events.
For a top quark mass of 175~$\GeVcc$, corresponding to a cross section of 6.7~pb, 
such events produce 19\% of the jets in the lepton~+~\met~sample with three or more 
jets, the sample used derive the fakes contribution to the background for
the cross section measurement.  Lepton~+~jets $t\bar{t}$ events make a negligible 
contribution to lepton~+~\met\ samples with one or two jets.
Single top quarks, produced through the electroweak interaction, can also generate
the required lepton, \met, and jets signature, but the small cross section and lower
jet multiplicity at leading order mean that their contribution is negligible.

Because of its size, the lepton~+ jets $t\bar{t}$ contribution deserves 
separate consideration.  Study of simulated $t\bar{t}$ events indicates
that over 90\% of the jets in these events come from quarks. This is a very 
different fraction than for any other sample considered, although 
many (roughly 2/3) of the jets are from heavy quarks, such as a $b$ from $t$ decay or
a $c$ from $W$ decay.  The track lepton fake rate for jets from a light
quark is about twice the rate for jets from a heavy quark, because the latter
typically produce a larger number of charged tracks, leading to a greater 
likelihood to fail the track isolation criteria.
As a result, the fake lepton is associated with the jet produced by a quark 
from $W$ decay in about 90\% of the simulated events with a fake lepton.

The fake rate designed for $W$~+~jets events is still usable.
First, the smaller probability that heavy quark jets will be manifested as isolated tracks 
partially compensates for the fact that there are fewer gluon jets in $t\bar{t}$ than in 
the photon plus jets data used to define the fake rate. Second, the contribution of 
$t\bar{t}$ is still relatively small compared to $W$~+~jets and the large systematic on 
the fake rate due to the discrepancy in the single jet bin is adequate to cover any remaining 
difference. We test the validity of these statements using the fake rates from the 
simulated $W$~+~jets sample, because there is an insufficient number of events in the simulated 
\mbox{$\gamma + \geq 3$} jet sample to make a meaningful comparison.  If the $W$~+~jets fake rate
is sufficiently similar to the one from $t\bar{t}$ lepton~+~jets events, we may still
use the same fake rate for both.  
The fake rate from the $W$~+~jets sample predicts 518~$\pm$~45 isolated tracks 
in a simulated $t\bar{t}$ lepton~+~jets sample, in which 424 are actually observed.  
The level of disagreement (18\%) is not egregious when compared to the statistical 
uncertainty,
and is comparable to the systematic uncertainty.  We therefore use the photon~+~jets fake 
rate to predict the number of fake leptons for all events passing
the $W$~+~jets selection, regardless of their source.


\subsubsection{Fraction of Events having Opposite Sign Leptons}\label{sec:fakes_os}

We must also estimate the fraction of events with a fake lepton 
in which the real and fake lepton have opposite sign.  
This fraction is different for $W$~+~jets and $t\bar{t}$ events, so they
must be considered separately.

First we will consider the \mbox{opposite-sign} fraction for $W$~+~jets events.
Leading diagrams for $W$~+~jets production have the $W$ recoiling against a
quark which becomes a jet in the event, so the charge of the tracks in that 
jet is expected to be anti-correlated with the charge of the $W$.  

In simulated $W$~+~jets events, the charge correlation is large, but falls off with 
increasing jet multiplicity.  We measure \mbox{opposite-sign} fractions of 
81~$\pm$~3\% for events with zero jets, 73~$\pm$~2\% for events with one jet,
and 75~$\pm$~5\% for events with at least two jets.  The difficulty with the numbers 
from the simulation is that the charge correlation, like the fake rate, is influenced by 
details of jet fragmentation that may not be fully taken into account in the Monte Carlo simulations.  
The \mbox{opposite-sign} fractions can be checked in candidate 
lepton~+~track events with no jets, where the contribution from $t\bar{t}$ is negligible
but where the expected contribution of events with a fake lepton is large.  
Prior to an \mbox{opposite-sign} requirement, the total number of predicted events   
agrees with the number observed, within the statistical uncertainty.  
When the predicted \mbox{opposite-sign} fraction from simulation is applied, more same-sign
events are observed than are predicted, and the number of \mbox{opposite-sign} events observed 
is correspondingly too small compared to the prediction.  This suggests that the simulation 
overestimates the fraction of events with a fake lepton of opposite charge.
We therefore obtain the opposite sign fraction from observed data, using the zero-jet candidate events.
We can further enhance the fraction of events with a fake lepton by requiring 
a significant amount of energy in the region of the hadronic calorimeter at which the track 
points.  This is about 70\% efficient for events with a fake lepton and \Ztt\ events, but reduces 
all other contributions by an order of magnitude.  We then subtract the estimated number of 
events from sources with two real leptons for all events and for those with opposite sign.  
We find that 67~$\pm$~3\% of remaining events are \mbox{opposite-sign}, which is taken to be 
the \mbox{opposite-sign} fraction for all events with zero jets.  

As a cross-check, we also measure the \mbox{opposite-sign} fraction using the zero-jet events
without the hadronic energy requirement described above,
and find 69~$\pm$~5\%, in good agreement with the number calculated with the requirement.
The larger uncertainty is due to the systematic uncertainties on the larger contributions 
to the zero-jet sample from the other processes.

We cannot apply the same procedure to derive the \mbox{opposite-sign} fraction for higher
jet multiplicities because the contribution from $t\bar{t}$ is non-negligible.
To obtain the charge correlation for events with more jets, we rely on the simulation to 
model the dependence of the correlation on the number of jets in the event.
The assumption made is that each additional jet dilutes the \mbox{opposite-sign} fraction
toward 50\%, the fraction corresponding to no correlation.  Defining a dilution factor $x$,
the \mbox{opposite-sign} fraction $f_{i+1}$ of events with $i+1$ jets can be expressed in terms 
of the fraction $f_i$ for events with $i$ jets as 
\begin{equation}
f_{i+1} = 0.5x + f_i(1-x)  \ .
\end{equation}
From the \mbox{(81 $\pm$ 3)\%} and \mbox{(73 $\pm$ 2)\%} figures, we find 
\mbox{$x = (26 \pm 12)$\%}.
Applying this to the \mbox{(67 $\pm$ 3)\%} fraction from the zero-jet events, we find 
an \mbox{opposite-sign} fractions of (63~$\pm$~3)\% for events with one jet.  Repeating
the procedure with the same $x$ and the one-jet \mbox{opposite-sign} fraction, we 
find (59~$\pm$~3)\% for events with two or more jets.

Turning to the case of $t\bar{t}$ lepton~+~jets events with a fake track lepton, 
we observe in simulation that 79\% of such events have \mbox{opposite-sign} leptons.  
It seems likely that the charge correlation here is attributable to the fact that
the summed charge of the quark pair produced by the $W$ will be the opposite sign 
of the charge of the lepton from the other $W$.  
We prefer not to use the number from simulation directly, since our results in the 
$W$~+~jets sample suggest that the fragmentation model in simulation tends to overestimate charge 
correlations.  We correct the result from simulation using the observed difference
between observed and simulated data for $W$~+~1 jet events where the jet is reconstructed as an 
isolated track, using the same method described in the previous paragraph.  
In simulation, 81\% of events are \mbox{opposite-sign}, compared to the 67\% in the CDF
data.  For $t\bar{t}$ lepton~+~jets events with three or more jets, we scale the 
fraction from simulation to \mbox{(67 $\pm$ 6)\%}, which is higher than the fraction for $W$~+~jets 
events with the same number of jets.  The uncertainty is taken to be half the difference 
between the original and rescaled numbers, or 9\%.
The statistical uncertainties on the fractions from simulation are less than a percent
and negligible in comparison.  

To obtain a fake lepton background prediction for events with two or more jets, we must
combine the two \mbox{opposite-sign} fractions obtained for $W$~+~jets and $t\bar{t}$ lepton~+~jets 
in proportion to the estimated number of fake leptons contributed by each process.  Because 
the predicted fraction of $t\bar{t}$ in the $W$~+~jets data is based on an acceptance
measured in simulation, this introduces a dependence on the $t\bar{t}$ cross section. 
We remove this dependence by including it explicitly in the pretag background 
calculation in the likelihood expression used to measure the cross section,
which will be described in more detail in Section~\ref{sec:likeli}.
The result is that, for the measured cross section of 9.6~pb, (27~$\pm$~4)\% of jets in the 
normalizing $W$~+~jets sample are predicted to be from $t\bar{t}$ for 
the final cross section value, and the re-weighted \mbox{opposite-sign} fraction is (61~$\pm$~3)\%.
Note that the 27\% is calculated from the measured $t\bar{t}$ cross section and is
in a sense the result of the fit, so the uncertainty on it is the observed change in the 
fraction when the cross section is varied by its calculated uncertainties. 

\subsubsection{Efficiency for Additional Selection}\label{sec:fakes_effs}

We measure the efficiency of the $Z$ veto and the track lepton $\Delta \varphi$ 
requirement in simulated $W$~+~jets events generated using \alpgen+\pythia\ Monte Carlo.
The efficiency for the two criteria is (82~$\pm$~2)\% for events 
with zero jets, (85~$\pm$~1)\% for events with one jet, and (89~$\pm$~2)\%
for events with two or more jets.
Scaling the event counts by these efficiencies yields the final
prediction of candidate events with a fake lepton as a function of the number of jets.

\subsubsection{Systematic Uncertainties}\label{sec:fakes_sys}

There are three sources of uncertainty in the estimate of the background from events
with a fake lepton: the statistical uncertainty on the fake rate, 
the systematic uncertainty on the overall normalization
of the estimate, and the uncertainty on the fraction of events with a fake lepton 
where the leptons have opposite sign.

The uncertainty on the overall normalization comes from the largest
observed discrepancy in the simulated and observed data tests described above.  
We have also argued that this 18\% uncertainty covers possible discrepancies
between the fake rate for jets in $W$~+~jets events and jets in $t\bar{t}$ lepton~+~jets events.
The uncertainty on the \mbox{opposite-sign} fraction is the statistical uncertainty on the 
fraction calculated from zero-jet candidate data, combined with the systematic 
uncertainty from the scaling of the fraction for $t\bar{t}$ lepton~+~jets events, and is 5\%.

Combining the 18\% and 5\% systematic uncertainties with the 6\% statistical uncertainty,
the total uncertainty on this background is 20\%.

\subsection{Diboson}

\subsubsection{Diboson Acceptance}

Diboson events ($WW$, $WZ$, and $ZZ$) events have small cross sections, comparable
to the $t\bar{t}$ cross section.  It is only recently that CDF and D\O\ have obtained
sufficient data to observe $WZ$ production~\cite{wz_observation, d0_wz_observation} and 
$ZZ$ production~\cite{zz_evidence, d0_zz_observation}.  Also, if one or both bosons decay to leptons, these 
events can mimic the $t\bar{t}$ signature, so it is
not possible to isolate a large sample of such events in the data. Therefore, 
the acceptance calculated from simulated events is used together with the theoretical 
production cross sections to estimate this background.  
We use event samples generated with the \pythia\ Monte Carlo generator
and apply the same corrections to the diboson acceptance as we did for the
signal acceptance, with one additional correction for the rate of jet production, described below.
The estimated number of events in the candidate sample is then
the corrected acceptance multiplied by the theoretical cross section 
and the integrated luminosity.  The theoretical cross sections for
these processes are 12.4~$\pm$~0.8~pb for $WW$, 3.7~$\pm$~0.3~pb for $WZ$, 
and 3.7~$\pm$~0.3~pb for $ZZ$~\cite{diboson_theory}.  
The $ZZ$ sample includes the $\gamma^*$ contribution,
with \mbox{$M_{Z/\gamma^*} > 2\ \GeVcc$} for both bosons.  These cross sections are calculated using 
the MCFM Monte Carlo program~\cite{diboson_theory} and the uncertainties are based on the $Q^2$ and PDF dependence 
of the cross sections.  

\subsubsection{Correction for Number of Jets}\label{sec:diboson_njetcorr}

\pythia\ is a leading order Monte Carlo program and so is not expected to correctly predict 
the fraction of events with extra high-$p_T$ jets in addition to the core process. 
A scale factor derived from a comparison of jet production in real and 
simulated $Z$~+~jets data is 
applied to correct the acceptance for events with two or more jets up to the observed level. 
A sample of $Z$'s are selected with two \mbox{opposite-charge}, fully reconstructed 
electrons or muons having an invariant mass in the interval \masswin{76}{106}.  
The fraction of events with two or more jets in simulation, 0.0142~$\pm$~0.0002,
is lower than the fraction observed in data, 0.0153~$\pm$~0.0008.
A scale factor, once again defined as the ratio of the fraction in observed data to the 
fraction in simulated data, is calculated for each jet multiplicity. The scale factors
found are 1.006~$\pm$~0.002 for 
events with zero jets, 0.940~$\pm$~0.018 for events with one jet, and 1.08~$\pm$~0.06 for 
events with two or more jets, where the uncertainties are statistical only.  
We then multiply the acceptance by the appropriate scale factor
for each jet multiplicity. In order to maintain the same 
overall normalization, we rescale all three acceptances by a common factor so that their 
sum is unchanged.  


\subsubsection{Systematic Uncertainties}

The systematic uncertainties relevant to the signal acceptance also apply here.  The 
lepton identification uncertainties are still 1.1\% for both fully reconstructed and
track leptons.  The uncertainty on the number-of-jets correction, 5.5\%, also
applies to the predicted number of events.  The jet energy scale uncertainty is also 
relevant here, and is evaluated in the same way, but has a larger effect: the energy 
spectrum of radiated jets falls sharply, so that small changes in the jet energy can 
lead to large changes in the event selection efficiency.  The resulting 
uncertainty is 5.8\%.  Finally, we also include the theoretical uncertainties on the
cross sections used to normalize the background prediction, which are 6\% for the $WW$
and 7\% for $WZ$ and $ZZ$.  The total uncertainty on the diboson background calculation is
11\%.

\subsection{Drell-Yan}\label{sec:dy}

Drell-Yan events with \met\ are a significant source of background for
$t\bar{t}$ lepton~+~track events since there are two real leptons in the final state and the 
inclusive cross section is large 
(\mbox{$\sigma (p\bar{p} \to Z/\gamma^\star) \times Br (Z/\gamma^\star \to \ell^+\ell^-) = 251 \pm 5$ pb}
 for \mbox{$66\ \GeVcc < M_{\ell\ell} < 116\ \GeVcc$}~\cite{CDFDetector}).  
In the case of \Ztt, the \met\ is mostly from the neutrinos
from the $\tau$ lepton decays, and the background calculation is based on simulation. 
For \Zeemm, there are no neutrinos in the
final state and any \met\ is the result of the flawed reconstruction of one or
more leptons or jets.  Such events are rare 
and difficult to distinguish from other sources of two leptons and \met, so it is 
difficult to verify that they are simulated accurately.  It is possible, however, 
to select a sample of events from the collision data with a high concentration of 
\Zeemm\ events with \met, and build an estimate using it, integrating information from simulation.
The drawback is that the precision of this hybrid method is limited by its
statistical uncertainty.  This background estimate carries the 
largest uncertainty of any input to the cross section measurement.

\subsubsection{\Ztt}

It is difficult to isolate \Ztt\ events in the data, but Monte Carlo simulation is expected to
do a reasonable job of modeling the event kinematics because real neutrinos are responsible 
for the \met\ in the 
final state.  Therefore the estimate of the \Ztt\ background is calculated in the same
way as the diboson backgrounds, including the rescaling to compensate for the deficit 
in generated extra jets.  Events are generated with $M(Z/\gamma^{\star}) > 30\ \GeVcc$; the 
corresponding cross section is \mbox{327 $\pm$ 7 pb~\cite{CDFDetector}}.  The fractional systematic 
uncertainties on the resulting background are also identical to those in the diboson 
case.

\subsubsection{Calculation of \Zeemm\ Background}

The estimate of the background from \Zeemm\ events is calculated as follows:
\begin{eqnarray}
N^i_{\txtss{DY}} &=& N^i_{\txtss{out}} + N^i_{\txtss{in}}\label{eq:dy1} \\
N^i_{\txtss{out}} &=& (n_{25} - \hat{n}_{25}) \, {f_i} \, {R_i}\label{eq:dy2} \\
N^i_{\txtss{in}} &=& (n_{40} - \hat{n}_{40}) \, {f_i}\label{eq:dy3} 
\end{eqnarray}
The total number of background events, $N^i_{\txtss{DY}}$, is the sum of 
the number inside and outside the ``$Z$ region'', defined as those 
event where the lepton~+~track invariant mass is between \mbox{76 and 106 $\GeVcc$}.
The label $i$ designates the number of jets, where $i$ may be zero, one, or two.
All events with two or more jets are included in \mbox{$i = 2$}.   
Outside of the $Z$ region, the \met\ minimum from the event selection is 25~$\GeV$, so 
the background estimate $N^i_{\txtss{in}}$ for that region is based on the number $n_{25}$
of lepton~+~track events with at least 25~GeV\ of \met\ in the CDF data in the $Z$ region.  
Inside the $Z$ region, the \met\ minimum is 40~$\GeV$, so we count the number of 
events $n_{40}$ in the same data with at least 40~\GeV\ of \met.  To isolate
the contribution of \Zeemm\ to those samples, we subtract the estimated number of 
events from other sources passing the selection, labeled $\hat{n}_{25}$ and $\hat{n}_{40}$ 
in the above.  The selection used for $n_{25}$ and $n_{40}$, and the calculation
of $\hat{n}_{25}$ and $\hat{n}_{40}$, are described in Section~\ref{sec:dy_datanorm}.  

These data include events with any number of jets, so we multiply the
number of events by the fraction $f_i$ expected to have a particular jet
multiplicity $i$.  Also, the background estimate for events outside the $Z$ region
is based on an event count inside the $Z$ region, so we multiply it by the
expected ratio $R_i$ of the number of background events outside the $Z$ region to 
the number inside the region.  
These fractions are measured in simulated \Zeemm\ events.  
See Section~\ref{sec:dy_mcratios} for details.  

The statistical and systematic uncertainties on this method are described in 
Section~\ref{sec:dy_sys}.

\subsubsection{Data Sample for Normalization}\label{sec:dy_datanorm}

To obtain a sample of events from the CDF data similar to the candidate sample but 
with a larger contribution from \Zeemm, we alter the event selection by restricting the 
sample to the $Z$ region (\masswin{76}{106}) and including events 
of all jet multiplicities.  The event selection is otherwise identical to that
of the main analysis.  In this sample, we count the number of events with
\met~$>$~25~GeV and \met~$>$~40~GeV, corresponding to the \met\ thresholds used
inside and outside the $Z$ region in the candidate selection.
The number of events in these samples are $n_{25}$ and $n_{40}$ in Equations
\ref{eq:dy1}-\ref{eq:dy3}.  

These data are expected to contain many \Zeemm\ events with \met, but may also contain
events from other sources, including $t\bar{t}$, $WW$, $WZ$, $ZZ$, \Ztt, and events 
with a fake lepton.  We calculate the contributions of each of these exactly as described 
for the main analysis, except that the event selection is the modified version described 
above.  The predicted number of events in both the \met~$>$~25~GeV and \met~$>$~40~GeV
samples, labeled $\hat{n}_{25}$ and $\hat{n}_{40}$ in Equations \ref{eq:dy1}-\ref{eq:dy3} 
are subtracted from the corresponding number of observed events in the data
to yield the number attributable to \Zeemm.  

Once again, a dependence on the $t\bar{t}$ cross section appears and must be treated 
with care.  As with the background from events with a fake lepton, we include this 
dependence explicitly in the likelihood used to calculate the cross section 
(See Section~\ref{sec:likeli}).  

\subsubsection{Application of Simulated Data}\label{sec:dy_mcratios}

We use calculations from simulated \Zeemm\ events to divide the Drell-Yan events among the 
jet multiplicity bins and estimate the number of events outside the $Z$ region.
To calculate the necessary ratios $f_i$ and $R_i$, we select events in the 
simulation using the criteria described in the previous section (\ref{sec:dy_datanorm}), 
except that the \met\ threshold is kept constant at 25~GeV and events both inside 
and outside the $Z$ region are included.  That is, the event selection is identical
to the main event selection except that there is no $Z$ veto.

To measure $R_i$, we count the number of events inside and outside the $Z$ region
for each jet multiplicity.  The ratio of the number outside to the number inside
is $R_i$.  We correct $R_i$ for the different invariant mass resolutions in observed 
and simulated data.  Comparing the fractions of events inside and outside the $Z$ region
for the different fully reconstructed lepton types, we find correction factors 
significantly different from unity only for electron~+~track pairs.  For CEM~+~track
$Z$ events, the mass region only includes 98\% as many events in real data as it does in
simulation.  For PHX~+~track events, the number is 94\%.  We multiply these numbers 
by $R_i$.  The uncertainties on these numbers are negligible
compared to other uncertainties on this background.  

To distribute the estimate among the zero, one, and two-or-more jet categories, we 
measure the fraction of events in the $Z$ region having each of these jet multiplicities.
These fractions, labeled $f_i$ in Equations \ref{eq:dy1}-\ref{eq:dy3}, depend on \pythia's 
modeling of the probability to produce extra jets, like the acceptances measured for diboson 
and \Ztt\ events.  Therefore, we apply the correction factors derived in 
Section~\ref{sec:diboson_njetcorr} here as well.  After correction, 
the fractions $f_i$ are rescaled by a common factor so that they sum to unity.  

The values of $R_i$ and $f_i$, after correction, are shown in Table~\ref{tab:dy_mc}
for each type of fully reconstructed lepton.
\begin{table*}[tbp]
 \centering
 \begin{tabular*}{\widetable}{c@{\hspace{0.3\textwidth}}c}
   \hline\hline
   & The fraction $f_i$ with each jet multiplicity \\ 
 \end{tabular*}
 \begin{tabular*}{\widetable}{@{\extracolsep{\fill}}cccc}
 & $i=0$ jets	& $i=1$ jet	& $i=2$ jets \\ 
   \hline
CEM 		& 0.63 $\pm$ 0.02	& 0.28 $\pm$ 0.02	& 0.09 $\pm$ 0.01	\\
CMUP+CMX	& 0.57 $\pm$ 0.02	& 0.32 $\pm$ 0.02	& 0.11 $\pm$ 0.01	\\
PHX 		& 0.68 $\pm$ 0.03	& 0.26 $\pm$ 0.02	& 0.06 $\pm$ 0.01	\\
   \hline
 \end{tabular*}
 \begin{tabular*}{\widetable}{c@{\hspace{0.3\textwidth}}c}
  & Ratio $R_i$ of number inside 76-106\ \GeVcc\ to number outside \\ 
 \end{tabular*}
 \begin{tabular*}{\widetable}{@{\extracolsep{\fill}}cccc}
 & $i=0$ jets	& $i=1$ jet	& $i=2$ jets \\ 
   \hline
CEM		& 1.22 $\pm$ 0.09	& 0.94 $\pm$ 0.10	& 0.89 $\pm$ 0.19	\\
CMUP+CMX	& 0.41 $\pm$ 0.04	& 0.31 $\pm$ 0.04	& 0.34 $\pm$ 0.08	\\
PHX		& 0.47 $\pm$ 0.05	& 0.47 $\pm$ 0.08	& 0.84 $\pm$ 0.29	\\
\hline\hline
 \end{tabular*}
 \caption{Inputs to the \Zeemm\ background estimate from simulation.  The index $i$ represents
   the number of jets in the events in which the quantity is measured, and the \mbox{$i = 2$}
     category includes all events with two or more jets.  Shown uncertainties 
 are statistical only; systematic uncertainties are discussed in Section~\ref{sec:dy_sys}}
 \label{tab:dy_mc}
\end{table*}

\subsubsection{Systematic Uncertainties}\label{sec:dy_sys}

The largest uncertainty on the \Zeemm\ background estimate is the statistical uncertainty, 
which is 20\%.  This uncertainty is due in approximately equal parts to the sizes of the 
real and simulated data samples.
The Monte Carlo samples used to calculate the ratios $R_i$ and $f_i$, described in 
the previous section (\ref{sec:dy_mcratios}), contain 13.8 million events. 
To generate enough events to significantly reduce the uncertainty is impractical, and  
even if enough events were generated to make the contribution 
from simulation negligible, the total statistical uncertainty would still be 13\%.

Since the scale factor that is used to correct the number of extra jets produced by \pythia\
is applied to $R_i$, the fraction of events with jet multiplicity $i$, 
the statistical uncertainty of 5.5\% on the correction factor also contributes here.

Finally, the reliability of the ratios $R_i$ and $f_i$ depends on the ability of the 
simulation to model the \met\ from mismeasured objects.  One way to make a quantitative 
comparison between observed and simulated data is to compare the fraction of events which 
exceed the 25~GeV \met\ threshold.  Since many processes will contribute to the high-\met\ 
``Drell-Yan'' data sample, we require the \met\ to be pointing at a jet or the track lepton
by inverting the corresponding $\Delta\varphi$ selection requirements.  This ensures that 
the comparison is mostly between real and simulated \Zeemm\ events in which the \met\ is 
due to a mismeasured jet or lepton.  The fraction of events with \mbox{\met\ $>$ 25 GeV} is then 
measured in the data, and the \met\ distribution from the simulation
is integrated to find the threshold that would give the same fraction of events above threshold.
The outcome is a shift of 1~GeV in the threshold, to 24~GeV. 
All of the ratios from the simulation are re-derived with the 24~GeV threshold and the background
is recalculated.  The recalculated background estimate is 13.5\% lower 
than the default estimate, and the full difference is taken as a systematic uncertainty.  

Combining the statistical and systematic uncertainties in quadrature yields a total 
uncertainty of 25\% on this background.

\subsection{Summary of Pretag Backgrounds}

Backgrounds to the lepton~+~track $t\bar{t}$ sample come from diboson, Drell-Yan, and
$W$~+~jets events.  Where possible, background estimates include information from control 
samples in the observed data.  In the case of $W$~+~jets with a fake lepton, the background
estimate is based almost entirely on data.  For the \Zeemm\ background, measurements in the 
data set the overall normalization but simulation is used to fill in the details.  
Diboson and \Ztt\ contributions are estimated using simulation alone, with corrections 
obtained from comparisons between real and simulated applied where relevant.  

The predicted number of background events for each of these sources is presented 
in Table~\ref{tab:pretagsummary}.
The systematic uncertainties on all of the backgrounds and the corresponding uncertainty 
on the cross section measurement are collected in Table \ref{tab:syserrs_bg}.  Some care must be taken
when combining the background uncertainties, due to correlations.  The systematic
uncertainties due to lepton and jet reconstruction are fully correlated between the diboson and 
\Ztt\ estimates, and must be summed directly rather than in quadrature.  
Similarly, the uncertainty on the jet multiplicity correction
is correlated between the Drell-Yan and diboson backgrounds.  All other uncertainties are
uncorrelated.
\begin{table*}[tp]
\centering
\begin{tabular*}{14cm}{@{\extracolsep{\fill}}lcc}
\hline\hline
				& Uncertainty   & Uncertainty   \\
Source                         	& on background & on cross section      \\      
\hline
Lepton identification           & 1.6\%   	& 0.3\%		\\
Jet energy scale                & 5.8\%         & 1.0\%         \\
Jet multiplicity		& 5.5\%         & 1.8\%         \\
Diboson normalizations		& 6-7\%		& 0.5\%		\\
\Zeemm                          & 25\%          & 3.6\%         \\
$W$ + fake lepton               & 20\%          & 7.9\%         \\
\hline\hline
\end{tabular*}
\caption{Summary table of systematic uncertainties on the pretag background estimate.  The 
\Zeemm\ uncertainty includes the statistical uncertainty and the systematic uncertainty on
the projection of the number of events to outside the $Z$ mass region.  The uncertainty
on the jet multiplicity correction is listed separately, as it applies to the diboson and
\Ztt\ backgrounds as well.}
\label{tab:syserrs_bg}			
\end{table*}


\section{\label{sec:tagbg}Background Estimation in Tagged Sample}

  The tagged background estimate differs substantially from the pretag background estimate.
  First, the nature of the background changes when a tagging requirement is added. 
  In the pretag analysis the dominant background is $W$+jets events with a fake lepton.
  In the tagged analysis backgrounds containing $b$-jets dominate.  This includes
  processes producing two leptons and one or more $b$-jets, such as $Z+b\bar{b}$
  events, as well as events from $t\bar{t}$ in the lepton~+~jets channel where
  one of the jets, either from the light quarks from the $W$ decay or
  from one of the $b$ quarks, is misidentified as a lepton.
  Second, in the tagged analysis, we are able to estimate all backgrounds,
  except for those arising from $t\bar{t}$ itself, using a single data-driven 
  technique discussed below.  The background from $t\bar{t}$ events with a fake
  lepton are estimated separately using a combination of real and simulated CDF data.

  The tagged background estimate is based upon jet tagging rates obtained in
  generic QCD multijet events.  We apply this tagging rate to the pretag
  candidate events, taking advantage of the fact that it has a large
  background component.  We then correct this for tagged events from $t\bar{t}$ decays
  in the pretag sample, and for $t\bar{t}$ lepton~+~jets events with a fake
  lepton.  Simulated events are used to estimate the size of the corrections.

\subsection{Data-Based Estimate of Background}

  The background of the tagged lepton~+~track sample can be organized into two
  parts. The first is made up of processes with a decay signature similar 
  to the signal, such as $Z+b\bar{b}$ events, or 
  any event passing pretag selection criteria that also has a mistagged jet.
  In the $Z+b\bar{b}$ case, the $b$ tag is legitimate and the mismeasurement of 
  some object in the event produces false \met. In the mistag case, a jet 
  is falsely identified as a $b$-jet.  
  Such backgrounds may be estimated using tag
  rate matrices, discussed in more detail below. The second category of background events are fakes from 
  $t\bar{t}$ decay in the lepton~+~jets channel, where a
  jet is falsely identified as a track lepton. For these events, the probability to
  tag the event will be underestimated by the matrix because there are two $b$-jets in the event.

  As stated in Section~\ref{ss:SVXAlg},
  positive tags are interpreted to be tags of long-lived $B$ hadrons
  and negative tags are interpreted as ``mistags'', or mistakes due to 
  material interactions or resolution
  effects. Positive and negative tag rates of generic QCD jets
  are parameterized in five quantities: jet $E_T$, the number of tracks in
  the jet, jet $\eta$, the number of primary vertices in the event, and the
  total scalar sum of the $E_T$ of all the jets in the event, $\Sigma{p_T}$.
  These parameterizations are termed ``tag matrices''. The generic
  QCD jet samples used to build the matrices contain real tags from $B$
  hadron decays, as well as mistags. 

  As a first step in estimating the backgrounds, we treat all  
  events in the pretag sample as if they are from background 
  sources that have the same relative proportion of heavy and light flavor
  jets as the generic multijet sample.  We apply the positive tag rate matrix 
  to all of the jets in the sample to obtain
  a first estimate of the expected number  $N_\txtss{matrix}$  of background events in the sample. 
  This estimate has to be corrected for the fact that the 
  sample is not entirely background and the fact that the jets do not have the
  same mix of heavy and light quarks as generic QCD multijet events. 

  In particular, $t\bar{t}$ events do not have the same tagging rate as generic
  QCD events, the rate represented by the tag rate matrices. 
  Top quark pair decays via the dilepton channel  make up a considerable
  portion of the pretag events by design, but these should not contribute 
  to the background estimate.  We estimate this number $N^\txtss{dil}_\txtss{matrix}$ 
  in Section~\ref{sec:DilAcc} and subtract it from $N_\txtss{matrix}$.
  Also, a portion of the fake lepton background is not due to generic QCD processes, but  
  arise from $t\bar{t}$ decays in the lepton~+~jets channel. 
  Recall that in this analysis $t\bar{t}$ decays in the lepton~+~jets
  channel decays are considered a background.
  Again, these $t\bar{t}$ events do not have the tagging rate 
  predicted by the tag rate matrices, but unlike the dilepton contribution,
  they are a background.  Therefore, we need to subtract their
  contribution $N^\txtss{LJ}_\txtss{matrix}$ from the matrix estimate, 
  and add back the correct contribution $N^\txtss{LJ}_\txtss{fakes}$. 
  ${N^\txtss{LJ}_\txtss{fakes}}$ is the proper estimate of the number of
  lepton~+~jets events which pass the lepton~+~track selection with a fake 
  lepton and are tagged because of the presence of $b$-jets.
  These numbers will both be derived in Section~\ref{sec:LJAcc}.
  Thus, the total tagged lepton~+~track background is given by:
  \begin{equation}
    \label{eq:bkgs}
          {N^\txtss{tag}_\txtss{bkg}} = N_\txtss{matrix} - {N^\txtss{dil}_\txtss{matrix}} - {N^\txtss{LJ}_\txtss{matrix}} + {N^\txtss{LJ}_\txtss{fakes}}\ .
  \end{equation}

  Note that the tagging rate for background
  events with a fake lepton from $W$~+~jets processes is well estimated by 
  the tag rate matrices, and
  is included in $N_\txtss{matrix}$. Also, we are now including jets 
  from $W/Z$~+~jet events in the
  category of generic jets though we pointed out earlier that they differ from QCD jets
  in average track multiplicity, which
  has a significant impact on the lepton fake rate. 
  The impact of these differences is taken into account in the tag rate
  matrix, which is parametrized as a function of track multiplicity. 
  As such, it can be applied
  equally well to jets in generic multijet processes and $W/Z$~+~jet events. 

\begin{table}[tp]
  \begin{center}
    \begin{tabular*}{8.6cm}{@{\extracolsep{\fill}}@{\hspace{1.5cm}}lr@{\extracolsep{0in}}@{.}l@{ $\pm$ }l@{\hspace{1.5cm}}}
        \hline \hline
	$N_\txtss{matrix}$   		  & 13&7   & 1.1 \\
	$N^\txtss{dil}_\txtss{matrix}$    &  8&6   & 2.3 \\
	$N^\txtss{LJ}_\txtss{matrix}$     &  0&7   & 0.1 \\
	\hline
	$N^\txtss{pretag}_\txtss{fakes}$  & 30&7   & 6.7 \\		
	$f^\txtss{tag}_{W \geq 3j}$ 	  &  0&239 & 0.008 \\
	$f^\txtss{top}_\txtss{tag}$       &  0&70  & 0.03 \\
	$N^\txtss{LJ}_\txtss{fakes}$      &  5&2   & 1.2 \\		
	\hline
	${N^\txtss{tag}_\txtss{bkg}}$     &  9&5   & 2.8 \\
	\hline\hline
    \end{tabular*}
    \caption{Details and results of the tagged background calculation. Uncertainties include
    	systematic contributions.}
    \label{tab:tagbkgs}
  \end{center}
\end{table}

\subsection{Correction for $t\bar{t}$ Dilepton Content in the Pretag Candidate Sample}
\label{sec:DilAcc}
  The pretag lepton~+~track sample has a large fraction of $t\bar{t}$ events, which
  should not counted in the background estimate.
  The contribution to $N_\txtss{matrix}$ from dilepton $t\bar{t}$ decays is
  estimated using simulated $t\bar{t}$ events. We derive the matrix tag rate,
  $\epsilon_\txtss{matrix}$
  by applying the tag rate matrix to all jets in the pretag candidate events in
  simulated $t\bar{t}$ events, and divide by the total number of pretag events.
  We find \mbox{$\epsilon_\txtss{matrix}$ = 0.122 $\pm$ 0.025}.   
  The total contribution
  to the background ${N^\txtss{dil}_\txtss{matrix}}$ from dilepton $t\bar{t}$ events is
  then given by 
  \begin{equation}
    {N^\txtss{dil}_\txtss{matrix}} = \epsilon_\txtss{matrix} \, ({N^\txtss{pretag}_\txtss{obs}} - {N^\txtss{pretag}_\txtss{bkg}}) \ ,
  \end{equation}
  where $\epsilon_\txtss{matrix}$ is the $t\bar{t}$ tag rate, and the estimated number of 
  $t\bar{t}$ events in the pretag candidate sample is the difference between the number
  of observed pretag candidates and the predicted background, 
  $N^\txtss{pretag}_\txtss{obs} - N^\txtss{pretag}_\txtss{bkg}$, which 
  were described in Section~\ref{sec:pretagbg}.

  We find ${N^\txtss{dil}_\txtss{matrix}}$ to be 8.6~$\pm$~2.3 events.  
  Values used in the calculation are found in Table~\ref{tab:tagbkgs}.

\subsection{Correction for $t\bar{t}$ Lepton + Jets in Pretag Sample}
\label{sec:LJAcc}

  As discussed in Section~\ref{sec:pretagbg} the fake lepton background
  originates from two processes. The first is the QCD radiation of extra 
  jets in $W$~+~jets events, for which we can estimate the tag rate
  using the matrix.  The second is the lepton~+~jets channel 
  decay of $t\bar{t}$ events, which is estimated separately.


  We treat the $t\bar{t}$ lepton~+~jets channel as
  a background source and these events are rejected from
  the acceptance for our selection. These events are present in the
  pretag sample, and therefore contribute to
  $N_\txtss{matrix}$. However, like dilepton $t\bar{t}$ events,
  lepton~+~jets $t\bar{t}$ events
  do not have the same tag rate as predicted by
  the tag rate matrices. As such, their contribution to $N_\txtss{matrix}$
  needs to be replaced by a more accurate estimate.
  We subtract the lepton~+~jets $t\bar{t}$
  contribution in the same manner as for dilepton events. The tag rate
  in lepton~+~jets events is slightly larger than in dilepton events because
  there are more jets per event. Using simulated
  $t\bar{t}$ lepton~+~jets events we use the matrix to find the tag rate, 
  and then multiply by the number of predicted pretag background events from 
  the $t\bar{t}$ process with a fake lepton:
  \begin{equation}
    N^\txtss{LJ}_\txtss{matrix} = {N^\txtss{pretag}_\txtss{fakes}} \, f^\txtss{LJ}_\txtss{fakes} \, \epsilon^\txtss{LJ}_\txtss{matrix}\,  ,
  \end{equation}
  where ${N^\txtss{pretag}_\txtss{fakes}}$ is the estimated number of fakes in the
  pretag sample and provides an overall normalization for the tagged estimate.
  $f^\txtss{LJ}_\txtss{fakes}$ is the fraction of pretag fakes that come from $t\bar{t}$ decays
  in the lepton~+~jets channel, derived from simulation. 
  We find ${N^\txtss{LJ}_\txtss{matrix}}$ to be 0.7~$\pm$~0.1.

  The actual contribution from the $t\bar{t}$ lepton~+~jets channel,
  $N^\txtss{LJ}_\txtss{fakes}$,
  now needs to be estimated and added back into the total background;
  see Eq.~\ref{eq:bkgs}. Because the pretag fake lepton background estimate is based on
  $W$~+~jets data, this is equivalent to finding the actual
  $t\bar{t}$ content of the events passing that selection which are also tagged.
  The quantity $N^\txtss{LJ}_\txtss{fakes}$ is factorized as
  \begin{equation}
    {N^\txtss{LJ}_\txtss{fakes}} = {N^\txtss{pretag}_\txtss{fakes}} \; f^\txtss{tag}_{W + \geq 3 \txtss{jets}} \; f^\txtss{top}_\txtss{tag} \ .
  \end{equation}
  Recall that in the pretag background estimate we used the fraction of
  events with a fake lepton that has the opposite charge of the primary
  lepton.  We need it here as well, but it 
  is different in the pretag fake lepton sample than in the tagged fake lepton
  from $t\bar{t}$ sample, so we correct for the different opposite sign fractions 
  in pretag and
  tagged fakes.    \mbox{$f^\txtss{tag}_{W + \geq 3 jets}$} is the fraction of
  \mbox{$W + \geq 3$ jet} events observed in data that are tagged, and $f^\txtss{top}_\txtss{tag}$
  is the fraction of tagged \mbox{$W + \geq 3$ jet} events which are from the 
  $t\bar{t}$ lepton~+~jets channel. As in the pretag case, we
  are concerned with \mbox{$W + \geq 3$ jet} events because we require at least two
  jets in the events selection, and so an additional jet is required to
  fake the track lepton.

  The fraction $f^\txtss{tag}_{W + \geq 3 jets}$ is calculated directly from the data,
  without using simulation.  We select
  \mbox{$W + \geq 3$ jet} events with the same criteria that defines the 
  lepton~+~\met~+~jets sample used to normalize the pretag fake lepton estimate
  (See Section~\ref{sec:fakes}).  The fraction of those events which are tagged
  is ${f^\txtss{tag}_{W + \geq 3jets}}$.

  The fraction $f^\txtss{top}_\txtss{tag}$ of tagged \mbox{$W + \geq 3$ jet} events which 
  are $t\bar{t}$ lepton~+~jets is estimated as
  \begin{equation}
    f^\txtss{top}_\txtss{tag} =\frac{ \alpha_\txtss{LJ}  \, \sigma_{t\overline{t}} \, \int \mathscr{L} dt } {N^\txtss{LJ}_\txtss{cand}} \ .
  \end{equation}
  In the above, $\alpha_\txtss{LJ}$ is the lepton~+~jets acceptance in simulated $t\bar{t}$ events 
  using the lepton~+~\met~+~$\geq 3$~jet selection, including the requirement that
  at least one jet be tagged, that defines the tagged \mbox{$W + \geq 3$ jet} sample described 
  above.  $\int \mathscr{L} dt$ is the total integrated luminosity, 
 $\sigma_{t \bar{t}}$ is the $t\bar{t}$ cross section, and
  $N^\txtss{LJ}_\txtss{cand}$ is the number of tagged lepton~+~jets events in the CDF data.
 Those events are selected using the same criteria as the events used to 
 find ${f^\txtss{tag}_{W + \geq 3jets}}$.
  So the number
  of tagged $W$~+~$\geq$~3 jets events which are from $t\bar{t}$ decays is estimated
  by multiplying the acceptance for the  $t\bar{t}$ lepton~+~jets channel
  by the integrated luminosity and the $t\bar{t}$ cross section.
  By dividing by the number of candidate lepton~+~jets events
  in the data we find the fraction of tagged events which are from $t\bar{t}$.

  Like some of the pretag backgrounds, this background depends on the 
  $t\bar{t}$ cross section, so we also include this dependence explicitly in 
  the likelihood calculation (See Section~\ref{sec:likeli}).  
  The final value of ${f^\txtss{top}_\txtss{tag}}$ is in Table~\ref{tab:tagbkgs}.

\subsection{Systematic Uncertainties on Tagged Background Estimate}

  The systematic uncertainty on the tagged background estimate consists of the 
  combined uncertainties from the two components of the background: the
  background estimated using the tag rate matrix and the
  background from lepton~+~jets channel events with a fake lepton.
  Statistical errors on quantities derived from simulation, such as
  $\epsilon^\txtss{top}_\txtss{matrix}$, and ${f^\txtss{tag}_{W + \geq 3}}$, are negligible.

  \subsubsection{Data-Based Background Prediction}

  The systematic uncertainty one the data-based 
  prediction method is 8\%.  This uncertainty applies to all predictions from 
  the mistag matrix.  It mostly arises from charm and light flavor contamination in the
  data used to derive the tag rate matrices.
  Because this systematic uncertainty is correlated among all predictions
  made using the matrix, we only apply the systematic for the matrix technique 
  to the physics background portion of $N_\txtss{matrix}$, the difference
  ${N_\txtss{matrix}} - {N^\txtss{dil}_\txtss{matrix}} - {N^\txtss{LJ}_\txtss{matrix}}$.

  Tagging predictions made by the tag rate matrices also have a statistical
  uncertainty due to the limited sample size for each
  entry in the matrix.  This uncertainty applies to each of the three numbers 
  calculated using the matrix, but is uncorrelated between them.  This 
  contributes uncertainties of 1.1 events to $N_\txtss{matrix}$, 1.8 events to 
  $N^\txtss{dil}_\txtss{matrix}$, and 0.1 events to ${N^\txtss{LJ}_\txtss{matrix}}$ (see also 
  Table~\ref{tab:tagbkgs}).  Combined, these contribute an uncertainty of
  2.1 events, or 47\%, to the predicted matrix background of 4.4 events.  
  
  The predicted contribution from $t\bar{t}$ dilepton events, $N^\txtss{dil}_\txtss{matrix}$,
  is computed from the number of predicted pretag $t\bar{t}$ events, which is based on
  the number of observed candidates and the predicted background in the pretag sample.  
  The uncertainty on this prediction contributes another 1.5 events to the
  uncertainty on $N^\txtss{dil}_\txtss{matrix}$, bringing its total uncertainty to
  2.3 events.  This is another 34\% uncertainty on the matrix 
  background prediction.  

  The total systematic uncertainty on the data-based background prediction
  is 2.6 events or 59\%.  
  
  \subsubsection{Lepton + Jets with a Fake Second Lepton}

  The estimate $N^\txtss{LJ}_\txtss{fakes}$ of the number of background events from $t\bar{t}$
  lepton~+~jets events with a fake second lepton has the same sources of systematic 
  uncertainty as the pretag fake lepton background estimate, upon which it is based.  
  This is a 20\% systematic uncertainty on the overall normalization and 
  a 9\% systematic uncertainty on the opposite sign fraction used for lepton~+~jets
  events.  See Section~\ref{sec:fakes} for details.  

  This background has a smaller additional contribution to the uncertainty which is unique
  to the tagged background estimate.  These are a 3\% statistical uncertainty 
  on ${f^\txtss{tag}_{W \geq 3j}}$ and a 4\% statistical uncertainty on $f^\txtss{top}_\txtss{tag}$.
  Combined, these add an extra 5\% uncertainty to the fake lepton background.  

  The total systematic uncertainty on the background from lepton~+~jets events with
  a fake lepton is 1.2 events.  

  The total tagged background systematic uncertainty is obtained by adding the total 
  uncertainty on the matrix and fakes predictions in quadrature.
  We find the overall systematic uncertainty on the background estimate to be
  2.8 events, or 30\%. 
  The systematic uncertainties on all of the backgrounds, and the corresponding 
  uncertainty contributed to the cross section measurement, are collected in 
  Table \ref{tab:tag_syserrs_bg}.
  \begin{table*}[tp]
    \centering
    \begin{tabular*}{14cm}{@{\extracolsep{\fill}}lccc}
      \hline\hline
      Source                    		& Uncertainty	& Uncertainty   & Uncertainty   \\
                                		& (in events)	& on background & on cross section      \\
      \hline
      tag matrix technique      		& 2.6		& 59\%          & 4\%         \\
$t\bar{t}$ (lepton + jets) + fake lepton        & 1.2		& 22\%          & 2\%         \\
      \hline
      total					& 2.8		& 30\%		& 5\%		\\
      \hline\hline
    \end{tabular*}
    \caption{Summary table of systematic uncertainties on the tagged background estimate.}
    \label{tab:tag_syserrs_bg}
  \end{table*}


\section{\label{sec:results}Results}


We first describe the likelihood used to derive the cross section results, 
including the treatment of uncertainties.
Then we summarize the predicted and observed event counts and 
present the cross sections for the two individual samples, 
the combined result, and selected kinematic distributions.

\subsection{Likelihood Fit}\label{sec:likeli}

To calculate the cross section results, we construct a likelihood function
describing the joint probability of finding a particular number of candidate
events in each sample given the predicted signal and backgrounds.
We vary the input parameters to find the cross section
value most likely to give the observed number of candidates in each sample.  

In order to combine the results, we must define two statistically independent
samples so that the number of candidates in each can be described by independent 
Poisson distributions.  Because the tagged events are a subset of the pretag 
events, we can divide the pretag candidate events into non-overlapping tagged
and untagged samples.  Although everything in this paper is described in 
terms of the pretag and tagged samples, the combined result is 
found from the tagged and untagged samples.  
The expected number of events in the tagged sample
has already been characterized in terms of the signal acceptance,
the event tagging efficiency, and the calculated background.
The expected number of events in the untagged sample may be derived from the
information about the tagged and pretag samples.  The acceptance is identical
to both the pretag and tagged samples, and the fraction of pretag events
that go into the untagged sample is approximately $1 - \epsilon\txtss{tag}$,
where $\epsilon\txtss{tag}$ is the event tagging efficiency.  The equality is 
not exact because there are some events in the pretag sample which cannot be 
tagged because the silicon tracking was not in usable condition when those 
events were recorded (data quality requirements are described in 
Section~\ref{sec:samples}).  Therefore, we calculate the number of expected 
untagged signal events as the difference between the predicted number of 
pretag and tagged signal events.  Similarly, the backgrounds in the untagged 
sample are calculated as the difference between the pretag and tagged 
backgrounds.  

The likelihood function has seven independent parameters.  One is the 
input cross section, and the other six are ``nuisance parameters'' corresponding 
to systematic uncertainties.  The likelihood $\mathcal{L}$ may be expressed as:
\begin{eqnarray}\label{eq:likeli}
\log\mathcal{L} &=& \log\mathscr{P}(N_u,N_u^\txtss{pred}) + \log\mathscr{P}(N_t,N_t^\txtss{pred}) \nonumber\\
		& &	+ \sum_{i=1}^6 \left(\frac{1}{2}\frac{(Q_i-Q_i^0)}{\delta Q_i}\right) 
\end{eqnarray}
where
\begin{equation}
N_t^\txtss{pred} = \sigma \mathcal{A} \epsilon\txtss{tag} \int\mathscr{L}dt + B_t(\sigma)   
\end{equation}
and 
\begin{eqnarray}
N_u^\txtss{pred} &=& N^\txtss{pred} - N_t^\txtss{pred}	\\
                 &=& \sigma \mathcal{A} \int\mathscr{L}dt + B(\sigma) - N_t^\txtss{pred}   .
\end{eqnarray}
A Poisson distribution $\mathscr{P}(N,N^\txtss{pred})$ describes the probability 
to find $N$ candidates given the mean number $N^\txtss{pred}$
predicted.  The number of tagged and untagged
candidates are each described as independent Poisson distributions.  
In the above, $N^\txtss{pred}$ is the number of candidates predicted in the pretag sample, 
$N_t^\txtss{pred}$ is the number predicted in the tagged sample, 
and $N_u^\txtss{pred}$ is the number predicted in the untagged sample.  
The corresponding numbers of observed candidates are $N$, $N_t$, and $N_u$.
$\mathcal{A}$ is the pretag acceptance, $\epsilon_\txtss{tag}$ is the event tagging 
efficiency, $\int\mathscr{L}dt$ is the integrated luminosity, and $\sigma$
is the $t\bar{t}$ cross section, for which we are fitting.  The pretag
and tagged background estimates, which depend on the signal cross section, are 
$B(\sigma)$ and $B_t(\sigma)$.
The probability distribution used for the nuisance parameters, such as the 
acceptance and backgrounds, is a Gaussian centered on the predicted value 
and having width equal to the relevant systematic uncertainty.  
This is shown as the sum in Equation~\ref{eq:likeli}, where $Q_i^0$ is the central
value of the nuisance parameter, $Q_i$ is the varied value, and $\delta Q_i$ the 
associated uncertainty.  

The systematic uncertainties treated as nuisance parameters are on the acceptance, 
the event tagging efficiency, the fake lepton background in the pretag and tag
samples, the remaining pretag background from Drell-Yan and diboson events, and
the remaining tagged background as estimated using the data-driven matrix method.  
These sources of uncertainty are independent from each other, but some of them are 
shared between the pretag and tagged measurements, and are varied together in the fit.
Table~\ref{tab:correlations} shows the cross section inputs and their uncertainties,
with the correlations between uncertainties shown.  
For example, the systematic uncertainty on the acceptance 
is correlated because the same number is used in both the tag and pretag samples.  
The number of expected tagged and untagged events vary up or down together with the
acceptance.  In contrast, the uncertainty on the event tagging efficiency
applies to the tagged sample but not the pretag sample. As the event tagging efficiency
is varied in the fit, the number of events predicted shifts between the tagged and untagged
samples but their sum, the number of pretag events, remains constant.  To put it 
another way, the number of 
predicted untagged signal events is correlated with the number of pretag and tagged events 
through the acceptance, but anticorrelated with the number of tagged signal events through
the event tagging efficiency.  The common 20\% systematic uncertainty on 
the background from fake leptons is treated similarly to the acceptance, because
the predicted pretag, tagged, and untagged background from fake leptons will all 
increase or decrease together.  This happens because the tagged fake lepton background
prediction is normalized to the pretag fake lepton background prediction.  All 
additional uncertainties on the pretag and tagged backgrounds 
are treated independently, because the calculations do not depend on each other,
and the sources of uncertainty are distinct. 
\begin{table*}[tbp]
  \centering
  \begin{tabular*}{14cm}{@{\extracolsep{\fill}}l*{2}{r@{\extracolsep{0in}}@{.}l@{ $\pm$ }l@{ $\pm$ }l@{\extracolsep{\fill}}}}
  \hline\hline
		& \multicolumn{4}{c}{Pretag} & \multicolumn{4}{c}{Tagged}\\
\hline
Acceptance (\%)				& 0&84 & 0.03 & 0.0	& 0&84 & 0.03 & 0.0	\\
Event Tagging Efficiency 		& \multicolumn{4}{c}{-}	& 0&67 & 0.0  & 0.04	\\
Background from Fake Leptons  (events)	& 29&9 & 5.9  & 0.0	& 5&2  & 1.2  & 0.05	\\
Other Pretag Backgrounds (events) 	& 24&0 & 0.0  & 3.1     & \multicolumn{4}{c}{-}	   \\
Other Tagged Backgrounds (events)	& \multicolumn{4}{c}{-}	& 4&4  & 0.0  & 2.6    \\
\hline\hline
  \end{tabular*}
  \caption{Likelihood inputs for the pretag and tagged samples, with systematic
  uncertainties.  Uncertainties are shown in the form
  (number) $\pm$ (correlated uncertainty) $\pm$ (uncorrelated uncertainty).
  The fake lepton background for tagged events includes only the leading $t\bar{t}$ lepton~+~jets 
  contribution, because the $W$~+~jets contribution is included in the tag-matrix-based
  background calculation, which is uncorrelated with the pretag background estimate.
  \label{tab:correlations}}
\end{table*}

Some of the background calculations depend on the $t\bar{t}$ cross section, 
the quantity we wish to measure.  In the fitting procedure, the number of predicted
events is calculated as a function of the cross section.  In addition to 
recalculating the number of expected signal events, we also recalculate the
number of background events for the cross section at each point in the fit.  
This removes any dependence of the measured cross section on the expected
value and allows the statistical uncertainties to be correctly calculated.

The seven parameters is allowed to float, and we find the combination
that maximizes the likelihood.  The cross section at the maximum
is our result.
To calculate the uncertainty on the combined cross section, we find the points
above and below the maximum value of the likelihood function at which the
logarithm of the likelihood function has decreased by 0.5.  

\begin{figure}[tp]
  \centering
  \includegraphics[width=0.5\textwidth]{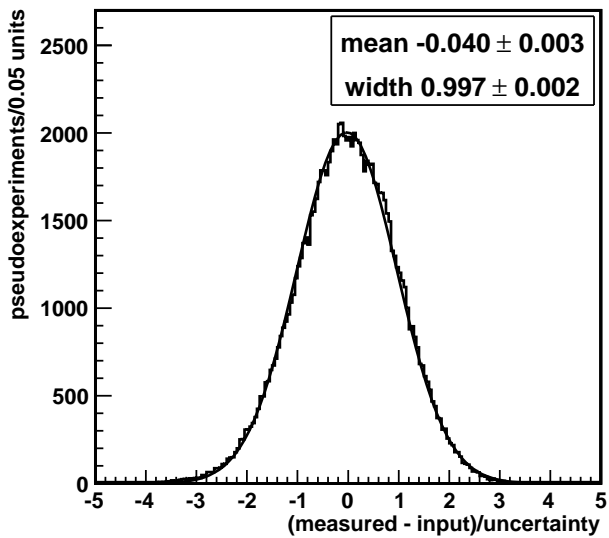}
  \caption{\label{fig:combined_pull} Pull distribution from pseudoexperiments run
    with the likelihood used to find the combined cross section.
	}
\end{figure}

To estimate the
expected improvement in precision of the combined cross section over the two
single measurements we perform pseudoexperiments with an input cross section 
of 6.7~pb. We find an expected improvement in
precision of 15\%, from 21\% to 18\%.  The pull distribution from these pseudoexperiments
is shown in Fig.~\ref{fig:combined_pull}.  The pull distribution width 
is one within the uncertainties,
demonstrating that the experimental uncertainties are correctly estimated.  
The slight bias in the mean
is due to the fact that the number of candidates is restricted to integer values.
This limits the possible values of the measured cross section for samples with small expected
numbers of events.  

\subsection{Pretag Sample}

Using the event selection described in Section~\ref{sec:evtsel}, we find
129 
pretag candidate events in the data, which has an acceptance-weighted integrated luminosity of
1070~$\pm$~60~\pb.
The background is calculated to be
53.8~$\pm$~6.7
events and the summed acceptance times efficiency is 
0.84~$\pm$~0.03\%. 
Using these, we calculate the cross section using the likelihood fit described earlier in this section.
Assuming \mbox{$m_{t} = 175$ \GeVcc}\
and \mbox{$BR(W\rightarrow \ell \nu) = 10.8$~\%}, we find
\[
\sigma_{t\overline t} = 8.3 \pm 1.3(\textrm{stat.}) \pm 0.8 (\textrm{sys.}) \pm 0.5 (\textrm{lum.})\ \textrm{pb ,}
\]
consistent with the standard model prediction of
\mbox{$6.6^{+0.3}_{-0.5}$ (scale) $^{+0.4}_{-0.3}$ (PDF) pb~\cite{ttbar_new}}.

\begin{figure}[tp]
  \centering
  \includegraphics[width=0.5\textwidth]{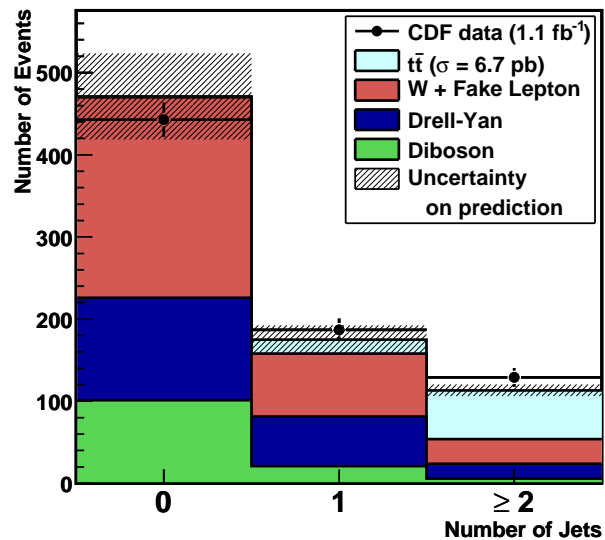}
  \caption{\label{fig:pretag_njet} Number of predicted pretag lepton~+~track events compared
	to the number observed in the CDF data.  The cross-hatched areas show
	the combined statistical and systematic uncertainties (one standard deviation) 
	on the prediction.}
\end{figure}

For the pretag sample, the signal and background predictions are summarized and compared
to the observed number of candidate events, for events with zero, one, and two or more jets,
in Table \ref{tab:pretagsummary}.  The zero and one jet event comparisons test the 
background predictions, because the contribution from $t\bar{t}$ in these jet multiplicities 
is very small.  The number of events predicted and observed agree for all jet multiplicities, 
although it should be noted
that the zero jet events are not as strong of a cross-check, since a subset of these is used 
to derive the \mbox{opposite-sign} fraction for $W$~+~jets events with a fake lepton 
(see Section \ref{sec:fakes}).  Figure \ref{fig:pretag_njet} is a visual 
representation of Table \ref{tab:pretagsummary}.  In both the table and the figure, the signal prediction
is shown at the theoretical cross section value of 6.7~pb, but the backgrounds which depend
on the $t\bar{t}$ cross section are calculated at the measured value of 8.3~pb.
\begin{table*}[tbp]
  \centering
  \begin{tabular*}{14cm}{c*{3}{@{\extracolsep{\fill}}r@{\extracolsep{0in}}@{ $\pm$ }r@{.}l}}
   \hline\hline
	& \multicolumn{3}{c}{0 jets} & \multicolumn{3}{c}{1 jet} & \multicolumn{3}{c}{$\geq 2$ jets}	\\ 
  \hline
$WW$	& 85.8 & 8&7	& 14.9 & 1&5	& 3.7 & 0&4	\\
$WZ$	&  9.3 & 1&0	&  4.3 & 0&5	& 1.3 & 0&2	\\
$ZZ$	&  6.0 & 0&6	&  1.6 & 0&2	& 0.8 & 0&1	\\
\hline
\Zee	& 71.3 & 15&7	& 25.5 & 6&0	& 7.6 & 2&2	\\
\Zmm	& 17.9 & 5&2 	&  8.4 & 2&7	& 3.2 & 1&1	\\
\Ztt	& 35.5 & 3&2	& 26.5 & 2&5	& 7.3 & 0&9	\\
\hline
Fakes	& 244.1 & 46&4	& 76.8 & 14&6	& 29.9 & 5&9	\\
\hline
All Backgrounds	& 469.9 & 52&5	& 157.9 & 17&2	& 53.8 & 6&7	\\
\hline
$t\bar{t}, \sigma=6.7$ pb	& 1.2 & 0&1	& 17.3 & 0&6	& 60.3 & 1&9	\\
\hline
Predicted	& 471.1 & 52&5	& 175.2 & 17&3	& 114.2 & 7&1	\\
Observed	& \multicolumn{3}{c}{443}	& \multicolumn{3}{c}{187}	& \multicolumn{3}{c}{129}	\\
   \hline\hline
  \end{tabular*}
  \caption{Predicted and observed pretag events in 1.1 fb$^{-1}$, with details of the background 
    contributions.  Systematic uncertainties on the predictions are included.}	
  \label{tab:pretagsummary}
\end{table*}

The cross section calculated at other values of $m_{t}$ is shown in Table~\ref{tab:pretag_bymass}.
The measured cross section decreases with increasing $m_t$ even though the number
of observed events is unchanged.  For higher top quark masses $t\bar{t}$ decay products 
are more energetic and therefore more likely to pass the kinematic selection, increasing 
the acceptance.  The background estimates also depend weakly on the top quark mass 
through the use of simulated $t\bar{t}$ events to calculate the prevalence of $t\bar{t}$
in control data samples and the background from $t\bar{t}$ lepton~+~jets events with a 
fake lepton.

\begin{figure}[tp]
  \centering
  \includegraphics[width=0.5\textwidth]{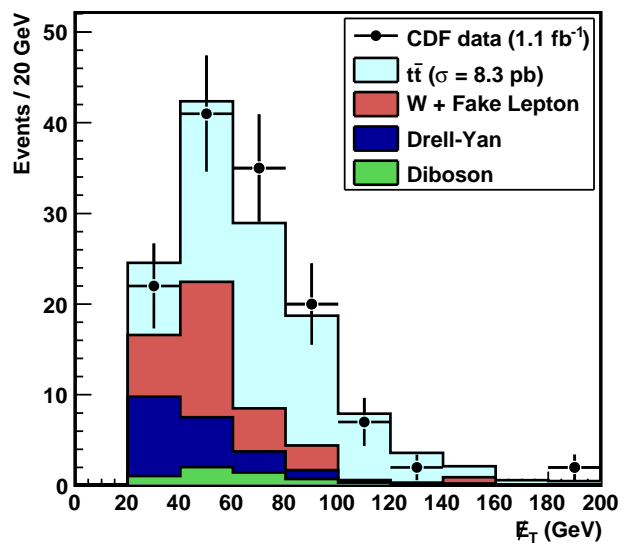}
  \caption{\label{fig:pretag_met} Missing transverse energy of 
      pretag lepton~+~track candidate events with two or more jets, compared to 
      the predicted distribution.
      The highest bin shown includes all events which would be past the right edge of the plot.
	}
\end{figure}

\begin{figure}[tp]
  \centering
  \includegraphics[width=0.5\textwidth]{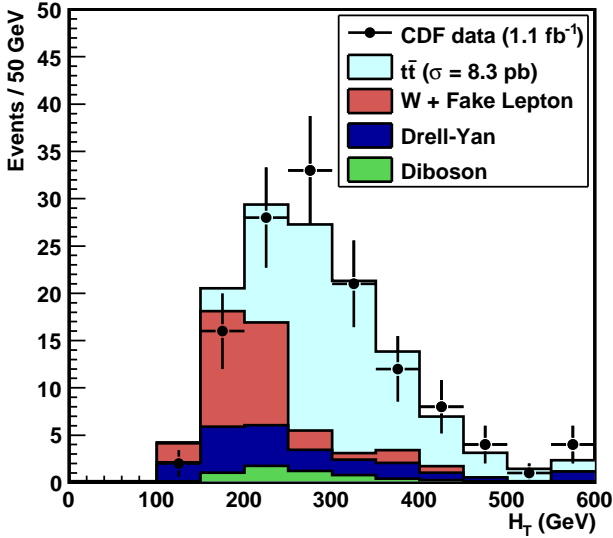}
  \caption{\label{fig:pretag_ht} Summed transverse energy ($H_T$) of
      pretag lepton~+~track candidate events with two or more jets, compared to 
      the predicted distribution.  The sum includes the fully reconstructed lepton,
	  the track lepton, the \met, and all jets passing the analysis selection.
      The highest bin shown includes all events which would be past the right edge of the plot.
	}
\end{figure}

\begin{figure}[tp]
  \centering
  \includegraphics[width=0.5\textwidth]{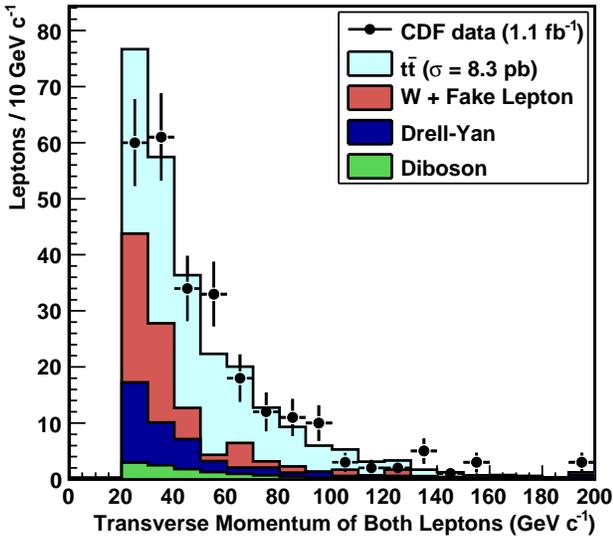}
  \caption{\label{fig:pretag_pt} Distribution of lepton transverse momenta in
    pretag lepton~+~track candidate events with two or more jets, compared to
    the predicted distribution.  There are two entries for each event; one each
    for the fully reconstructed lepton and the isolated track.  For fully reconstructed
    electrons, the $E_T$ is used to estimate the $p_T$.
      The highest bin shown includes all events which would be past the right edge of the plot.
	}
\end{figure}

\begin{figure}[tp]
  \centering
  \includegraphics[width=0.5\textwidth]{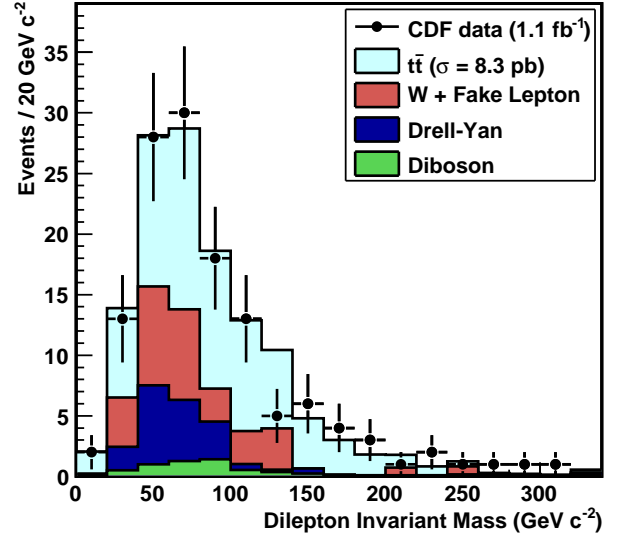}
  \caption{\label{fig:pretag_mll} Reconstructed invariant mass of the fully reconstructed
    lepton and isolated track pair in pretag candidate events with two or more jets,
    compared to the predicted distribution.  
      The highest bin shown includes all events which would be past the right edge of the plot.
	}
\end{figure}

We compare the kinematic features of the observed pretag lepton~+~track candidates to 
the expected distributions.  In each of these figures, the $t\bar{t}$ contribution is 
normalized to the measured cross section, so only the shapes of the distributions are 
to be compared, not the normalization.  
One of the most prominent features of dilepton $t\bar{t}$ events in particular is the \met\
from the two neutrinos.  Figure \ref{fig:pretag_met} compares the \met\ spectrum
of candidate events to the summed spectra expected for signal and background as
predicted by simulation.  
Because the top quark is so massive, the $H_T$ distribution of $t\bar{t}$ events,
defined as the scalar sum of the primary lepton $E_T$, track
lepton $p_T$, \met, and the $E_T$ of all jets in the event, is also distinctive.
The $H_T$ distribution, shown in Fig.~\ref{fig:pretag_ht}, is skewed toward
higher values for the $t\bar{t}$ signal than for its backgrounds.
Turning to the charged leptons in the event, we show the
$p_T$ of the fully reconstructed lepton and the track lepton in Fig.~\ref{fig:pretag_pt},
and their invariant mass in Fig.~\ref{fig:pretag_mll}.  Fig.~\ref{fig:pretag_pt}
is useful for comparing the signal to the background contribution from events with a fake 
lepton, since the latter produces a much softer lepton $p_T$ distribution, due to the
exponentially falling jet $p_T$ spectrum.  Similarly, Fig.~\ref{fig:pretag_mll} is
useful for comparing the signal to the background from Drell-Yan events, since the invariant
mass distribution is more peaked for this background than for the signal.
The agreement between the predicted and observed distributions suggests that the 
content of the candidate sample is well-understood within the uncertainties.


\subsection{Tagged Sample}

In the tagged sample, we find 69 candidate events, and measure 9.5~$\pm$~2.8 background
events. The predicted and observed numbers of events in the sample, with the prediction
divided by source, are shown in Table~\ref{tab:tagsummary}.  
The inputs to the tagged cross section calculation
are summarized in Table~\ref{tab:tagresults}. Using data with an
integrated luminosity of 1000~$\pm$~60~\pb, and assuming \mbox{$m_{t} = 175$ \GeVcc}\ and
\mbox{$BR(W\rightarrow \ell \nu) = 10.8$~\%}  we find: 
\[
\sigma_{t\overline t} = 10.5\,^{+1.4}_{-1.3}\;(\textrm{stat.})\;^{+0.8}_{-0.7}\;(\textrm{sys.}) \pm 0.6 (\textrm{lum.})\ \textrm{pb .}
\]

The tagged cross section as a function of assumed top quark mass is included in 
Table~\ref{tab:pretag_bymass}.
Similar to the pretag measurement, the background estimate is relatively insensitive to the value 
of the top quark mass.  The event tagging efficiency, for top quark masses between \mbox{170 and
180~\GeVcc,} is consistent with that measured at \mbox{$m_t$ = 175~\GeVcc}.  Therefore, the only change
to the measurement as a function of the top quark mass is the acceptance for $t\bar{t}$ events.

The tagged cross section has a combined lower uncertainty of 1.6~pb, including the
uncertainty on the luminosity.  This translates to an excess above the standard
model prediction of 2.4~in units of the calculated uncertainty on the measurement.  
However, the tagged measurement is consistent with the pretag measurement.
The two measurements differ by the observed 2.2~pb in about 10\% of pseudoexperiments, where
the exact fraction depends on the assumed true cross section.  The kinematic features of
the observed candidates are also consistent with the standard model expectation, which for 
the tagged sample are predominantly $t\bar{t}$.

\begin{figure}[tp]
  \centering
  \includegraphics[width=0.5\textwidth]{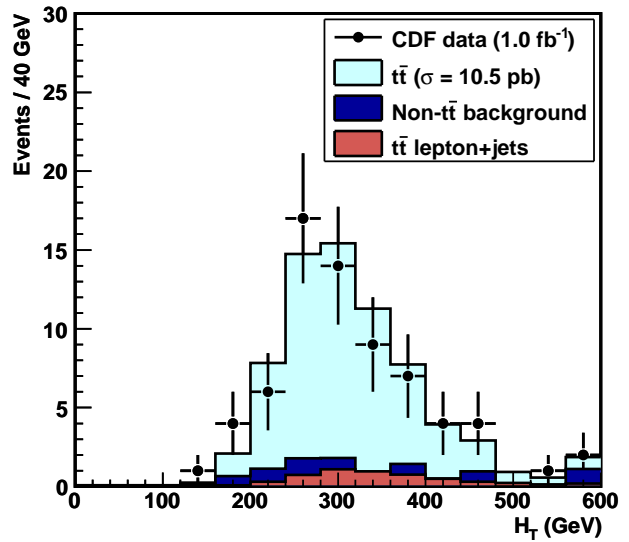}
  \caption{\label{fig:tag_Ht} Summed scalar energy ($H_T$) of the fully reconstructed lepton, the
    isolated track, the \met\ and all jets in the tagged candidate sample, compared
    to the combined signal and background predictions.
	}
\end{figure}

\begin{figure}[tp]
  \centering
  \includegraphics[width=0.5\textwidth]{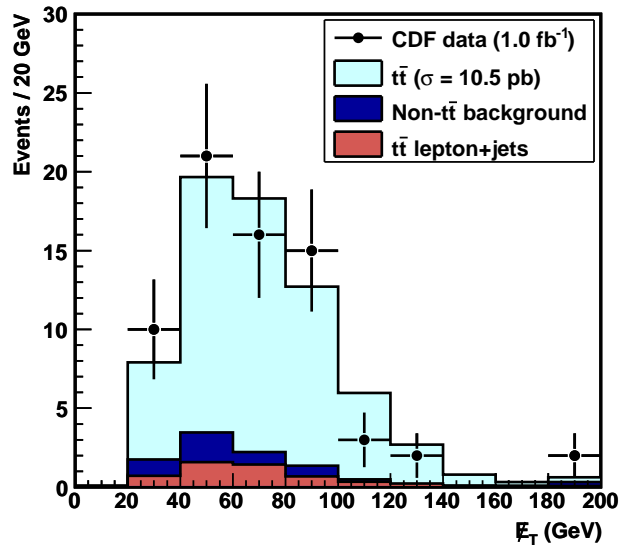}
  \caption{\label{fig:tag_Met} Missing transverse energy of the lepton~+~track tagged 
    candidate sample, compared to the combined signal and background predictions.  
	}
\end{figure}

\begin{figure}[tp]
  \centering
  \includegraphics[width=0.5\textwidth]{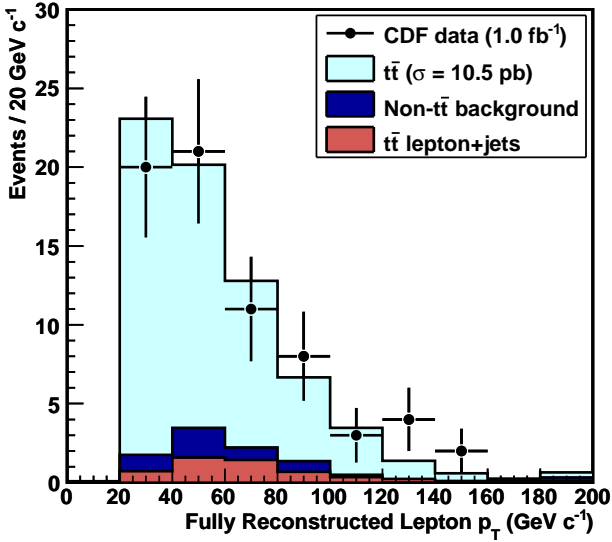}
  \caption{\label{fig:tag_TCL_et} Transverse momentum of the fully reconstructed lepton
    in the tagged lepton~+~track candidate sample, compared to the combined signal and 
    background predictions.  The $E_T$ is used as an estimate of the 
    lepton $p_T$ for fully reconstructed electron candidates.
	}
\end{figure}

\begin{figure}[tp]
  \centering
  \includegraphics[width=0.5\textwidth]{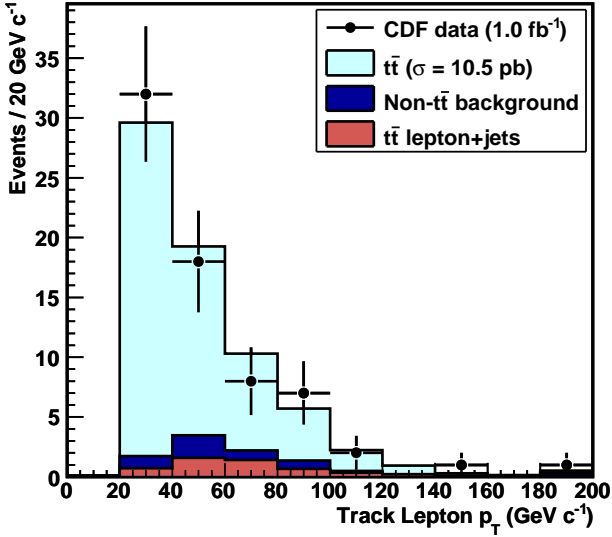}
  \caption{\label{fig:tag_TL_pt} Transverse momentum of the track lepton (isolated track)
    in the tagged lepton~+~track candidate sample, compared to the combined signal and 
    background predictions.  
	}
\end{figure}

\begin{figure}[tp]
  \centering
  \includegraphics[width=0.5\textwidth]{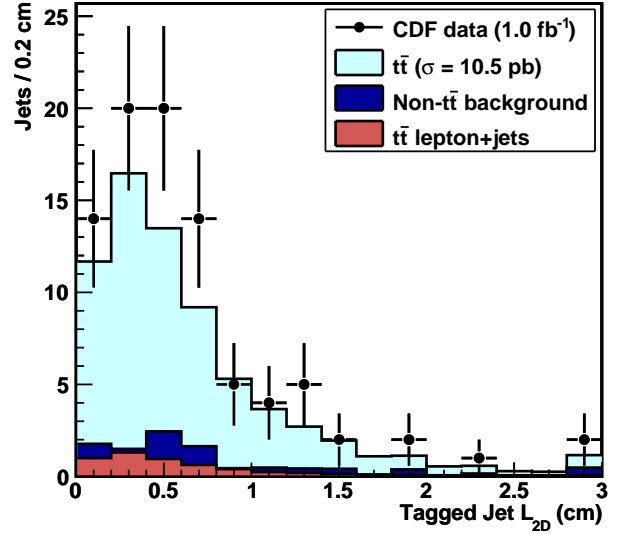}
  \caption{\label{fig:tag_L2D} Distance along the jet axis to the reconstructed 
    secondary vertex in tagged jets in the lepton~+~track candidate sample,
    compared to the combined signal and background predictions.  
	}
\end{figure}

Figures~\ref{fig:tag_Ht}-\ref{fig:tag_L2D} display some of the kinematic features of 
the tagged candidate sample, compared the expected combined signal and background contribution.  
In all of these figures, the signal is normalized to the measured cross section.
The $H_T$ and \met\ of the tagged candidate events are shown in Fig.~\ref{fig:tag_Ht}
and Fig.~\ref{fig:tag_Met}, respectively.  These both have distinctive distributions
for $t\bar{t}$ events; see the discussion above on pretag kinematic distributions for
details.  Fig.~\ref{fig:tag_TCL_et} and~\ref{fig:tag_TL_pt} show the transverse
momentum distributions of the fully reconstructed leptons and the isolated tracks.
Fig.~\ref{fig:tag_L2D} shows a unique feature of the tagged events, the distance
along the tagged jet axis from the interaction point to the reconstructed secondary
vertex, which corresponds to the distance traveled by the $b$ hadron before decaying.
In all of the figures, the last bin on the right includes all events which would be past
the right edge of the plot.

In all of the distributions, good agreement is observed between the candidate events 
from data and the expected distributions.
Comparing the distributions from the tagged sample to the ones from the pretag sample,
the improvement in sample purity from the $b$-tag requirement is evident.  The 
agreement between the predicted and observed distributions shows that although 
the measured cross section is on the high side, the observed candidates are 
consistent with the expected $t\bar{t}$ signature.

\begin{table}[tbp]
  \centering
  \begin{tabular*}{8.6cm}{@{\extracolsep{\fill}}cr@{\extracolsep{0in}}@{ $\pm$ }l}
  \hline\hline
Source			& \multicolumn{2}{c}{Number of events} \\
\hline
From Matrix (e.g. $Z+b\bar{b}$) & ~~~~ 4.4 & 2.6\\
Fakes	                    	& ~~~~ 5.2 & 1.2 \\
All Backgrounds	           	& ~~~~ 9.5 & 2.8 \\
$t\bar{t}, \sigma=6.7$ pb   	& ~~~~37.7 & 2.4 \\
\hline
Predicted	            & ~~~~47.3 & 3.7 \\
Observed	            & \multicolumn{2}{c}{69} \\
\hline\hline
  \end{tabular*}
  \caption{Predicted and observed events with two or more jets, at least one of which 
    is tagged, in 1.0 fb$^{-1}$, with details of the background contributions.}
  \label{tab:tagsummary}
\end{table}

\begin{table}[tp]
  \begin{center}
  \begin{tabular*}{8.6cm}{@{\hspace{1cm}}cr@{.}l@{ $\pm$ }l}
        \hline\hline
        Input 					& \multicolumn{3}{c}{\hspace{2cm}Value} \\
        \hline
        $N^\txtss{tag}$ 			& \multicolumn{3}{c}{\hspace{2cm}69} \\
        $N^\txtss{tag}_\txtss{bkgd}$ 		& \hspace{2.5cm} 9&5   & 2.8  \\
        $\mathcal{A} \times \int\mathscr{L}dt $ & \hspace{2.5cm} 8&4   & 0.03 \pb \\
        $\epsilon_\txtss{tag}$ 			& \hspace{2.5cm} 0&669 & 0.037 \\
        \hline\hline
    \end{tabular*}
    \caption{Predicted background and observed events in 1.0 fb$^{-1}$, with inputs to the cross 
      section calculation for the
      tagged analysis. Systematic uncertainties are included in the prediction numbers.}
    \label{tab:tagresults}
  \end{center}
\end{table}


\subsection{Combined Cross Section Results}\label{sec:results_combo}

\begin{figure}[tp]
  \centering
  \includegraphics[width=0.5\textwidth]{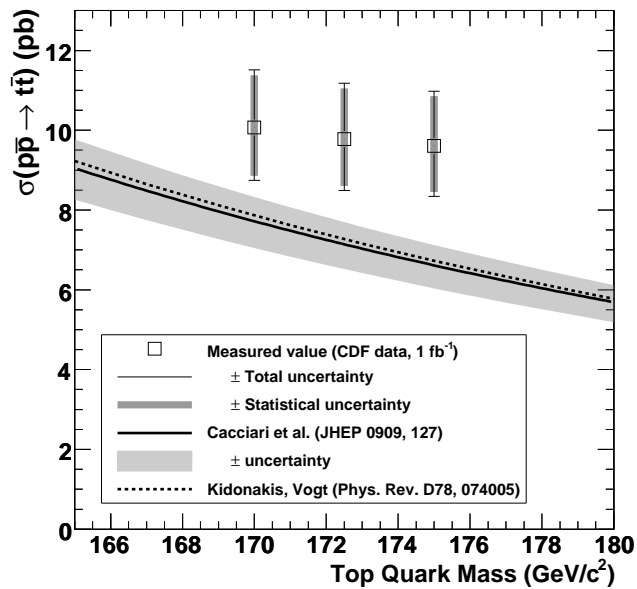}
  \caption{\label{fig:xsec_v_mass} Measured cross section as a function of the 
    assumed top quark mass, compared to theoretical predictions for the cross section.
	}
\end{figure}

Using the likelihood fitter, we find a combined cross section of
\[
\sigma_{t\bar{t}} = 9.6_{-1.3}^{+1.4}(\textrm{stat+sys})\pm0.6(\textrm{lum})\ \textrm{pb }
\]
or
\[
\sigma_{t\bar{t}} = 9.6\pm1.2(\textrm{stat})_{-0.5}^{+0.6}(\textrm{sys})\pm0.6(\textrm{lum})\ \textrm{pb.}
\]
The 14\% combined statistical and systematic uncertainty is an improvement in 
precision on either of the individual measurements.
The combined cross section is also shown as a function of the assumed top quark 
mass in Table~\ref{tab:pretag_bymass} and Fig.~\ref{fig:xsec_v_mass}.  In both,
the measured cross section is compared to the theoretical prediction.
Both the predicted and measured cross section depend on the top quark mass, but
the dependence is stronger for the predicted cross section.  The predicted
cross section drops off with increasing top quark mass because of the
increased collision energy needed to exceed the kinematic threshold for pair
production of top quarks.  The measured cross section depends on the assumed
mass more weakly, through the increased acceptance at higher top quark masses
because of the increased average transverse momentum of the decay products.

To facilitate comparison of this result with other measurements and calculations, 
which may be performed at different assumed top quark masses, we fit the combined 
cross section results shown in Table~\ref{tab:pretag_bymass}
to the functional form
\[
\sigma_{t\bar{t}}(m_t) = A + B(m_t-175) + C(M_t-175)^2\ \textrm{,}
\]
in the spirit of Ref.~\cite{ttbar_new}.  The top quark mass $m_t$ is in \GeVcc, 
and the fit yields the coefficients 
$A = 9.6$~pb, $B = 4.4 \times 10^{-2}$~pb/(\GeVcc), and 
$C = 9.6 \times 10^{-3}$~pb/(\GeVcc)$^2$.  

We compare the measured cross section to the standard model prediction at 
the current world average measurement, 173.1~\GeVcc~\cite{topmass}.  
Using the fit described in the previous paragraph, we measure 
$9.7^{+1.5}_{-1.4}$~pb, where the uncertainty includes the statistical, systematic,
and luminosity uncertainties.  The prediction for this top quark mass from
Ref.~\cite{ttbar_new} is $7.0^{+0.5}_{-0.6}$~pb.  
The uncertainty on the difference between the two, 1.5~pb, includes the 
uncertainty on the measurement and the theoretical prediction.  The significance of the 
the difference is therefore \mbox{$2.7\ \textrm{pb} / 1.5\ \textrm{pb} = 1.8$}.		
Comparison of the kinematics of the pretag and tagged candidate samples
to the standard model expectation shows that the content of the sample is reasonably well-understood.
Also, the measurement agrees with the D\O\ measurement in the dilepton channel,
\mbox{$7.4 \pm 1.4$ (stat.) $\pm 0.9$ (sys.) $\pm 0.5$ (lum.) pb \cite{d0_run2_dil_prd}}, 
as well as with the published measurements from other channels
cited in the Introduction, most of which are also above the predicted theoretical value.

\begin{table*}[tbp]
\centering
\begin{tabular*}{14cm}{@{\extracolsep{\fill}}ccc*{3}{r@{\extracolsep{0in}}@{.}l@{\extracolsep{\fill}}}}
\hline\hline
Input 		& Theoretical 	& Pretag		& \multicolumn{6}{c}{Measured cross section (pb)}	\\
$m_t$ (\GeVcc)	& $\sigma$ (pb)	& $t\bar{t}$ Acceptance & \multicolumn{2}{c}{Pretag} & \multicolumn{2}{c}{Tagged} & \multicolumn{2}{c}{Combined}  \\
\hline
\rule[-1.5ex]{0pt}{4.5ex}170	& $7.7_{-0.7}^{+0.6}$	& 0.80 $\pm$ 0.02 \%	
	&  8&8 $_{-1.6}^{+1.7}$
	& 11&0 $_{-1.5}^{+1.7}$
	& 10&1 $_{-1.3}^{+1.4}$	\\
\rule[-1.5ex]{0pt}{4.5ex}172.5	& $7.1_{-0.6}^{+0.6}$	& 0.83 $\pm$ 0.02 \%	
	&  8&5 $_{-1.5}^{+1.7}$
	& 10&7 $_{-1.5}^{+1.7}$
	&  9&8 $_{-1.3}^{+1.4}$	\\
\rule[-1.5ex]{0pt}{4.5ex}175	& $6.6_{-0.6}^{+0.5}$	& 0.84 $\pm$ 0.03 \%	
	&  8&3 $_{-1.5}^{+1.6}$
	& 10&5 $_{-1.5}^{+1.6}$
	&  9&6 $_{-1.3}^{+1.4}$	\\
\hline\hline
\end{tabular*}
\caption{\label{tab:pretag_bymass} 
 	 The pretag and tagged cross section as calculated at several input top masses.
	   Theoretical prediction from Ref.~\cite{ttbar_new}.  
	The statistical and systematic uncertainties are combined, and a common uncertainty 
	of 6\%, due to the uncertainty on the integrated luminosity, is omitted.}
\end{table*} 


\section{Conclusion}

We have measured the $t\bar{t}$ production cross section in the dilepton 
channel using events selected with one fully reconstructed lepton and one
isolated track, both with and without the requirement that at least one
jet in the event be tagged as a $b$.  The combined result of these measurements is
\[
\sigma_{t\bar{t}} = 9.6\pm1.2(\textrm{stat})_{-0.5}^{+0.6}(\textrm{sys})\pm0.6(\textrm{lum})\ \textrm{pb}
\]
for a top quark mass of 175~\GeVcc.
This is the first dilepton cross section result from CDF which uses $b$-tagging 
information.  We have also improved the
estimation of the pretag backgrounds with respect to the previous publication, 
particularly for the background from events in which a jet has been 
reconstructed as an isolated track.  These changes, combined with the 
integration of more data, result in a more precise measurement of the cross 
section in the dilepton channel compared to other published results.  

The cross sections measured are high compared to the standard model 
prediction, but the consistency between the tagged and pretag measurements, their 
agreement with other published measurements, and the consistency of the candidate
event kinematics with the standard model all support the hypothesis that the high 
cross sections observed are consistent with an upward fluctuation
in the number of $t\bar{t}$ events accepted by the lepton~+~track selection.

\begin{acknowledgments}

We thank the Fermilab staff and the technical staffs of the participating institutions for their vital contributions. This work was supported by the U.S. Department of Energy and National Science
Foundation; the Italian Istituto Nazionale di Fisica Nucleare; the Ministry of Education, Culture, Sports, Science and Technology of Japan; the Natural Sciences and Engineering Research Council of
Canada; the National Science Council of the Republic of China; the Swiss National Science Foundation; the A.P. Sloan Foundation; the Bundesministerium f\"ur Bildung und Forschung, Germany; the Korean
Science and Engineering Foundation and the Korean Research Foundation; the Science and Technology Facilities Council and the Royal Society, UK; the Institut National de Physique Nucleaire et Physique
des Particules/CNRS; the Russian Foundation for Basic Research; the Ministerio de Ciencia e Innovaci\'{o}n, and Programa Consolider-Ingenio 2010, Spain; the Slovak R\&D Agency; and the Academy of
Finland. 

\end{acknowledgments}

\bibliography{ltrk_prd}

\end{document}

%% file: July_2008_Authors_Visitors.tex
\affiliation{Institute of Physics, Academia Sinica, Taipei, Taiwan 11529, Republic of China} 
\affiliation{Argonne National Laboratory, Argonne, Illinois 60439} 
\affiliation{University of Athens, 157 71 Athens, Greece} 
\affiliation{Institut de Fisica d'Altes Energies, Universitat Autonoma de Barcelona, E-08193, Bellaterra (Barcelona), Spain} 
\affiliation{Baylor University, Waco, Texas  76798} 
\affiliation{Istituto Nazionale di Fisica Nucleare Bologna, $^y$University of Bologna, I-40127 Bologna, Italy} 
\affiliation{Brandeis University, Waltham, Massachusetts 02254} 
\affiliation{University of California, Davis, Davis, California  95616} 
\affiliation{University of California, Los Angeles, Los Angeles, California  90024} 
\affiliation{University of California, San Diego, La Jolla, California  92093} 
\affiliation{University of California, Santa Barbara, Santa Barbara, California 93106} 
\affiliation{Instituto de Fisica de Cantabria, CSIC-University of Cantabria, 39005 Santander, Spain} 
\affiliation{Carnegie Mellon University, Pittsburgh, PA  15213} 
\affiliation{Enrico Fermi Institute, University of Chicago, Chicago, Illinois 60637}
\affiliation{Comenius University, 842 48 Bratislava, Slovakia; Institute of Experimental Physics, 040 01 Kosice, Slovakia} 
\affiliation{Joint Institute for Nuclear Research, RU-141980 Dubna, Russia} 
\affiliation{Duke University, Durham, North Carolina  27708} 
\affiliation{Fermi National Accelerator Laboratory, Batavia, Illinois 60510} 
\affiliation{University of Florida, Gainesville, Florida  32611} 
\affiliation{Laboratori Nazionali di Frascati, Istituto Nazionale di Fisica Nucleare, I-00044 Frascati, Italy} 
\affiliation{University of Geneva, CH-1211 Geneva 4, Switzerland} 
\affiliation{Glasgow University, Glasgow G12 8QQ, United Kingdom} 
\affiliation{Harvard University, Cambridge, Massachusetts 02138} 
\affiliation{Division of High Energy Physics, Department of Physics, University of Helsinki and Helsinki Institute of Physics, FIN-00014, Helsinki, Finland} 
\affiliation{University of Illinois, Urbana, Illinois 61801} 
\affiliation{The Johns Hopkins University, Baltimore, Maryland 21218} 
\affiliation{Institut f\"{u}r Experimentelle Kernphysik, Universit\"{a}t Karlsruhe, 76128 Karlsruhe, Germany} 
\affiliation{Center for High Energy Physics: Kyungpook National University, Daegu 702-701, Korea; Seoul National University, Seoul 151-742, Korea; Sungkyunkwan University, Suwon 440-746, Korea; Korea Institute of Science and Technology Information, Daejeon, 305-806, Korea; Chonnam National University, Gwangju, 500-757, Korea} 
\affiliation{Ernest Orlando Lawrence Berkeley National Laboratory, Berkeley, California 94720} 
\affiliation{University of Liverpool, Liverpool L69 7ZE, United Kingdom} 
\affiliation{University College London, London WC1E 6BT, United Kingdom} 
\affiliation{Centro de Investigaciones Energeticas Medioambientales y Tecnologicas, E-28040 Madrid, Spain} 
\affiliation{Massachusetts Institute of Technology, Cambridge, Massachusetts  02139} 
\affiliation{Institute of Particle Physics: McGill University, Montr\'{e}al, Qu\'{e}bec, Canada H3A~2T8; Simon Fraser University, Burnaby, British Columbia, Canada V5A~1S6; University of Toronto, Toronto, Ontario, Canada M5S~1A7; and TRIUMF, Vancouver, British Columbia, Canada V6T~2A3} 
\affiliation{University of Michigan, Ann Arbor, Michigan 48109} 
\affiliation{Michigan State University, East Lansing, Michigan  48824}
\affiliation{Institution for Theoretical and Experimental Physics, ITEP, Moscow 117259, Russia} 
\affiliation{University of New Mexico, Albuquerque, New Mexico 87131} 
\affiliation{Northwestern University, Evanston, Illinois  60208} 
\affiliation{The Ohio State University, Columbus, Ohio  43210} 
\affiliation{Okayama University, Okayama 700-8530, Japan} 
\affiliation{Osaka City University, Osaka 588, Japan} 
\affiliation{University of Oxford, Oxford OX1 3RH, United Kingdom} 
\affiliation{Istituto Nazionale di Fisica Nucleare, Sezione di Padova-Trento, $^z$University of Padova, I-35131 Padova, Italy} 
\affiliation{LPNHE, Universite Pierre et Marie Curie/IN2P3-CNRS, UMR7585, Paris, F-75252 France} 
\affiliation{University of Pennsylvania, Philadelphia, Pennsylvania 19104}
\affiliation{Istituto Nazionale di Fisica Nucleare Pisa, $^{aa}$University of Pisa, $^{bb}$University of Siena and $^{cc}$Scuola Normale Superiore, I-56127 Pisa, Italy} 
\affiliation{University of Pittsburgh, Pittsburgh, Pennsylvania 15260} 
\affiliation{Purdue University, West Lafayette, Indiana 47907} 
\affiliation{University of Rochester, Rochester, New York 14627} 
\affiliation{The Rockefeller University, New York, New York 10021} 
\affiliation{Istituto Nazionale di Fisica Nucleare, Sezione di Roma 1, $^{dd}$Sapienza Universit\`{a} di Roma, I-00185 Roma, Italy} 

\affiliation{Rutgers University, Piscataway, New Jersey 08855} 
\affiliation{Texas A\&M University, College Station, Texas 77843} 
\affiliation{Istituto Nazionale di Fisica Nucleare Trieste/Udine, I-34100 Trieste, $^{ee}$University of Trieste/Udine, I-33100 Udine, Italy} 
\affiliation{University of Tsukuba, Tsukuba, Ibaraki 305, Japan} 
\affiliation{Tufts University, Medford, Massachusetts 02155} 
\affiliation{Waseda University, Tokyo 169, Japan} 
\affiliation{Wayne State University, Detroit, Michigan  48201} 
\affiliation{University of Wisconsin, Madison, Wisconsin 53706} 
\affiliation{Yale University, New Haven, Connecticut 06520} 
\author{T.~Aaltonen}
\affiliation{Division of High Energy Physics, Department of Physics, University of Helsinki and Helsinki Institute of Physics, FIN-00014, Helsinki, Finland}
\author{J.~Adelman}
\affiliation{Enrico Fermi Institute, University of Chicago, Chicago, Illinois 60637}
\author{T.~Akimoto}
\affiliation{University of Tsukuba, Tsukuba, Ibaraki 305, Japan}
\author{B.~\'{A}lvarez~Gonz\'{a}lez$^t$}
\affiliation{Instituto de Fisica de Cantabria, CSIC-University of Cantabria, 39005 Santander, Spain}
\author{S.~Amerio$^z$}
\affiliation{Istituto Nazionale di Fisica Nucleare, Sezione di Padova-Trento, $^z$University of Padova, I-35131 Padova, Italy} 

\author{D.~Amidei}
\affiliation{University of Michigan, Ann Arbor, Michigan 48109}
\author{A.~Anastassov}
\affiliation{Northwestern University, Evanston, Illinois  60208}
\author{A.~Annovi}
\affiliation{Laboratori Nazionali di Frascati, Istituto Nazionale di Fisica Nucleare, I-00044 Frascati, Italy}
\author{J.~Antos}
\affiliation{Comenius University, 842 48 Bratislava, Slovakia; Institute of Experimental Physics, 040 01 Kosice, Slovakia}
\author{G.~Apollinari}
\affiliation{Fermi National Accelerator Laboratory, Batavia, Illinois 60510}
% ...

\collaboration{CDF Collaboration\footnote{With visitors from $^a$University of Massachusetts Amherst, Amherst, Massachusetts 01003,
$^b$Universiteit Antwerpen, B-2610 Antwerp, Belgium, 
$^c$University of Bristol, Bristol BS8 1TL, United Kingdom,
$^d$Chinese Academy of Sciences, Beijing 100864, China, 
$^e$Istituto Nazionale di Fisica Nucleare, Sezione di Cagliari, 09042 Monserrato (Cagliari), Italy,
$^f$University of California Irvine, Irvine, CA  92697, 
$^g$University of California Santa Cruz, Santa Cruz, CA  95064, 
$^h$Cornell University, Ithaca, NY  14853, 
$^i$University of Cyprus, Nicosia CY-1678, Cyprus, 
$^j$University College Dublin, Dublin 4, Ireland,
$^k$University of Edinburgh, Edinburgh EH9 3JZ, United Kingdom, 
$^l$University of Fukui, Fukui City, Fukui Prefecture, Japan 910-0017
$^m$Kinki University, Higashi-Osaka City, Japan 577-8502
$^n$Universidad Iberoamericana, Mexico D.F., Mexico,
$^o$University of Iowa, Iowa City, IA  52242,
$^p$Queen Mary, University of London, London, E1 4NS, England,
$^q$University of Manchester, Manchester M13 9PL, England, 
$^r$Nagasaki Institute of Applied Science, Nagasaki, Japan, 
$^s$University of Notre Dame, Notre Dame, IN 46556,
$^t$University de Oviedo, E-33007 Oviedo, Spain, 
$^u$Texas Tech University, Lubbock, TX  79609, 
$^v$IFIC(CSIC-Universitat de Valencia), 46071 Valencia, Spain,
$^w$University of Virginia, Charlottesville, VA  22904,
$^x$Bergische Universit\"at Wuppertal, 42097 Wuppertal, Germany,
$^{ff}$On leave from J.~Stefan Institute, Ljubljana, Slovenia, 
}}
\noaffiliation